\RequirePackage[OT1]{fontenc}
\documentclass[journal, twoside]{IEEEtran}

\newcommand\secondaddition[1]{#1}

\newif\ifshowframe\showframefalse
\newif\ifexternalize\externalizefalse
\newif\iffinal\finaltrue
\newif\ifcompileplots\compileplotstrue
\compileplotsfalse

\usepackage{lipsum}

\newcommand\plotplaceholder{
    \begin{tikzpicture}
        \draw
            (0, 0) 
            coordinate (origin)
            rectangle 
            (0.98*\columnwidth, 0.98*\columnwidth/1.6) 
            coordinate (temp);
        \draw
            ($(origin)!0.5!(temp)$)
            node[align=center]{
                This is a placeholder image.\\
                Uncomment \texttt{compileplotstrue}\\
                close to line 6 on \texttt{main.tex}\\
                to see the actual plots\\
                (compilation slows down).
            };
    \end{tikzpicture}
}

\usepackage{textcomp}
\usepackage{siunitx}
    \DeclareSIUnit\voltampere{VA}
    \DeclareSIUnit\var{var}
    \DeclareSIUnit\pu{pu}
    \DeclareSIUnit\norm{norm.}
\usepackage{xspace}

\usepackage[dvipsnames]{xcolor}
\usepackage{graphicx}
\usepackage{pdfpages}
\usepackage{tikz}
    \usetikzlibrary{calc}
    \usetikzlibrary{shapes.misc}
    \usetikzlibrary{patterns}
    \usetikzlibrary{patterns.meta}
    \usetikzlibrary{decorations.text}
    \usetikzlibrary{decorations.markings}
    \usetikzlibrary{spy}
    \usetikzlibrary{perspective}
    \usetikzlibrary{positioning}
    \usetikzlibrary{shapes.geometric}
\usepackage{circuitikz}
\usepackage{pgfplots}
    \pgfplotsset{compat=newest}
    \usepgfplotslibrary{colormaps}
    \usepgfplotslibrary{units}
    \usepgfplotslibrary{colorbrewer}
    \usepgfplotslibrary{groupplots}
    \usepgfplotslibrary{fillbetween}

\ifexternalize
    \usepgfplotslibrary{external}
\fi

\usepackage{amsmath}
\usepackage{amsfonts}
\usepackage{amssymb}
\usepackage{mathtools}
\usepackage{array}
\usepackage{nicematrix}
\usepackage{stfloats}
\usepackage{bm}

\usepackage{ifthen}
\usepackage{etoolbox}
\usepackage{afterpage}

\usepackage{xltabular}
\usepackage[referable]{threeparttablex}
\usepackage{threeparttable}
\usepackage{booktabs}

\makeatletter
\let\MYcaption\@makecaption
\makeatother
    \usepackage{subcaption}
    \DeclareCaptionLabelFormat{r-parens}{#2} 
    
\makeatletter
\let\@makecaption\MYcaption
\makeatother

\newcommand\mycaption[4][]{%
    \ifthenelse{\equal{#1}{}}{%
        \caption[#2]{\linespread{1.213}#2#3 #4.}%
    }{%
        \caption[#2]{\linespread{1.213}#2#3 #4. \textit{#1.}}%
    }%
}

\usepackage[intoc]{nomencl}
\makenomenclature

\renewcommand{\nomgroup}[1]{%
    \ifthenelse{\equal{#1}{A}}{%
        \item[\textbf{Mathematics}]%
    }{%
        \ifthenelse{\equal{#1}{B}}%
            {\item[\textbf{Control and optimization}]%
        }{%
            \ifthenelse{\equal{#1}{C}}{%
                \item[\textbf{Electricity and stability}]%
            }{%
            }%
        }%
    }%
}

\usepackage[acronym, toc, shortcuts]{glossaries}
\glsdisablehyper

\definecolor{regionBlue}{HTML}{D7E2E7}

\makeglossaries
\newignoredglossary{phantom}
\glsaddkey*
    {in}
    {\glsentrytext{in\glslabel}}
    {\glsentryin}
    {\Glsentryin}
    {\glsin}
    {\Glsin}
    {\GLSin}
\glsaddkey*
    {er}
    {\glsentrytext{\glslabel{}er}}
    {\glsentryer}
    {\Glsentryer}
    {\glser}
    {\Glser}
    {\GLSer}
\glsaddkey*
    {ers}
    {\glsentrytext{\glslabel{}ers}}
    {\glsentryers}
    {\Glsentryers}
    {\glsers}
    {\Glsers}
    {\GLSers}
\newcommand\newac[3][]{%
    \newglossaryentry{#2}{%
        type=\acronymtype,
        name={#2},
        description={#3},
        first={#3 (#2)},
        firstplural={#3s (#2s)},
        short={#2},
        shortplural={#2s},
        long={\MakeLowercase{#3}},
        longplural={\MakeLowercase{#3}s},
        #1
    }
}
\newcommand\newdefinedac[4][]{
    \newglossaryentry{g#2}{%
        name={#3 (#2)},
        text={#3},
        description={#4},
        long={#3},
        longplural={\MakeLowercase{#3}s},
    }
    \newac[{#1,
            first={#3 (#2)\glsadd{g#2}},
            firstplural={#3s (#2s)\glsadd{g#2}},
            long={\MakeLowercase{#3}},
            longplural={\MakeLowercase{#3}s},
            see=[Glossary:]{g#2}}]{#2}{#3}
}
\newcommand\addshortterm[5][]{
    \ifthenelse{\equal{#2}{y}}{
        \newdefinedac[#1]{#3}{#4}{#5}
    }{
        \newac[#1]{#3}{#4}
    }
}
\newcommand\addterm[5][]{
    \ifthenelse{\equal{#2}{y}}{%
        \newglossaryentry{#3}{%
            name={\MakeLowercase{#4}},
            description={#5},
            first={\MakeLowercase{#4}\glsadd{#3}},
            long={\MakeLowercase{#4}},
            longplural={\MakeLowercase{#4}s},
            #1
        }
    }{%
        \newglossaryentry{#3}{%
            type=phantom,
            name={\MakeLowercase{#4}},
            description={#5},
            first={\MakeLowercase{#4}\glsadd{#3}},
            long={\MakeLowercase{#4}},
            longplural={\MakeLowercase{#4}s},
            #1
        }
    }
}

\newcommand\eg{e.g.,\xspace}
\newcommand\ie{i.e.,\xspace}
\newcommand\foreign[1]{\textit{#1}\xspace}
\renewcommand\emph[1]{\textit{#1}\xspace}
\newcommand\term[1]{\textit{#1}\xspace}

\newcommand\incorrect[1]{``#1''\xspace}
\newcommand\sic[1]{``#1''}

\addshortterm{n}{AC}{Alternating Current}{}
\addshortterm{n}{DN}{Distribution Network}{}
\addshortterm{n}{HV}{High Voltage}{}
\addshortterm{n}{LV}{Low Voltage}{}
\addshortterm{n}{MV}{Medium Voltage}{}
\addshortterm{n}{SCADA}{Supervisory Control and Data Acquisition}{}
\addshortterm{n}{T-D}{Transmission-Distribution}{}
\addshortterm{n}{TN}{Transmission Network}{}

\addterm{y}{AR}{Area}{}
\addterm[longplural=boundary buses, plural=boundary buses, firstplural=boundary buses]{y}{BB}{Boundary Bus}{}
\addterm{n}{FR}{Feasible Region}{}
\addterm[in=long-term voltage instability]{y}{LTVS}{Long-Term Voltage Stability}{}
\addterm{n}{STVS}{Short-Term Voltage Stability}{}
\addterm[in=strong area]{y}{WA}{Weak Area}{}
\addterm{n}{VC}{Voltage Collapse}{}
\addterm[in=voltage instability]{y}{VS}{Voltage Stability}{}
\addterm{n}{PSST}{Power System Stability}{}
\addterm{n}{DIST}{Disturbance}{}
\addterm{n}{LR}{Load Restoration}{}

\addterm{n}{AC-OPF}{\MakeUppercase{AC} Optimal Power Flow}{}
\addterm{n}{AS}{Ancillary Service}{}
\addterm[er=centralized controller]{n}{CEC}{Centralized Control}{}
\addterm{y}{COO}{Coordination}{}
\addterm[er=distributed controller]{n}{DC}{Distributed Control}{}
\addshortterm[er=local controller, ers=local controllers]{y}{LC}{Local Control}{}
\addterm[er=coordinated controller, ers=coordinated controllers]{y}{CC}{Coordinated Control}{} 
\addterm{y}{LS}{Load Shedding}{}
\addshortterm{y}{MPC}{Model Predictive Control}{}

\addshortterm{n}{AVR}{Automatic Voltage Regulator}{}
\addterm{n}{BESS}{Battery Energy Storage System}{}
\addshortterm{y}{DER}{Distributed Energy Resource}{}
\addterm{n}{DG}{Distributed Generator}{}
\addterm{n}{FL}{Flexible Load}{}
\addshortterm{y}{LTC}{Load Tap Changer}{}
\addshortterm{n}{OEL}{OverExcitation Limiter}{}
\addshortterm{n}{PMU}{Phasor Measurement Unit}{}
\addterm{n}{PV}{Photovoltaic}{}
\addterm{n}{SCB}{Shunt Capacitor Bank}{}
\addterm{n}{SVC}{Static var Compensator}{}
\addterm{n}{TCL}{Thermostatically Controlled Load}{}
\addshortterm{n}{UPS}{Uninterrupted Power Supply}{}

\addterm{n}{AIEE}{American Institute of Electrical Engineers}{}
\addshortterm{n}{IEEE}{Institute of Electrical and Electronic Engineers}{}
\addshortterm[long=European Network of Transmission System Operators for Electricity]{n}{ENTSO-E}{European Network of Transmission System Operators for Electricity}{}

\addterm{n}{FSA}{Finite-State Automaton}{}
\addshortterm{n}{LIVES}{Local Identification of Voltage Emergency Situations}{}
\addshortterm{n}{NLI}{New LIVES Index}{}
\addshortterm{n}{QP}{Quadratic Program}{}
\addshortterm{n}{UKGDS}{United Kingdom Generic Distribution System}{}
\addterm{n}{CO}{Aggregator}{}
\addterm{n}{DSO}{Distribution System Operator}{}
\addterm{n}{TSO}{Transmission System Operator}{}

\newlength\mylabelsep
\newlength\plotwidth

\def\schematicthickness{semithick}
\def\busunitlength{0.2}
\newcommand\drawbus[6][none]{
    \ifthenelse{
        \equal{#4}{vertical}
    }{
        \gdef\direction{90}
    }{
        \ifthenelse{
            \equal{#4}{horizontal}
        }{
            \gdef\direction{180}
        }{
            \gdef\direction{#4}
        }
    }
    \def\infbuswidth{0.3}
    \ifthenelse{\equal{#1}{inf-right}}{
        \edef\busoldlength{\busunitlength}
        \def\busunitlength{2*\busoldlength}
        \path[fill=gray!30]
            (#3)
            \foreach \x in {1, ..., #5}{
                --++(\direction:\busunitlength)
                --++(\direction-90:\infbuswidth)
                --++(\direction-180:\busunitlength)
                --++(\direction-270:\infbuswidth)
                  ++(\direction:\busunitlength)
            }
        ;
        \path[fill=gray!30]
            (#3)
            \foreach \x in {1, ..., #5}{
                --++(\direction-180:\busunitlength)
                --++(\direction-90:\infbuswidth)
                --++(\direction:\busunitlength)
                --++(\direction+90:\infbuswidth)
                  ++(\direction+180:\busunitlength)
            }
        ;
    }{
    }
    \ifthenelse{\equal{#1}{inf-left}}{
        \edef\busoldlength{\busunitlength}
        \def\busunitlength{2*\busoldlength}
        \path[fill=gray!30]
            (#3)
            \foreach \x in {1, ..., #5}{
                --++(\direction:\busunitlength)
                --++(\direction+90:\infbuswidth)
                --++(\direction+180:\busunitlength)
                --++(\direction+270:\infbuswidth)
                  ++(\direction:\busunitlength)
            }
        ;
        \path[fill=gray!30]
            (#3)
            \foreach \x in {1, ..., #5}{
                --++(\direction-180:\busunitlength)
                --++(\direction+90:\infbuswidth)
                --++(\direction:\busunitlength)
                --++(\direction-90:\infbuswidth)
                  ++(\direction-180:\busunitlength)
            }
        ;
    }{
    }
    \draw[\schematicthickness]
        (#3) coordinate (#2 0)
        \foreach \x in {1, ..., #5}{
            -- ++(\direction:\busunitlength)
            coordinate(#2 \x)
        }
        (#3)
        \foreach \x in {1, ..., #6}{
            -- ++(\direction+180:\busunitlength)
            coordinate(#2 -\x)
        }
    ;
}
\def\loadhorizontal{0.5}
\def\loadvertical{1}
\newcommand\drawload[3]{
    \tikzset{>=latex}
    \ifthenelse{
        \equal{#3}{right}
    }{
        \gdef\direction{0}
    }{
        \gdef\direction{180}
    }
    \draw[->, \schematicthickness]
        (#2)
        -- ++(\direction:\loadhorizontal)
        -- ++(-90:\loadvertical)
        coordinate(#1)
    ;
}
\def\traforadius{0.2}
\def\trafosep{1.3*\traforadius}
\def\tapangle{70}
\def\taplength{0.8}
\newcommand\drawtransformer[4][false]{%
    \path ($(#3)!0.5!(#4)$) coordinate (cm)
        coordinate (#2)
    ;
    \coordinate (A) at (#3);
    \coordinate (B) at (#4);
    \pgfmathanglebetweenpoints{\pgfpointanchor{A}{center}}
                              {\pgfpointanchor{B}{center}}
    \edef\rotationangle{\pgfmathresult}
    \draw[\schematicthickness]
        (cm)
        ++(\rotationangle+180:\trafosep/2) circle(\traforadius)
        ++(\rotationangle+180:\traforadius) coordinate(input #2)
        ++(\rotationangle:\traforadius)
        ++(\rotationangle:\trafosep) circle(\traforadius)
        ++(\rotationangle:\traforadius) coordinate(output #2)
        ++(\rotationangle+180:\traforadius)
        (#3) -- (input #2)
        (#4) -- (output #2)
    ;
    \ifthenelse{\equal{#1}{true}}{
        \draw[\schematicthickness, -latex]
            (cm)
            ++(\rotationangle+180:1.4*\trafosep/2)
            ++(\rotationangle+180+\tapangle:\taplength/2)
            --
            ++(\rotationangle+\tapangle:\taplength)
        ;
    }{}
}
\newcommand\drawtransmissionline[3]{
    \draw[\schematicthickness]
        (#2)
        -- (#3)
    ;
    \path ($(#2)!0.5!(#3)$) coordinate (#1);
}
\def\generatorradius{0.25}
\newcommand\drawgenerator[2]{
    \draw[\schematicthickness]
        (#2)
        circle(\generatorradius)
        +(-0.2*\generatorradius/0.3, 0.) coordinate(c1)
        +( 0. , \generatorradius) coordinate(c2)
        +( 0. , -\generatorradius) coordinate(c3)
        +( 0.2*\generatorradius/0.3, 0.) coordinate(c4)
        (c1) .. controls (c2)
                     and (c3) .. (c4)
        (#2) ++(0:\generatorradius) coordinate(#1 east)
        (#2) ++(90:\generatorradius) coordinate(#1 north)
        (#2) ++(180:\generatorradius) coordinate(#1 west)
        (#2) ++(270:\generatorradius) coordinate(#1 south)
    ;
}
\newcommand\drawcapacitor[3]{
    \tikzset{>=latex}
    \ifthenelse{
        \equal{#3}{right}
    }{
        \gdef\direction{0}
    }{
        \gdef\direction{180}
    }
    \draw[\schematicthickness]
        (#2)
        -- ++(\direction:\loadhorizontal)
        ++(-90:\loadvertical)
        to[C, /tikz/circuitikz/bipoles/length=0.75cm]
        ++(90:0.45*\loadvertical) coordinate(temp1)
        --
        ++(90:0.55*\loadvertical)
        ++(-90:\loadvertical) coordinate(temp2)
        ($(temp1)!0.5!(temp2)$) coordinate(#1)
        (temp2) ++(-90:0) coordinate(#1 south)
        (temp2)
    ;
}

\tikzset{
    every node/.style={
        font=\footnotesize
    },
    curve/.style={
        semithick
    },
    join=round,
    >=latex,
}

\tikzset{
    aggregated der/.prefix style={
        draw,
        rounded corners=2pt,
        inner sep=6pt,
        region,
    },
    line/.style={
        black
    },
    signal/.style={
        line,
        -latex,
    },
    block/.style={
        draw,
        rectangle,
        minimum height=0.9cm,
        minimum width=1.25cm
    },
    binary operation/.style={
        draw,
        circle,
        minimum height=1.1em,
        inner sep=0pt
    },
    sum/.style={
        binary operation,
        node contents={}
    },
    division/.style={
        binary operation,
        node contents={$\div$}
    },
    multiplication/.style={
        binary operation,
        node contents={$\times$}
    },
    connection/.style={
        fill=black,
        circle,
        inner sep=0pt,
        minimum size=2.25pt,
        node contents={}
    },
    input/.style={
        coordinate,
        node contents={}
    },
    output/.style={
        coordinate,
        node contents={}
    },
    axis/.style={
        help lines
    },
    blockplot/.style={
        black,
        semithick,
        draw
    },
    deadband/.style={
        block,
        node contents={%
            \blockplot{
                \path[blockplot]
                    (0, 0) -- ++(180:0.2*\xlen)
                    -- ++(-180+\limitangle:0.3*\xlen)
                    (0, 0) -- ++(0:0.2*\xlen)
                    -- ++(\limitangle:0.3*\xlen);
            }
        }
    },
    rate/.style={
        block,
        node contents={%
            \blockplot{
                \path[blockplot]
                    (0, 0) ++(\limitangle-180:\satdiagonal)
                    -- ++(\limitangle:2*\satdiagonal);
            }
        }
    },
    saturation/.style={
        block,
        node contents={%

            \newcommand\protrusion{0.25*\xlen}%
            \blockplot{
                \path[blockplot]
                    (0, 0) -- ++(\limitangle-180:\satdiagonal)
                    -- ++(180:\protrusion)
                    (0, 0) -- ++(\limitangle:\satdiagonal)
                    -- ++(0:\protrusion);
            }
        }
    },
    hyst/.style={
        block,
        node contents={%
            \newcommand\windowheight{0.5}%
            \newcommand\windowwidth{0.4}%
            \newcommand\branchprot{0.2}%
            \noindent
            \begin{tikzpicture}
                \coordinate (origin) at (0, 0);
                \fill[MyLightGray]
                    (origin) ++(0:\branchprot)
                    rectangle ++(\windowwidth, -\windowheight);

                \draw
                    (origin) -- ++(0:\branchprot)
                    coordinate (temp);
                \draw[-latex, shorten >=-2pt,shorten <=-2pt]
                    (temp) -- ++(0:\windowwidth/2)
                    coordinate (temp);
                \draw
                    (temp) -- ++(0:\windowwidth/2) -- ++(-90:\windowheight)
                    ++(0:\branchprot) coordinate (temp);
                \draw
                    (temp) -- ++(180:\branchprot)
                    coordinate (temp);
                \draw[-latex, shorten >=-2pt]
                    (temp) -- ++(180:\windowwidth/2)
                    coordinate (temp);
                \draw
                    (temp) -- ++(180:\windowwidth/2) -- ++(90:\windowheight);
            \end{tikzpicture}%
        }
    },
    region/.style={
        fill=regionBlue,
        draw=MyDarkGray,
        very thin
    },
    unused input/.style={
        draw,
        dotted
    },
    used input/.style={
        draw,
        line
    },
    hv/.style={
        semithick
    },
    mv/.style={
        thin
    },
    lv color/.style={
        gray
    },
    lv/.style={
        thin,
        lv color
    },
    substation bus/.style={
        very thick,
        line cap=round,
        line join=round
    },
    disc/.style={
        mv,
        densely dotted
    },
    no coordinator/.style={
        semithick,
        densely dotted,
        mark=none,
        join=round
    },
    coordinator/.style={
        semithick,
        mark=none,
        join=round
    },
    subtle grid/.style={
        semithick,
        gray
    },
    zeroline/.style={
        help lines,
        axescolor
    },
    callout color/.style={
        MyDarkGray
    },
    callout/.style={
        callout color,
        -latex,
    },
    axis/.style={
        draw,
        help lines,
        ->
    },
    tick/.style={
        draw
    },
    foreseen/.style={
        region,
        draw=none
    },
    measured/.style={
        draw,
        curve
    },
    predicted/.style={
        draw,
        dotted,
        curve
    },
    control input/.style={
        Dark2-A
    },
    control output/.style={
        Dark2-B
    },
    control reference/.style={
        Dark2-C
    }
}

\pgfplotscreateplotcyclelist{ExtendedDark2}{%
    {Dark2-A},
    {Dark2-B},
    {Dark2-C},
    {Dark2-D},
    {Dark2-E},
    {Dark2-F},
    {Dark2-G},
    {Dark2-H},
    {Paired-C!80},
    {Paired-E},
    {Paired-I}%
}
\definecolor{MyLightGray}{rgb}{0.941, 0.941, 0.941}
\definecolor{MyMediumGray}{rgb}{0.741, 0.741, 0.741}
\definecolor{MyDarkGray}{rgb}{0.388, 0.388, 0.388}
\colorlet{axescolor}{MyDarkGray}
\makeatletter
    \def\pgfplotsdataxmin{\pgfplots@data@xmin}
    \def\pgfplotsdataxmax{\pgfplots@data@xmax}
    \def\pgfplotsdataymin{\pgfplots@data@ymin}
    \def\pgfplotsdataymax{\pgfplots@data@ymax}
\makeatother
\newif\iftufte\tuftetrue
\newif\iftuftex\tuftextrue
\newif\iftuftey\tufteytrue
\pgfplotsset{
    range frame/.style={
        tick align=outside,
        axis line style={opacity=0},
        after end axis/.code={
            \iftufte
                \iftuftex
                    \draw[axescolor]
                        ({rel axis cs:0,0}-|{axis cs:\pgfplotsdataxmin,0})
                        --
                        ({rel axis cs:0,0}-|{axis cs:\pgfplotsdataxmax,0});
                \fi
                \iftuftey
                    \draw[axescolor]
                        ({rel axis cs:0,0}|-{axis cs:0,\pgfplotsdataymin})
                        --
                        ({rel axis cs:0,0}|-{axis cs:0,\pgfplotsdataymax});
                \fi
            \fi
        }
    },
    enlarge x limits=0.05,
    enlarge y limits=0.05,
    width=0.95\columnwidth,
    height=0.95\columnwidth/1.62,
    axis x line*=bottom,
    axis y line*=left,
    every tick/.append style={axescolor},
    cycle list name=ExtendedDark2,
    colormap/RdBu,
    precise x/.style={
        x tick label style={
            /pgf/number format/.cd,
            fixed,
            fixed zerofill,
            precision=2,
            /tikz/.cd
        }
    },
    precise y/.style={
        y tick label style={
            /pgf/number format/.cd,
            fixed,
            fixed zerofill,
            precision=2,
            /tikz/.cd
        }
    },
    semi precise y/.style={
        y tick label style={
            /pgf/number format/.cd,
            fixed,
            fixed zerofill,
            precision=1,
            /tikz/.cd
        }
    },
    not precise y/.style={
        y tick label style={
            /pgf/number format/.cd,
            fixed,
            fixed zerofill,
            precision=0,
            /tikz/.cd
        }
    },
    legend image code/.code={
        \draw plot coordinates {(0cm, 0cm) (0.5cm, 0cm)};
    }
}

\pgfdeclaredecoration{ignore}{final}
{
\state{final}{}
}

\pgfdeclaremetadecoration{middle}{initial}{
    \state{initial}[
        width={(\pgfmetadecoratedpathlength - \the\pgfdecorationsegmentlength)/2},
        next state=middle
    ]
    {\decoration{curveto}}

    \state{middle}[
        width={\the\pgfdecorationsegmentlength},
        next state=final
    ]
    {\decoration{curveto}}

    \state{final}
    {\decoration{ignore}}
}

\tikzset{middle segment/.style={decoration={middle},decorate, segment length=#1}}

\tikzset{
    line/.style={
        black
    },
    signal/.style={
        line
    },
    hv/.style={
        semithick
    },
    mv/.style={
        thin
    },
    lv color/.style={
        gray
    },
    lv/.style={
        thin,
        lv color
    },
    substation bus/.style={
        very thick,
        line cap=round,
        line join=round
    },
    disc/.style={
        mv,
        densely dotted
    },
    controller node/.prefix style={
        draw,
        rectangle,
        rounded corners=3pt,
        region
    },
    signal/.prefix style={
        ->
    },
    reverse signal/.prefix style={
        <-
    },
    aggregated der/.prefix style={
        draw,
        rounded corners=2pt,
        inner sep=6pt
    },
    current/.prefix style={
        sloped,
        pos=#1,
        inner sep=0pt
    },
    current right/.prefix style={
        current=#1,
        node contents={
            \tikz \draw[-latex] (-1pt, 0) -- (1pt, 0);
        }
    },
    current left/.prefix style={
        current=#1,
        node contents={
            \tikz \draw[-latex] (1pt, 0) -- (-1pt, 0);
        }
    },
    bus/.pic={
        \draw[substation bus]
            (0, 0) coordinate (#1)
            ++(-90:\busheight/2) coordinate (#1 bottom)
            -- ++(90:\busheight) coordinate (#1 top);
        \foreach \x in {0, 0.1, ..., 1.1}{
            \pgfmathsetmacro\pos{int(\x*10}
            \edef\temp{%
                \noexpand\coordinate
                    (#1 \pos) at ($(#1 top)!\x!(#1 bottom)$);
            }
            \temp
        }
    },
    pics/transformer/.style args={#1/#2/#3/#4/#5}{
        code={
            \pgfmathsetmacro\traforad{0.3}
            \pgfmathsetmacro\centersep{1.5*\traforad}
            \pgfmathsetmacro\tapangle{70}
            \pgfmathsetmacro\taplength{3.2*\traforad}
            \coordinate (origin) at (0, 0);
            \path
                (#2) pic{bus={#1 HV}}
                (#3) pic{bus={#1 LV}};
            \draw
                ($(#2)!#5!(#3)$)
                ++(180:\centersep/2)
                circle (\traforad)
				coordinate (center HV winding)
                ++(180:\traforad)
                --
                (#1 HV)
                (center HV winding)
                ++(0:\centersep)
                circle (\traforad)
				coordinate (center LV winding)
                ++(0:\traforad)
                --
                (#1 LV);
			\draw[->]
				(center HV winding)
				++(\tapangle:-\taplength/2)
				--
				++(\tapangle:\taplength);
			\coordinate (tap) at (center HV winding);
			\ifdrawratio
			\path
				($(center HV winding)!0.5!(center LV winding)$)
				++(-90:0.65)
				node{$#4:1$};
			\fi
        }
    },
    generator/.pic={
        \draw[dn line]
            (0, 0) -- ++(180:\gendist) coordinate (temp);
        \draw[dn line]
            (temp) ++(180:\genrad) circle (\genrad);
    },
    >=latex,
    load/.pic={
        \draw[dn line, ->] (0, 0) -- ++(0:\loadh) -- ++(-90:\loadv)
        coordinate (end load);
    },
    region/.prefix style={
        fill=regionBlue
    },
    line/.style={
        black
    },
    tn line/.prefix style={
        semithick
    },
    dn line/.prefix style={
        thin
    }
}

\tikzset{
    action/.style={
        block,
    },
    base node/.style={
        align=center,
        draw,
        sharp corners,
    },
    block/.style={
        base node,
        rectangle,
        region,
    },
    decision/.style={
        base node,
        diamond,
        align=center,
        aspect=1.7,
        inner sep=2pt,
    },
    joining/.style={
        base node,
        circle, 
        inner sep=1pt,
        region,
    },
    opposite flow/.prefix style={
        -latex,
    },
    flow/.style={
        -latex,
    },
    optimization/.style={
        region
    },
    terminator/.style={
        base node,
        region, 
        rounded rectangle,
    },
    yesno/.style={
        near start,
        font=\scriptsize
    },
}

\tikzset{
    join=round,
    no MPC/.style={
        semithick,
        densely dotted,
        mark=none,
        join=round
    },
    MPC/.style={
        semithick,
        mark=none,
        join=round
    },
    every node/.style={
        font=\footnotesize
    },
    subtle grid/.style={
        semithick,
        gray
    },
    generation/.style={
        fill=MyLightGray,
        draw=none
    },
    zeroline/.style={
        help lines,
        axescolor
    },
    callout/.style={
        -latex,
        MyDarkGray
    }
}

\makeatletter
\def\pgfplots@getautoplotspec into#1{%
    \begingroup
    \let#1=\pgfutil@empty
    \pgfkeysgetvalue{/pgfplots/cycle multi list/@dim}\pgfplots@cycle@dim
    \let\pgfplots@listindex=\pgfplots@numplots
    \pgfkeysgetvalue{/pgfplots/cycle list set}\pgfplots@listindex@set
    \ifx\pgfplots@listindex@set\pgfutil@empty
    \else
        \c@pgf@counta=\pgfplots@listindex
        \c@pgf@countb=\pgfplots@listindex@set
        \advance\c@pgf@countb by -\c@pgf@counta
        \globaldefs=1\relax
        \edef\setshift{%
            \noexpand\pgfkeys{
                /pgfplots/cycle list shift=\the\c@pgf@countb,
                /pgfplots/cycle list set=
            }
        }%
        \setshift%
    \fi
    \pgfkeysgetvalue{/pgfplots/cycle list shift}\pgfplots@listindex@shift
    \ifx\pgfplots@listindex@shift\pgfutil@empty
    \else
        \c@pgf@counta=\pgfplots@listindex\relax
        \advance\c@pgf@counta by\pgfplots@listindex@shift\relax
        \ifnum\c@pgf@counta<0
            \c@pgf@counta=-\c@pgf@counta
        \fi
        \edef\pgfplots@listindex{\the\c@pgf@counta}%
    \fi
    \ifnum\pgfplots@cycle@dim>0
        \c@pgf@counta=\pgfplots@cycle@dim\relax
        \c@pgf@countb=\pgfplots@listindex\relax
        \advance\c@pgf@counta by-1
        \pgfplotsloop{%
            \ifnum\c@pgf@counta<0
                \pgfplotsloopcontinuefalse
            \else
                \pgfplotsloopcontinuetrue
            \fi
        }{%
            \pgfkeysgetvalue{/pgfplots/cycle multi list/@N\the\c@pgf@counta}\pgfplots@cycle@N
            \pgfplotsmathmodint{\c@pgf@countb}{\pgfplots@cycle@N}%
            \divide\c@pgf@countb by \pgfplots@cycle@N\relax
            \expandafter\pgfplots@getautoplotspec@
                \csname pgfp@cyclist@/pgfplots/cycle multi list/@list\the\c@pgf@counta @\endcsname
                {\pgfplots@cycle@N}%
                {\pgfmathresult}%
            \t@pgfplots@toka=\expandafter{#1,}%
            \t@pgfplots@tokb=\expandafter{\pgfplotsretval}%
            \edef#1{\the\t@pgfplots@toka\the\t@pgfplots@tokb}%
            \advance\c@pgf@counta by-1
        }%
    \else
        \pgfplotslistsize\autoplotspeclist\to\c@pgf@countd
        \pgfplots@getautoplotspec@{\autoplotspeclist}{\c@pgf@countd}{\pgfplots@listindex}%
        \let#1=\pgfplotsretval
    \fi
    \pgfmath@smuggleone#1%
    \endgroup
}

\pgfplotsset{
    cycle list set/.initial=
}
\makeatother

\newlength{\powerthreshold}
\tikzset{
    faulted/.style={
        Red
    },
    400 kV lines/.style={
        semithick
    },
    220 kV lines/.style={
        thin,
        black!70
    },
    130 kV lines/.style={
        220 kV lines
    },
    power bus/.style n args={3}{
        node contents={%
            \pgfmathsetmacro\genradius{sqrt(#2)/60}%
            \pgfmathsetmacro\loadradius{sqrt(#3)/60}%
            \noindent
            \begin{tikzpicture}%
                \ifdim\genradius pt < \powerthreshold
                    \ifdim\loadradius pt < \powerthreshold
                        \node at (0, 0) [transit];
                    \else
                        \path[nordic load] (0, 0) circle (\loadradius cm);
                    \fi
                \else
                    \ifdim\loadradius pt < \powerthreshold
                        \path[generation] (0, 0) circle (\genradius cm);
                    \else
                        \pgfmathsetmacro\powerdif{sqrt(abs((\loadradius)^2-(\genradius)^2))}
                        \ifdim\powerdif pt < -0.129pt 
                            \path[net transit] (0, 0) circle (\loadradius cm);
                        \else
                            \ifdim\loadradius pt > \genradius pt
                                \path[nordic load]
                                    (0, 0) circle (\loadradius cm);
                                \path[overlaid generation]
                                     (0, 0) circle (\genradius cm);
                            \else
                                \path[generation] (0, 0) circle (\genradius cm);
                                \path[overlaid load]
                                    (0, 0) circle (\loadradius cm);
                            \fi
                        \fi
                    \fi
                \fi
            \end{tikzpicture}%
        },
        #1,
        name=#1
    },
    voltage bus/.style n args={4}{
        node contents={%
            \noindent
            \begin{tikzpicture}
                \pgfmathsetmacro\xminmeta{0.925}
                \pgfmathsetmacro\xmaxmeta{1 + (1 - \xminmeta)}
                \pgfmathsetmacro\thisvoltage{1000*(#4-\xminmeta)/(\xmaxmeta-\xminmeta)}
                \pgfplotscolormapdefinemappedcolor{\thisvoltage}
                \draw[fill=mapped color] (0, 0) circle (0.2cm);
            \end{tikzpicture}%
        },
        #1,
        name=#1
    },
    simple bus/.style n args={4}{
        transit,
        rounded rectangle,
        node contents={#1},
        #1,
        name=#1
    },
    nordic bus/.style n args={4}{
        power bus={#1}{#2}{#3},
    },
    greece bus/.style={
        transit,
        circle,
        inner sep=0pt,
        minimum size=0.1pt,
        name={#1}
    },
    load bus/.style n args={4}{
        nordic bus={#1}{#2}{#3}{#4},
    },
    boundary bus/.style n args={4}{
        nordic bus={#1}{#2}{#3}{#4},
    },
    standard bus/.style n args={4}{
        nordic bus={#1}{#2}{#3}{#4},
    },
    nordic line/.style={
        ultra thick
    },
    nordic transformer/.style={
        nordic line
    },
    boundary/.style={
        ultra thick,
        densely dashed
    },
    region label/.style={
        draw=none,
        fill=none,
        font=\small
    },
    generation/.style={
        fill=Pastel2-A,
        draw=Dark2-A
    },
    overlaid pattern/.style={
        pattern={Lines[%
                    distance=1.5mm,%
                    angle=135,%
                    line width=0.6mm%
                 ]}
    },
    overlaid generation/.style={
        generation,
        overlaid pattern,
        pattern color=Dark2-A
    },
    nordic load/.style={
        fill=Pastel2-B,
        draw=Dark2-B
    },
    overlaid load/.style={
        nordic load,
        overlaid pattern,
        pattern color=Dark2-B
    },
    transit/.style={
        fill=Pastel2-H,
        draw=Dark2-H,
        node contents={\phantom{l\,l}}
    },
    net transit/.style={
        draw=Dark2-D,
        fill=Pastel2-D
    },
    nordic region/.style={
        font=\bfseries\sffamily\Huge,
        white
    },
    4011/.style={},
    4012/.style={},
    4021/.style={},
    4022/.style={},
    4031/.style={},
    4032/.style={},
    4041/.style={},
    4042/.style={},
    4043/.style={},
    4044/.style={},
    4045/.style={},
    4046/.style={},
    4047/.style={},
    4051/.style={},
    4061/.style={},
    4062/.style={},
    4063/.style={},
    4071/.style={},
    4072/.style={},
    2031/.style={},
    2032/.style={},
    1011/.style={},
    1012/.style={},
    1013/.style={},
    1014/.style={},
    1021/.style={},
    1022/.style={},
    1041/.style={},
    1042/.style={},
    1043/.style={},
    1044/.style={},
    1045/.style={},
    4071_4072_A/.style={},
    4071_4072_B/.style={},
    4071_4011/.style={},
    4071_4012/.style={},
    4012_4011/.style={},
    4011_4021/.style={},
    4012_4022/.style={},
    4022_4011/.style={},
    4022_4031_A/.style={},
    4022_4031_B/.style={},
    4031_4041_A/.style={},
    4031_4041_B/.style={},
    4031_4032/.style={},
    4021_4032/.style={},
    4021_4042/.style={},
    4032_4042/.style={},
    4042_4043/.style={},
    4043_4046/.style={},
    4046_4047/.style={},
    4043_4047/.style={},
    4032_4044/.style={faulted},
    4042_4044/.style={},
    4041_4044/.style={},
    4043_4044/.style={},
    4041_4061/.style={},
    4061_4062/.style={},
    4062_4063_A/.style={},
    4062_4063_B/.style={},
    4045_4062/.style={},
    4044_4045_A/.style={},
    4044_4045_B/.style={},
    4045_4051_A/.style={},
    4045_4051_B/.style={},
    1041_1043_A/.style={},
    1041_1043_B/.style={},
    1041_1045_A/.style={},
    1041_1045_B/.style={},
    1042_1045/.style={},
    1042_1044_A/.style={},
    1042_1044_B/.style={},
    1043_1044_A/.style={},
    1043_1044_B/.style={},
    1021_1022_A/.style={},
    1021_1022_B/.style={},
    1011_1013_A/.style={},
    1011_1013_B/.style={},
    1012_1014_A/.style={},
    1012_1014_B/.style={},
    1013_1014_A/.style={},
    1013_1014_B/.style={},
    2031_2032_A/.style={},
    2031_2032_B/.style={},
    4031_2031/.style={},
    4011_1011/.style={},
    1012_4012/.style={},
    4022_1022/.style={},
    1044_4044_A/.style={},
    1044_4044_B/.style={},
    1045_4045_A/.style={},
    1045_4045_B/.style={},
    double circuit/.style={
        nordic line,
        postaction={
            decorate,
            decoration={
                markings,
                mark=between positions 0 and 1 step 1.5mm with {
                    \draw[nordic line, solid, ultra thin] (0, 0) -- (0, 0.8mm);
                }
            }
        }
    },
    under construction/.style={
        dashed
    },
    greek border/.style={
        draw,
        400 kV lines
    }
}

\newcommand\drawline[3][]{
    \draw[nordic line, #2_#3]
    (#2.center)
        \foreach \x in {#1}{
            -- (\x)
        }
    --
    (#3.center);
}
\newcommand{\mytraforad}{3pt}
\newcommand{\mytrafosep}{1.75pt}
\newcommand\drawtrafo[3][]{
    \ifthenelse{\equal{#1}{}}{
        \xdef\mystyle{#2_#3}
    }{
        \xdef\mystyle{#1}
    }
    \draw[nordic transformer, \mystyle]
        ($(#2)!0.5!(#3)$) coordinate (temp)
        ($(temp)!\mytrafosep!(#2)$) coordinate (temp 1)
        circle (\mytraforad)
        ($(temp 1)!\mytraforad!(#2)$) -- (#2.center)
        ($(temp)!\mytrafosep!(#3)$) coordinate (temp 2)
        circle (\mytraforad)
        ($(temp 2)!\mytraforad!(#3)$) -- (#3.center);
}
\newcommand\drawlines[3][2pt]{%
    \draw[nordic line, #2_#3_A, double circuit]
        ($(#2)$) -- ($(#3)$);
}
\newcommand\drawtrafos[3][3pt]{
    \path
        ($(#2)!#1!90:(#3)$)     coordinate (start A)
        ($(#3)!#1!-90:(#2)$)    coordinate (end A)
        ($(#2)!#1!-90:(#3)$)    coordinate (start B)
        ($(#3)!#1! 90:(#2)$)    coordinate (end B);
    \drawtrafo[#2_#3_A]{start A}{end A}
    \drawtrafo[#2_#3_B]{start B}{end B}
}

\pgfdeclarelayer{map}
\pgfdeclarelayer{lines}
\pgfsetlayers{map, lines, main}

\newcommand\drawnordic[1][]{
    \ifthenelse{\equal{#1}{}}{
        {
            \tikzset{every path/.style={}}
            \clip
                (0, 0) rectangle (1, 1);
        }
    }{

    }
    \node at ($(origin) + (0.710, 0.854)$) [standard bus={4011}{668.5}{0}{1.0224}];
    \node at ($(origin) + (0.570, 0.821)$) [standard bus={4012}{600}{0}{1.0235}];
    \node at ($(origin) + (0.630, 0.651)$) [standard bus={4021}{250}{0}{1.0488}];
    \node at ($(origin) + (0.508, 0.651)$) [standard bus={4022}{0}{0}{0.9947}];
    \node at ($(origin) + (0.383, 0.540)$) [standard bus={4031}{310}{0}{1.0367}];
    \node at ($(origin) + (0.488, 0.540)$) [standard bus={4032}{0}{0}{1.0487}];
    \node at ($(origin) + (0.263, 0.331)$) [boundary bus={4041}{0}{540}{1.0506}];
    \node at ($(origin) + (0.424, 0.384)$) [boundary bus={4042}{630}{400}{1.0428}];
    \node at ($(origin) + (0.422, 0.328)$) [standard bus={4043}{0}{900}{1.0370}];
    \node at ($(origin) + (0.274, 0.269)$) [boundary bus={4044}{0}{0}{1.0395}];
    \node at ($(origin) + (0.323, 0.115)$) [standard bus={4045}{0}{0}{1.0533}];
    \node at ($(origin) + (0.543, 0.331)$) [load bus={4046}{0}{700}{1.0357}];
    \node at ($(origin) + (0.494, 0.263)$) [load bus={4047}{1080}{100}{1.0590}];
    \node at ($(origin) + (0.395, 0.073)$) [load bus={4051}{600}{800}{1.0659}];
    \node at ($(origin) + (0.197, 0.239)$) [load bus={4061}{0}{500}{1.0387}];
    \node at ($(origin) + (0.239, 0.147)$) [load bus={4062}{530}{300}{1.0560}];
    \node at ($(origin) + (0.280, 0.041)$) [load bus={4063}{1060}{590}{1.0536}];
    \node at ($(origin) + (0.883, 0.827)$) [standard bus={4071}{300}{300}{1.0484}];
    \node at ($(origin) + (0.965, 0.735)$) [standard bus={4072}{2137.4}{2000}{1.0590}];

    \node at ($(origin) + (0.272, 0.540)$) [standard bus={2031}{0}{100}{1.0279}];
    \node at ($(origin) + (0.136, 0.540)$) [standard bus={2032}{750}{200}{1.0695}];

    \node at ($(origin) + (0.730, 0.920)$) [standard bus={1011}{0}{200}{1.0618}];
    \node at ($(origin) + (0.461, 0.792)$) [standard bus={1012}{600}{300}{1.0634}];
    \node at ($(origin) + (0.658, 0.937)$) [standard bus={1013}{300}{100}{1.0548}];
    \node at ($(origin) + (0.512, 0.875)$) [standard bus={1014}{550}{0}{1.0611}];
    \node at ($(origin) + (0.264, 0.651)$) [standard bus={1021}{400}{0}{1.0311}];
    \node at ($(origin) + (0.391, 0.651)$) [standard bus={1022}{200}{280}{1.0512}];
    \node at ($(origin) + (0.356, 0.168)$) [load bus={1041}{0}{600}{1.0124}];
    \node at ($(origin) + (0.449, 0.204)$) [load bus={1042}{360}{330}{1.0145}];
    \node at ($(origin) + (0.344, 0.213)$) [load bus={1043}{180}{260}{1.0274}];
    \node at ($(origin) + (0.383, 0.263)$) [load bus={1044}{0}{840}{1.0066}];
    \node at ($(origin) + (0.442, 0.140)$) [load bus={1045}{0}{720}{1.0111}];

    \begin{pgfonlayer}{lines}
        \tikzset{
            nordic line/.style={
                400 kV lines
            }
        }
        \drawlines{4071}{4072}
        \drawline{4071}{4011}
        \drawline{4071}{4012}
        \drawline{4012}{4011}
        \drawline{4011}{4021}
        \drawline{4012}{4022}
        \drawline{4022}{4011}
        \drawlines{4022}{4031}
        \drawlines{4031}{4041}
        \drawline{4031}{4032}
        \drawline{4021}{4032}
        \drawline{4021}{4042}
        \drawline{4032}{4042}
        \drawline{4042}{4043}
        \drawline{4043}{4046}
        \drawline{4046}{4047}
        \drawline{4043}{4047}
        \drawline{4032}{4044}
        \drawline{4042}{4044}
        \drawline{4041}{4044}
        \drawline{4043}{4044}
        \drawline{4041}{4061}
        \drawline{4061}{4062}
        \drawlines{4062}{4063}
        \drawline{4045}{4062}
        \drawlines{4044}{4045}
        \drawlines{4045}{4051}
        \tikzset{
            nordic line/.style={
                220 kV lines
            }
        }
        \drawlines{1041}{1043}
        \drawlines{1041}{1045}
        \drawline{1042}{1045}
        \drawlines{1042}{1044}
        \drawlines{1043}{1044}
        \drawlines{1021}{1022}
        \drawlines{1011}{1013}
        \drawlines{1012}{1014}
        \drawlines{1013}{1014}
        \drawlines{2031}{2032}
        \drawtrafo{4031}{2031}
        \drawtrafo{4011}{1011}
        \drawtrafo{1012}{4012}
        \drawtrafo{4022}{1022}
        \drawtrafos{1044}{4044}
        \drawtrafos{1045}{4045}
    \end{pgfonlayer}
}

\newcommand\dummymatrix{\mymatrix{A}}
\newcommand\dummyvectorletter{x}

\newcommand\dummyvector{\myvector{\dummyvectorletter}}

\newcommand\dummyonevariablefunction{f}

\newcommand\dummysignal{f}
\newcommand\dummyphasor{\voltage}

\newcommand\reals[1][]{%
    \ifthenelse{\equal{#1}{}}{%
        \mathbb{R}%
    }{%
        \mathbb{R}^{#1}%
    }%
}

\newcommand\setk{
    \mathbb Z_{\geq0}%
}

\newcommand\setpointaccent{\ast}
\newcommand\incrementsymbol{\Delta}

\newcommand\initialsymbol{0}
\newcommand\average[1]{\overline{#1}}
\newcommand\constantscalar[1]{\MakeUppercase{#1}}
\newcommand\equipmentname[1]{\MakeUppercase{#1}}
\newcommand\mymatrix[1]{\mathbf{\MakeUppercase{#1}}}
\newcommand\myset[1]{\mathcal{#1}}
\newcommand\myvector[1]{\mathbf{\MakeLowercase{#1}}}
\newcommand\normalized[1]{#1}
\newcommand\estimated[1]{\hat{#1}}

\newcommand\phasor[1]{\widetilde{\MakeUppercase{#1}}}
\newcommand\setpoint[1]{#1^\setpointaccent}
\newcommand\variablescalar[1]{\MakeLowercase{#1}}

\newcommand\conttime{t}
\newcommand\ssep{:}
\DeclarePairedDelimiter\abs{\lvert}{\rvert}%

\newcommand\differential{h}
\newcommand\tapdifferential{\differential}
\newcommand\increment[1]{\incrementsymbol #1}

\newcommand\boundarybus{b}
\newcommand\boundarybuscount{\constantscalar{B}}
\newcommand\dummyindex{\ell}
\newcommand\horizonindex{k}
\newcommand\substation{j}
\newcommand\substationcount{\constantscalar{H}}

\newcommand\transmission{T}
\newcommand\distribution{D}
\newcommand\atdistribution{^\text{\distribution}}
\newcommand\attransmission{^\text{\transmission}}
\newcommand\atsubstation[1][]{_{\substation#1}}
\newcommand\othersubstation{\dummyindex}
\newcommand\atothersubstation{_{\othersubstation\neq\substation}}

\newcommand\fromactualDER{^{\text{\gls{DER}}}}
\newcommand\fromaggregatedDER{^{\text{\gls{DER}}}}
\newcommand\fromload{^{\,\text{\load}}}
\newcommand\atboundarybus{_\boundarybus}

\newcommand\identity{\mymatrix{I}}

\newcommand\mynorm[1]{\left\lVert#1\right\rVert}

\newcommand\transpose[1]{#1^\intercal}
\newcommand\vectorleq{\preceq}
\newcommand\weightednorm[3]{%
    \mynorm{#3}_{#1}^{#2}%
}

\newcommand\displaysensitivity[2]{%
    \frac{\partial #1}{\partial #2}%
}
\newcommand\inlinesensitivity[2]{%
    \partial #1/\partial #2%
}

\newcommand\derivative[1]{\dot{#1}}

\newcommand\timeopening{(}
\newcommand\timeclosing{)}
\newcommand\attime[1]{\timeopening#1\timeclosing}
\newcommand\attimegiven[2]{\timeopening#1\,|\,#2\timeclosing}
\newcommand\controlhorizon{\horizon_\controlsignal}
\newcommand\controlsignal{u}
\newcommand\horizon{\constantscalar{N}}
\newcommand\outputsignal{y}
\newcommand\predictionhorizon{\horizon_\outputsignal}

\newcommand\channels[2]{\left\langle #1, #2\right\rangle}

\newcommand\signalsymbol{\Sigma}
\newcommand\signal[1][]{%
    \ifthenelse{\equal{#1}{}}{%
        \ensuremath{\signalsymbol}%
    }{%
        \ifthenelse{\equal{#1}{p}}{%
            \def\subindex{\activepower}%
        }{%
            \def\subindex{\reactivepower}%
        }%
        \ensuremath{\signalsymbol_\subindex}%
    }%
}
\newcommand\samplingperiod{\tau_s}

\newcommand\sizevalueset{\upperbound{\abs{\Sigma}}}

\newcommand\costfunction{\variablescalar{f}}

\newcommand\lowerbound[1]{#1_\mathrm{min}}
\newcommand\slack{s}
\newcommand\lowerslacks{\myvector\slack_1}
\newcommand\upperslacks{\myvector\slack_2}
\newcommand\upperbound[1]{#1_\mathrm{max}}
\newcommand\weight{\mymatrix{W}}
\newcommand\weightchanges{\weight_{\!\incrementsymbol}}
\newcommand\weightu{\weight_{\!\controlsignal}}

\newcommand\weightslacks{\weight_{\!\slack}}
\newcommand\measuredoutputmatrix{\mymatrix{F}}
\newcommand\measuredinputmatrix{\mymatrix C_1}
\newcommand\inputincrementmatrix{\mymatrix C_2}
\newcommand\sensitivitymatrix{\bm\Phi}

\newcommand\halfdeadband{\delta}
\newcommand\activepower{p}

\newcommand\avnli{\average{\text{\acs{NLI}}}}
\newcommand\conductance{g}
\newcommand\current{i}

\newcommand\nli{\text{\acs{NLI}}}
\newcommand\nlicut{\myset{C}}

\newcommand\nlivector{n}
\newcommand\normfieldcurrent{\normalized{i}^{F}}

\newcommand\reactivepower{q}
\newcommand\apparentpower{S}

\newcommand\tapratio{r}
\newcommand\voltage{v}
\newcommand\pexponent{\alpha}
\newcommand\qexponent{\beta}

\newcommand\generator{\equipmentname{g}}
\newcommand\load{\equipmentname{l}}

\newcommand\objective{cost\xspace}

\newcommand\cvxopt{\textsc{cvxopt}\xspace}
\newcommand\dera{DER\_A\xspace}

\newcommand\python{\textsc{python}\xspace}
\newcommand\ramses{\textsc{ramses}\xspace}

\newcommand\ieeestandard{\acs{IEEE} Std. 1547-2018}

\renewcommand{\Re}{\operatorname{Re}}

\DeclareMathOperator*{\minimize}{minimize}
\DeclareMathOperator{\sgn}{sgn}
\DeclareMathOperator{\conj}{conj}

\makeatletter
    \newcommand\verarray[2][r]{%
        \gdef\atbeginning{1}
        \begin{bmatrix}\myrows{#2}\end{bmatrix}%
    }
    \def\myrows#1{\xmyrows#1;lastline;}
    \def\lastline{lastline}
    \def\xmyrows#1;{%
        \def\temp{#1}%
        \ifx\temp\lastline
        \else
            \ifnum\atbeginning=1%
                \gdef\atbeginning{0}
            \else%
                \@arraycr%
            \fi
            #1%
            \expandafter\xmyrows
        \fi
    }
\makeatother
\makeatletter
    
    \def\lastline{lastline}
    \def\xmycols#1;{%
        \def\temp{#1}%
        \ifx\temp\lastline
        \else
            \ifnum\atbeginning=1%
                \gdef\atbeginning{0}
            \else%
                &%
            \fi
            #1%
            \expandafter\xmycols
        \fi
    }
\makeatother

\makeatletter
    \newcommand\horarrayofvectors[1]{%
        \gdef\atbeginning{1}
        \transpose{
            \begin{bmatrix}
                \@horarrayofvectors{#1}
            \end{bmatrix}
        }
    }
    \def\@horarrayofvectors#1{\x@horarrayofvectors#1;lastelement;}
    \def\lastelement{lastelement}
    \def\x@horarrayofvectors#1;{\def\temp{#1}%
        \ifx\temp\lastelement
        \else
            \ifnum\atbeginning=1\gdef\atbeginning{0}
                \expandafter\@firstoftwo
            \else
                \expandafter\@secondoftwo
            \fi
              {}{&}
            \transpose{#1}%
            \expandafter\x@horarrayofvectors
        \fi}
\makeatother

\newcommand\bighorarray[3]{%
    \transpose{%
        \begin{bmatrix}%
            #1 & #2 & \cdots & #3\\%
        \end{bmatrix}%
    }%
}

\newcommand\bigverarray[3]{
    \verarray{#1; #2; \vdots; #3}
}

\DeclareMathOperator{\avg}{avg}
\def\prevoltages{}
\def\pretaps{}
\def\prefieldcurrents{}
\def\preNLIs{}
\def\prereactivepowers{}

\def\postvoltages{}
\def\posttaps{}
\def\postfieldcurrents{}
\def\postNLIs{}
\def\postactivepowers{}
\def\postaxis{}
\def\postscope{}
\newif\iftaps\tapstrue
\pgfmathsetmacro\tapheight{0.05}
\pgfmathsetmacro\tapwidth{0.11}
\pgfmathsetmacro\tapfraction{0.5} 
\pgfdeclareplotmark{tapdownmark}{%
    \path 
        (\tapwidth/2, \tapfraction*\tapheight) 
        coordinate (A)
        (-\tapwidth/2, \tapfraction*\tapheight) 
        coordinate (B)
        (0, \tapfraction*\tapheight-\tapheight)
        coordinate (C);
    \path[fill] (A) -- (B) -- (C) -- cycle;
}
\pgfdeclareplotmark{tapupmark}{%
    \path 
        (\tapwidth/2, \tapfraction*\tapheight - \tapheight) 
        coordinate (A)
        (-\tapwidth/2, \tapfraction*\tapheight - \tapheight) 
        coordinate (B)
        (0, \tapfraction*\tapheight)
        coordinate (C);
    \path (A) -- (B) -- (C) -- cycle;
}
\pgfplotsset{
    tap up/.style={
        draw,
        mark=tapupmark,
    },
    tap down/.style={
        draw,
        mark=tapdownmark,
    },
    at 1-1041/.style={
        Dark2-A,
    },
    at 1-1041 image/.style={
    },
    at 2-1042/.style={
        Dark2-B,
    },
    at 3-1043/.style={
        Dark2-C,
    },
    at 4-1044/.style={
        Dark2-D,
    },
    at 5-1045/.style={
        Dark2-E,
    },
    at 41-4041/.style={
        Dark2-F,
    },
    at 42-4042/.style={
        Dark2-G,
    },
    at 43-4043/.style={
        Dark2-H,
    },
    at 46-4046/.style={
        Paired-C!80,
    },
    at 47-4047/.style={
        Paired-E,
    },
    at 51-4051/.style={
        Paired-I,
    },
}
\pgfplotsset{
	voltage 1/.style={
    },
	voltage 2/.style={
    },
	voltage 3/.style={
    },
	voltage 4/.style={
    },
	voltage 5/.style={
    },
	voltage 41/.style={
    },
	voltage 42/.style={
    },
	voltage 43/.style={
    },
	voltage 46/.style={
    },
	voltage 47/.style={
    },
	voltage 51/.style={
    },
    current g6/.style={
    },
    current g7/.style={
    },
    current g13/.style={
    },
    current g14/.style={
    },
    current g15/.style={
    },
    current g16/.style={
    },
}
\newif\iffield\fieldtrue
\newif\ifpowers\powerstrue
\newif\ifactivepower\activepowerfalse
\newif\ifnli\nlitrue
\newif\ifaggregated\aggregatedtrue
\newlength\groupvsep
\def\groupsize{5}
\newcounter{panelnumber}
\newcommand\placepanelnumber{\stepcounter{panelnumber}(\alph{panelnumber})\xspace}
\newcommand\plotresults[6]{%

    \begin{figure}[t]
        \centering
        \ifcompileplots
        \setcounter{panelnumber}{0}
        \begin{tikzpicture}[
            every path/.style={
                line cap=round,
                line join=round
            },
        ]
            \begin{scope}[
                spy using outlines={
                    circle, 
                    magnification=7, 
                    connect spies,
                },
                every spy on node/.append style={
                    semithick,
                },
            ]

            \begin{groupplot}[
                range frame,
                width=0.89\columnwidth,
                height=0.532\columnwidth,
                group style={
                    group name=my plots,
                    group size=1 by \groupsize,
                    vertical sep=\groupvsep, %
                    xlabels at=edge bottom,
                },
                xlabel={Time},
                x unit=\si\minute,
                xtick={0, 1, ..., 8},
                clip=false,
                xmin=0,
                xmax=8, 
                legend cell align=right,
                legend columns=1,
                legend style={
                    fill=none,
                    draw=none,
                    at={(0.99, 1)},
                    inner sep=0pt,
                    anchor=north west,
                    column sep=1pt,
                    nodes={
                        scale=0.8,
                        transform shape,
                    }
                },
                semi precise y,
                clip=true,
            ]

                \nextgroupplot[
                    ylabel={\placepanelnumber Voltage},
                    y unit=\si{\pu},
                    ymin=0.947,
                    ymax=1.025,
                    ytick={0.94, 0.95, ..., 1.04},
                    precise y,
                    ylabel shift=-0.5pt
                ]

                    \prevoltages

                    \path[region, draw=none, opacity=0.5]
                        (axis cs: 0, 0.975)
                        rectangle
                        (axis cs: 8, 1.025)
                        node[opacity=1, below left, zeroline]{%
                            $1\pm\SI{0.025}{\pu}$%
                        };

                    \path[region, draw=none]
                        (axis cs: 0, 0.99)
                        coordinate (LTC bottom left)
                        rectangle
                        (axis cs: 8, 1.01)
                        coordinate (LTC top right);
                    \draw[zeroline]
                        (LTC bottom left) -- (LTC bottom left -| LTC top right)
                        (LTC bottom left |- LTC top right) --  (LTC top right)
                        node [below left] {\gls{LTC} deadband $=1\pm\SI{0.010}{\pu}$};

                    \foreach \x in {1, 2, 3, 4, 5, 41, 42, 43, 46, 47, 51}{
                        \edef\temp{
                            \noexpand\addplot+[curve, voltage \x] table[
                                x expr=\noexpand\thisrowno{0}/60,
                                y expr=\noexpand\thisrowno{1}
                            ] {#3/time_voltage/time_voltage_\x.txt};
                            \noexpand\addlegendentry{\x}
                        }
                        \temp
                    }

                    \postvoltages

                    \def\pgfplotsdataymin{0.950}
                    \def\pgfplotsdataymax{1.025}

                \iftaps

                    \nextgroupplot[
                        ylabel={\placepanelnumber \gls{LTC} actions},
                        ymin=-0.05,
                        ymax=1.15,
                        clip=true,
                        ylabel shift=0.73cm,
                        axis y line=left,
                        ytick=\empty,
                        tick align=outside,
                        axis line style={opacity=0},
                        after end axis/.code={
                            \iftufte
                                \iftuftex
                                    \draw[axescolor]
                                        ({rel axis cs:0,0}-|{axis cs:\pgfplotsdataxmin,0})
                                        --
                                        ({rel axis cs:0,0}-|{axis cs:\pgfplotsdataxmax,0});
                                \fi
                            \fi
                        },
                        legend image code/.code={%
                            \node at (0, 0){%
                                \pgfmathsetmacro\xscale{0.175*0.9}%
                                \pgfmathsetmacro\yscale{1.25*0.9}%
                                \begin{tikzpicture}
                                    \path[draw]
                                        (\xscale*\tapwidth/2, \yscale*\tapfraction*\tapheight) 
                                        coordinate (A)
                                        (-\xscale*\tapwidth/2, \yscale*\tapfraction*\tapheight)
                                        coordinate (B)
                                        (0, \yscale*\tapfraction*\tapheight - \yscale*\tapheight)
                                        coordinate (C);
                                    \path[draw, fill] (A) -- (B) -- (C) -- cycle;
                                \end{tikzpicture}%
                                \textcolor{black}{\,/\,}%
                                \begin{tikzpicture}
                                    \path[draw]
                                        (\xscale*\tapwidth/2, \yscale*\tapfraction*\tapheight - \yscale*\tapheight) 
                                        coordinate (A)
                                        (-\xscale*\tapwidth/2, \yscale*\tapfraction*\tapheight - \yscale*\tapheight) 
                                        coordinate (B)
                                        (0, \yscale*\tapfraction*\tapheight)
                                        coordinate (C);
                                    \path[draw] (A) -- (B) -- (C) -- cycle;
                                \end{tikzpicture}%
                            };
                        },
                    ]

                        \pretaps

                        \addplot+[draw=none, at 1-1041] coordinates {(0.5, 0.5)};
                        \addplot+[draw=none, at 2-1042] coordinates {(0.5, 0.5)};
                        \addplot+[draw=none, at 3-1043] coordinates {(0.5, 0.5)};
                        \addplot+[draw=none, at 4-1044] coordinates {(0.5, 0.5)};
                        \addplot+[draw=none, at 5-1045] coordinates {(0.5, 0.5)};
                        \addplot+[draw=none, at 41-4041] coordinates {(0.5, 0.5)};
                        \addplot+[draw=none, at 42-4042] coordinates {(0.5, 0.5)};
                        \addplot+[draw=none, at 43-4043] coordinates {(0.5, 0.5)};
                        \addplot+[draw=none, at 46-4046] coordinates {(0.5, 0.5)};
                        \addplot+[draw=none, at 47-4047] coordinates {(0.5, 0.5)};
                        \addplot+[draw=none, at 51-4051] coordinates {(0.5, 0.5)};

                        \input{#3/tap_movements/tap_movements.tex}

                        \addlegendentry{1}
                        \addlegendentry{2}
                        \addlegendentry{3}
                        \addlegendentry{4}
                        \addlegendentry{5}
                        \addlegendentry{41}
                        \addlegendentry{42}
                        \addlegendentry{43}
                        \addlegendentry{46}
                        \addlegendentry{47}
                        \addlegendentry{51}

                        \posttaps

                        \def\pgfplotsdataymin{-0.1}
                        \def\pgfplotsdataymax{1.05}

                \fi

                \iffield
                    \nextgroupplot[
                        ylabel={\placepanelnumber $\normfieldcurrent$},
                        y unit=\si\norm,
                        legend cell align=left,
                        ymin=0.4,
                        ymax=1.4,
                        ytick={0.4,  0.6, 0.8,
                               1.0, 1.2, 1.4},
                        ylabel shift=2.3pt
                    ]

                    \prefieldcurrents

                    \path[region, draw=none, opacity=0.5]
                        (axis cs: 0, 0.4)
                        rectangle
                        (axis cs: 8, 1);

                    \draw[zeroline]
                        (axis cs: 0, 1)
                        --
                        (axis cs: 8, 1);
                    \draw[zeroline, latex-]
                        (axis cs: 6.5, 1.05)
                        --
                        ++(axis direction cs: 0.2, 0.1)
                        node[above, yshift=0.1cm]{Thermal limit};

                    \foreach \x/\y in {g6/{$\noexpand\generator_{6}$},
                                       g7/{$\noexpand\generator_{7}$},
                                       g13/{$\noexpand\generator_{13}$},
                                       g14/{$\noexpand\generator_{14}$},
                                       g15/{$\noexpand\generator_{15}$},
                                       g16/{$\noexpand\generator_{16}$}}{
                        \edef\temp{
                            \noexpand\addplot+[curve, current \x] table[
                                x expr=\noexpand\thisrowno{0}/60,
                                y expr=\noexpand\thisrowno{1},
                            ] {#3/time_fieldcurrentnorm/time_fieldcurrentnorm_\x.txt};
                            \noexpand\addlegendentry{\y}
                        }
                        \temp
                    }

                    \postfieldcurrents

                \fi

                \ifnli
                \nextgroupplot[
                    ylabel={\placepanelnumber $\avnli$},
                    y unit=\si{\pu\per\pu},
                    ymin=-1.5,
                    ymax=1.5,
                    ylabel shift=-3.6pt,
                ]

                    \preNLIs

                    \path[region, draw=none, opacity=0.5]
                        (axis cs: 0, 0)
                        rectangle
                        (axis cs: 8, 1.5)
                        ;
                    \draw[zeroline]
                        (axis cs: 8, 1.5)
                        node [below left, yshift=-0.1cm]
                        {Stable region};

                    \draw[zeroline]
                        (axis cs: \pgfplotsdataxmin, 0)
                        --
                        (axis cs: \pgfplotsdataxmax, 0);

                    \foreach \x in {4041, 4042}{
                        \edef\temp{
                            \noexpand\addplot+[curve] table[
                                x expr=\noexpand\thisrowno{0}/60,
                                y expr=\noexpand\thisrowno{1}
                            ] {#3/time_NLI/time_NLI_\x.txt};
                            \noexpand\addlegendentry{\x}
                        }
                        \temp
                    }

                    \postNLIs

                    \def\pgfplotsdataymin{-1.5}
                    \def\pgfplotsdataymax{1.5}
                \fi

                \ifactivepower
                    \nextgroupplot[
                        ylabel={\placepanelnumber $\activepower\fromactualDER$},
                        y unit={\si{\mega\watt}},
                        ymin=0,
                        ymax=250,
                        ytick={0, 50, ..., 250},
                        ylabel shift=0.9pt,
                        not precise y,
                    ]

                        \prereactivepowers

                        \foreach \x in {1, 2, 3, 4, 5, 41, 42, 43, 46, 47, 51}{
                            \edef\temp{
                                \noexpand\addplot+[curve] table[
                                    x expr=\noexpand\thisrowno{0}/60,
                                    y expr=\noexpand\thisrowno{1}
                                ] {#3/time_DERApower/new_time_DERApowerP_\x.txt};
                                \noexpand\addlegendentry{\x}
                            }
                            \temp
                        }

                        \def\pgfplotsdataymin{0}
                        \def\pgfplotsdataymax{250}
                        \postactivepowers

                \fi

                \ifpowers
                    \ifaggregated
                        \nextgroupplot[
                            ylabel={\placepanelnumber $\reactivepower\fromactualDER$},
                            y unit={\si{\mega\var}},
                            ymin=0,
                            ymax=65,
                            ytick={0, 10, ..., 60},
                            ylabel shift=5pt,
                            not precise y,
                        ]

                            \prereactivepowers

                            \foreach \x in {1, 2, 3, 4, 5, 41, 42, 43, 46, 47, 51}{
                                \edef\temp{
                                    \noexpand\addplot+[curve] table[
                                        x expr=\noexpand\thisrowno{0}/60,
                                        y expr=\noexpand\thisrowno{1}
                                    ] {#3/time_DERApower/new_time_DERApowerQ_\x.txt};
                                    \noexpand\addlegendentry{\x}
                                }
                                \temp
                            }

                            \def\pgfplotsdataymin{0}

                    \else
                        \nextgroupplot[
                            ylabel={\iffield(d)\else(c)\fi{} $\reactivepower\fromactualDER$},
                            y unit={\si{\mega\var}},
                            ymin=0,
                            ymax=2.5,
                            ytick={0, 0.5, 1.0, 1.5, 2.0, 2.5},
                            ylabel shift=0pt,
                            not precise y,
                        ]

                            \prereactivepowers

                            \foreach \x in {1, 2, ..., 50}{
                                \edef\temp{
                                    \noexpand\addplot+[curve] table[
                                        x expr=\noexpand\thisrowno{0}/60,
                                        y expr=\noexpand\thisrowno{1}
                                    ] {#3/time_DERApower/new_time_DERApowerQ_\x.txt};
                                }
                                \temp
                            }

                            \def\pgfplotsdataymin{0}
                            \def\pgfplotsdataymax{2.5}
                            \postreactivepowers

                    \fi
                \fi
            \end{groupplot}
            \postaxis
            \end{scope}
            \postscope
        \end{tikzpicture}
        \else
        \plotplaceholder
        \fi
        \vspace{-\myvsepunderplot}
        \mycaption{\nordic #1}{,}{%
        following the post-fault outage of
        \gls{HV} line $4032$--$4044$ at $\conttime=\SI{1}{\second}$%
        #6%
        }
        \vspace{-0pt}
        \label{#2}
    \end{figure}
}
\newcommand\uppointing{%
    \pgfmathsetmacro\xscale{0.9}%
    \pgfmathsetmacro\yscale{3.3*0.9}%
    \tikz{
        \path[draw]
            (\xscale*\tapwidth/2, \yscale*\tapfraction*\tapheight) 
            coordinate (A)
            (-\xscale*\tapwidth/2, \yscale*\tapfraction*\tapheight)
            coordinate (B)
            (0, \yscale*\tapfraction*\tapheight - \yscale*\tapheight)
            coordinate (C);
        \path[fill] (A) -- (B) -- (C) -- cycle;
    }%
}
\newcommand\downpointing{%
    \pgfmathsetmacro\xscale{0.9}%
    \pgfmathsetmacro\yscale{3.3*0.9}%
    \tikz{
        \path[draw]
            (\xscale*\tapwidth/2, \yscale*\tapfraction*\tapheight - \yscale*\tapheight) 
            coordinate (A)
            (-\xscale*\tapwidth/2, \yscale*\tapfraction*\tapheight - \yscale*\tapheight) 
            coordinate (B)
            (0, \yscale*\tapfraction*\tapheight)
            coordinate (C);
        \path[draw] (A) -- (B) -- (C) -- cycle;
    }%
}
\newcommand\tapcaption[1][b]{%
    Panel (#1) depicts reductions (\uppointing) and increases (\downpointing)
    in~$\tapratio$ at transformers with \gls{MV}-side buses given in the
    legend%
}
\newcommand\simhorizon{T}
\newcommand\timeaverage[1]{%
    \langle #1 \rangle_{\simhorizon}%
}
\newcommand\doublemean[1]{%
    \avg_\dummyindex\,{\timeaverage{#1}}%
}

\newcommand\VoltageIntegralFormula{%
    \doublemean{\abs{v_\dummyindex(t) - v_\dummyindex(0)}}%
}
\newcommand\NLIFormula{%
    \doublemean{\avnli_\dummyindex(t)}
}
\newcommand\ControlEffortPFormula{%
    \doublemean{\activepower\fromactualDER_\dummyindex(t) - \activepower\fromactualDER_\dummyindex(0)}
}
\newcommand\ControlEffortQFormula{%
    \doublemean{\reactivepower\fromactualDER_\dummyindex(t) - \reactivepower\fromactualDER_\dummyindex(0)}
}
\newcommand\ReactiveMarginFormula{%
    \doublemean{1 - \normfieldcurrent_\dummyindex(t)}
}
\newcommand\PowerReserveFormula{%
    \doublemean{\apparentpower\fromactualDER_{\dummyindex,\mathrm{max}} - \apparentpower\fromactualDER_\dummyindex(t)}
}

\usepackage{cite}
\usepackage{amsmath}
\usepackage{amsfonts}
\usepackage{array}
\usepackage{stfloats}
\usepackage{balance}
\usepackage{microtype}
\usepackage{orcidlink}
\usepackage{hyperref}
\hypersetup{
    hidelinks,
    pdfproducer={LaTeX},
    pdfcreator={pdflatex}
}
\makeatletter
    \AtBeginDocument{
      \hypersetup{
        pdftitle={\@title},
        pdfproducer= {LaTeX},
        pdfcreator = {pdflatex},
        bookmarks  = {true},
      }
    }
\makeatother

\usepackage{cleveref}
\crefname{figure}{Fig.}{Figs.}
\Crefname{figure}{Figure}{Figures}
\crefname{equation}{}{}
\crefname{table}{Table}{Tables}
\crefname{section}{Section}{Sections}

\newlength\Fcolumnseprule
\newif\ifcolumndivision
\columndivisiontrue
\columndivisionfalse
\ifshowframe
    \usepackage{showframe}
    \makeatletter
    \def\@outputdblcol{%
      \if@firstcolumn
        \global \@firstcolumnfalse
        \global \setbox\@leftcolumn \box\@outputbox
      \else
        \global \@firstcolumntrue
        \setbox\@outputbox \vbox {%
                             \hb@xt@\textwidth {%
                               \hb@xt@\columnwidth {%
                                 \box\@leftcolumn \hss}%
                               \vrule \@width\Fcolumnseprule\hfil
                                    {\normalcolor\vrule \@width\columnseprule}
                               \hfil\vrule \@width\Fcolumnseprule
                               \hb@xt@\columnwidth {%
                                 \box\@outputbox \hss}%
                                                 }%
                                  }%
        \@combinedblfloats
        \@outputpage
        \begingroup
          \@dblfloatplacement
          \@startdblcolumn
          \@whilesw\if@fcolmade \fi
            {\@outputpage
             \@startdblcolumn}%
        \endgroup
      \fi
    }
    \makeatother
\fi

\newcommand\postmortem{\foreign{postmortem}}
\newcommand\nordic{Nordic test system\xspace}

\title{%
    Emergency Control of Transmission Voltages by Coordinating DERs From
    Multiple Substations%
}

\begin{document}

\bstctlcite{BSTcontrol} 

\author{%
    Francisco~Escobar\,\orcidlink{0000-0002-6235-5689},%
        ~\IEEEmembership{Member,~IEEE,}
    and~Gustavo~Valverde\,\orcidlink{0000-0002-2506-3505},%
        ~\IEEEmembership{Senior Member,~IEEE}%
    \thanks{%
        This work was supported by the University of Costa Rica, project
        B9217-22.%
    }%
    \thanks{%
        The authors are with the Power Systems Laboratory, 
        ETH Zurich, 8092 Zurich, Switzerland 
        (e-mail: fescobar@ethz.ch; valverde@eeh.ee.ethz.ch).%
    }
}


\maketitle

\begin{abstract}
    Voltage stability can be endangered in \aclp{TN} that operate under high strain
and exhibit a large separation between generators and loads.
During emergencies associated to \glsin{VS}, one promising supporting
countermeasure is to control large populations of \glspl{DER} connected to
medium- and low-voltage networks, so that they locally reduce the net load by
modifying their power generation or consumption.
However, it is challenging to achieve this control while anticipating
interactions with legacy voltage regulators, coordinating efforts across
substations, and exchanging small amounts of information with \glspl{DER}.
In this paper, we propose a control scheme that coordinates \gls{DER} aggregations
with the \aclp{LTC} of substation transformers, asks them to contribute power
depending on the substation they are connected to, and relies on lightweight
communication.
Our simulations show that the proposed control can successfully coordinate
\glspl{DER} and \aclp{LTC} to enhance \gls{LTVS}.

\end{abstract}

\glsresetall


\IEEEpeerreviewmaketitle


\section{Introduction}
\label{sec:introduction}
\IEEEPARstart{I}{n recent years}, environmental and societal changes have drawn
attention to the stability of mature power~systems.
One change is the electrification of the heat and transport sectors, which
places more stress on the aging transmission system infrastructure.
One type of system stability that is affected by these changes is long-term
voltage stability~\cite{hatziargyriou2021}, as the combined generation and
transmission system might no longer be able to supply the power demanded by
loads after the network has been weakened by contingencies.
One recent example of long-term voltage instability is the grid incident in
South-East Europe on June 21, 2024~\cite{2025}.

To improve system stability while deferring costs, the power-engineering
community is trying to harness \glspl{DER}.
We use this term for all controllable devices not directly connected to the
bulk power system that are able to modify their active-power generation or
consumption, \ie including flexible loads.
Examples of \glspl{DER} are~\cite{kroposki2020} \gls{PV} systems, fuel cells,
micro-turbines, gensets, \glspl{BESS}, electric vehicles, smart appliances, and
electric heat pumps, among others.
The production coming from power-generating \glspl{DER} can meet, during
certain periods, most of the demand of large communities~\cite{wirth2021}.
Such penetration levels facilitate the provision of \glspl{AS} to \glspl{DN}
and even \glspl{TN}.

Despite their potential, controlling large populations of \glspl{DER} to
improve \gls{TN} stability poses three major challenges.
The first one is the interaction between distributed and fast-acting
\glspl{DER} with discrete and slow legacy voltage regulators, such as
\glspl{LTC}~\cite{kraiczy2018} and \glspl{SCB}.
The second challenge, which is particularly relevant for voltage-related
phenomena~\cite{sun2019,jaramillo2023}, is the need for
\secondaddition{intersubstation~coordination.}
Here,
\gls{COO} refers to the 
scaling and signing of remedial
actions based on system-wide knowledge.
The \postmortem analyses of major blackouts listed the lack of \gls{COO} as one
of their root causes, and warned that the \gls{COO} problem would change
\sic{completely}~\cite[p.\,78]{zotero-5591} and
\sic{dramatically}~\cite[p.\,53]{zotero-1137} with the proliferation of
\glspl{DER}.
Finally, the third challenge is the constraint on the amount of information
that can be transmitted and processed in real time, since a control
architecture that relies on direct communication to and from millions of
\glspl{DER} could be overwhelmed 
in
an~emergency~\cite{kroposki2020}.

The literature has scarcely reported on schemes for controlling \glspl{DER} to
enhance \gls{LTVS} while (a)~\secondaddition{coordinating} \glspl{DER} and
\glspl{LTC}, (b)~\secondaddition{achieving intersubstation coordination,} and
(c)~keeping the communication requirements low.
The authors in~\cite{aristidou2017} control the active and reactive power of
large-scale \glspl{DG} while anticipating \gls{LTC} movements, but the control
decisions are not coordinated among substations, and they rely on bidirectional
communication to send dedicated setpoints and retrieve individual measurements. 
These drawbacks are also present in~\cite{tran2022a}. 
The work in~\cite{escobar2022} proposes a rule-based \secondaddition{control}
of thousands of \glspl{DER} to support the grid during emergency situations. 
However, no \gls{COO} with \glspl{LTC} or other substations was considered.
The approaches from~\cite{morin2018, prionistis2021} act directly on
\glspl{LTC} and \secondaddition{achieve coordination,} but they also rely on
dedicated setpoints.
References~\cite{pabonospina2020, mandoulidis2022a,pabonospina2021} all propose
\glsers{LC} (hence lightweight communication) that take into account the
\gls{LTC}, but, being local, they cannot achieve
\secondaddition{intersubstation \gls{COO}.}

In this paper, we propose a \secondaddition{hierarchical control} that
addresses the previous gaps.
At the transmission level, it controls the \glspl{LTC} of step-down
transformers and sends optimal power requests to the corresponding \glspl{DN}.
These actions keep \gls{LTVS} according to a numerical indicator that is
monitored through \glspl{PMU}. 
Then, at the distribution level, an \secondaddition{\gls{CO}} translates the
power requests into signals that are broadcast to all \glspl{DER} connected
downstream. 
Finally, a controller inside each individual \gls{DER} translates the signals
into active- and reactive-power contributions, aligned with the work
in~\cite{escobar2023a}. The disaggregation of power requests to individual \glspl{DER} is beyond the scope of this paper.

Related work has resorted to a broad range of control techniques.
These include rule-based control~\cite{pabonospina2020}, \gls{LC} steered
by centralized \gls{AC-OPF} solutions~\cite{prionistis2022a}, control dispatch
through dynamic programming~\cite{wu2001}, data-augmented
\gls{MPC}~\cite{ma2014a}, agent-based control~\cite{robitzky2018a}, and
data-based learning and control~\cite{cai2020}. 
This paper uses \gls{MPC} for three main reasons.
Firstly, \gls{MPC} handles multiple inputs and outputs, making it suitable to
control actuators from different substations and thus achieve \gls{COO}.
Secondly, it accommodates constraints, so that prescribed limits on voltages
and stability indicators can be explicitly embedded into the control law.
Finally, aided by a model and a feedback mechanism, it outputs the scale and
sign of \incorrect{good} control actions, thus contributing to the \gls{COO}.

The proposed \ac{MPC} coordinates resources from multiple substations. 
It also embeds a stability indicator whose sensitivity to manipulated variables
is estimated offline using a~static analysis.  
Being centralized~\cite{antoniadou-plytaria2017}, it does not require a
consensus~strategy.

A key feature of the proposed scheme is that it recognizes that some \glspl{DN}
must contribute more power than others and that the \glspl{LTC} must move in
(possibly) counter-intuitive directions to achieve a common goal.
Furthermore, since the \gls{MPC} profits from both a \gls{TN} model and
\gls{TN} measurements, it can outperform a comparable model-free \gls{LC}
scheme.

The main contributions of this paper are:
\begin{enumerate}
    \item A qualitative description of a long-term instability mechanism in the
        presence of \glspl{DER} and \glspl{LTC}, reflecting the need for
        coordination during voltage emergencies. 
    \item An optimization-based control scheme that takes into account the
        \glspl{DER} and \glspl{LTC} from multiple substations.  
        It is robust to model inaccuracies and system uncertainties.  
        To enhance the \gls{LTVS}, it embeds a model of a stability indicator
        in the prediction horizon.  
        Furthermore, it smoothly drives the voltages of controlled \gls{TN} and
        \gls{DN} buses within a desired band of operation.
    \item A comparison with a fully \gls{LC} scheme to demonstrate the
        benefits of \secondaddition{intersubstation coordination.}
        The performance of the controllers is assessed in terms of voltage
        deviations, usage of \glspl{LTC} and \gls{DER} power reserves, as well
        as reliance on nearby synchronous~machines.
\end{enumerate}

The rest of the paper is organized as follows.  
\Cref{sec:background} revisits the intricacies of supporting transmission
voltages from \glspl{DN}. 
It provides a comprehensive description of the mechanisms that govern long-term
voltage instability in the presence of high \gls{DER} penetrations, focusing on
device interactions and their impact on \ac{TN} and \ac{DN} voltages.
The proposed control scheme is introduced in \Cref{sec:proposed-control},
followed by a benchmarked case study in \Cref{sec:simulation-results}.
Finally, \cref{sec:conclusion} concludes the paper and discusses future work.

\section{Emergency Control of Transmission Voltages}  

\newcommand\widthTN{1.0cm}
\newcommand\widthDN{0.0cm}
\newcommand\DNsep{4.75cm}
\newcommand\bussep{0.1}
\pgfmathsetmacro\DERAsep{3}

\newcommand\surfrad{0.35}
\newcommand\busvsep{1.2}
\newcommand\bushsep{5}
\newcommand\busprot{0.3}
\newcommand\busheight{0.45}
\newcommand\gendist{0.5}
\newcommand\genrad{4pt}
\newcommand\loadh{0.35}
\newcommand\loadv{0.4}
\newcommand\dnsep{2.5}
\newcommand\tnsep{0.9}

\label{sec:background}
We first revisit the interplay of voltages, \glspl{LTC}, and powers in
\glspl{TN}, which justifies the need for \gls{COO}.
We then turn our attention to \glsin{VS} detection in real time.
In this paper, we denote constant (resp.\ time-varying) scalars using italic
uppercase (resp.\ lowercase) letters. 
We distinguish phasors with a tilde, as in $\phasor\dummyphasor$\!, and average
values with a bar, as in $\avnli$. 
Matrices (resp.\ column vectors) are written in boldface uppercase (resp.\
lowercase). 
The complex conjugate of $\phasor\dummyphasor$ is $\conj\phasor\dummyphasor$
and the setpoint of a signal~$\dummysignal$ is $\setpoint\dummysignal\!$.

\subsection{The Need for Coordination}
\label{sub:coordination}

\begin{figure}
    \centering
    \includegraphics{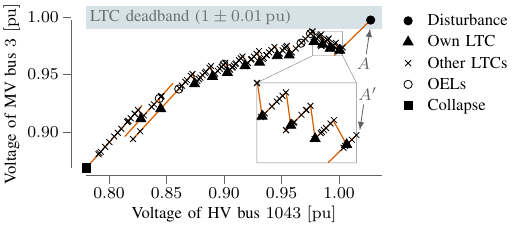}%
    \caption{Response of the \nordic to a fault near \gls{HV} bus $4032$ and 
             the subsequent tripping of line $4032$--$4044$.}
    \label{fig:collapse}
\end{figure}

The need to coordinate efforts \secondaddition{from \glspl{DER} and voltage
regulators across substations} becomes clear when analyzing a system that is
subject to a pure \gls{LC} scheme.
We take as an example the \nordic~\cite{vancutsem2020}.
After a fault in the \gls{TN} and the corresponding line tripping, the voltages
at both sides of a step-down transformer ($3$--$1043$), \ie its \gls{HV} and
\gls{MV} sides, evolve as shown in~\cref{fig:collapse}.
The solid line represents the trajectory of those two voltages plotted against
each other, with time as an implicit variable, whereas the markers indicate the
occurrence of discrete events throughout the system.
The large electromechanical oscillations due to the fault have been removed
from the plot for legibility.

Initially, the system is pushed away from the pre-disturbance long-term
equilibrium~$A$ towards the short-term equilibrium~$A^\prime$.
Once the electromechanical oscillations die out, the system evolves under the
action of \glspl{LTC} and \glspl{OEL} of large synchronous machines,
culminating in a \gls{VC}.
While some voltage drops are caused by the \glspl{OEL}, the collapse is mainly
due to the \glspl{LTC}.
Indeed, in their effort to restore their \gls{MV} voltage, the \glspl{LTC} also
restore load powers, thus inadvertently depressing their own \gls{HV} voltage
and causing some machines to operate beyond the thermal limit of the field
winding.
Moreover, since there is no \secondaddition{intersubstation} \gls{COO} and the
\glspl{LTC} focus on their own \gls{MV} side, they do not realize that their
actions also affect the \gls{HV} and \gls{MV} voltages of other~substations.

\begin{figure}
    \centering
    \includegraphics{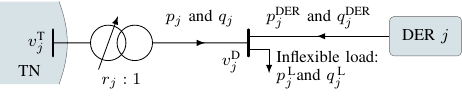}%
    \caption{Transformer with an \gls{LTC} feeding an inflexible, but
             voltage-sensitive, load in parallel with a \gls{DER}.}
    \label{fig:variables}
\end{figure}

With a view to proposing an effective \gls{CC}, we take a closer look at the
variables at play in the previous scenario.
For the $\substation$th step-down transformer of \cref{fig:variables}, we are
interested in its tap ratio~$\tapratio\atsubstation\in\reals$ and stack it into
the vector
\begin{equation}
    \label{eq:vector-tap-ratio}
    \myvector\tapratio
    =
    \bighorarray{\tapratio_1}
                {\tapratio_2}
                {\tapratio_{\substationcount}}\in\reals[\substationcount],
\end{equation}
where~$\substationcount$ is the number of step-down transformers in the
\gls{WA}. 
Similarly, the voltages of the \gls{HV} and \gls{MV} sides are grouped
resp.\,into $\myvector\voltage\attransmission\in\reals[\substationcount]$
and~$\myvector\voltage\atdistribution\in\reals[\substationcount]$\!, while the
active and reactive powers delivered by the transformers are grouped into
$\myvector\activepower\in\reals[\substationcount]$ and
$\myvector\reactivepower\in\reals[\substationcount]$\!. 
In this discussion, we consider for simplicity that the transformers feed
aggregate inflexible loads%
    \footnote{%
        We consider inflexible loads to be the complement of the set of
        \glspl{DER}, \ie inflexible loads  encompass non-controllable generators
        (as negative loads).%
    }
that consume $\myvector\activepower\fromload\in\reals[\substationcount]$ and
$\myvector\reactivepower\fromload\in\reals[\substationcount]$, together with
aggregate \glspl{DER} that inject
$\myvector\activepower\fromaggregatedDER\in\reals[\substationcount]$
and~$\myvector\reactivepower\fromaggregatedDER\in\reals[\substationcount]$.

\begin{figure}
    \centering
    \input{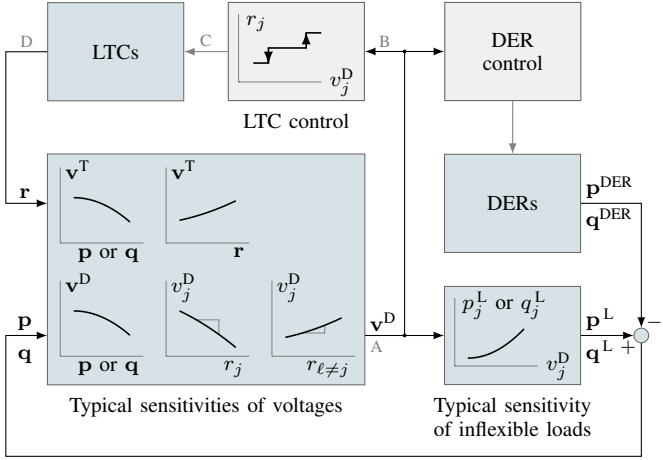}%
    \caption{Typical qualitative relationships between network voltages, tap
             ratios, and load and \gls{DER} consumed powers in a generic
             inductive network.}
    \label{fig:challenges}
\end{figure}

The qualitative interplay of the selected variables is depicted in
\cref{fig:challenges}.
The smooth curves summarize the physics of inductive networks, and the
step-wise curve stands for the discrete, timed control of \glspl{LTC}.
The powers generated by the \glspl{DER} are subtracted bus-wise from the
voltage-sensitive powers of the inflexible loads to give
$\myvector\activepower$ and~$\myvector\reactivepower$.
As expected, when operating in the stable branch of PV and QV
curves~\cite{cutsem1998}, an increase in either $\myvector\activepower$ or
$\myvector\reactivepower$ pulls down the voltage at both sides of any
transformer, which we express by saying that
$\myvector\voltage\attransmission$\!  and~$\myvector\voltage\atdistribution$\!
are decreasing functions of $\myvector\activepower$
and~$\myvector\reactivepower$.
On the other hand, the effect of $\myvector\tapratio$ on
$\myvector\voltage\attransmission$ and~$\myvector\voltage\atdistribution$ must
be differentiated. 
While the former is always an increasing function of $\myvector\tapratio$, the
latter can be either decreasing (desirable) or increasing (undesirable).
This is because the favorable effect of $\tapratio\atsubstation$ on
$\voltage\atdistribution\atsubstation$ might be outweighed by the detrimental,
cumulative effect of the remaining $\tapratio_\dummyindex$ with
$\dummyindex\neq\substation$ (which restore their voltage-sensitive loads).
We refer to this phenomenon, which is characteristic of weakened networks, as a
\term{loss of diagonal dominance}~\cite{vournas2008}.

\newcommand\myloop[4]{\textsc{#1}\textsc{#2}\textsc{#3}\textsc{#4}\textsc{#1}}

This perspective helps further understand the \glsin{LTVS} of the \nordic. 
The gradual loss of diagonal dominance with each \gls{OEL} intervention
essentially turns the loop~\myloop{a}{b}{c}{d} (labeled in
\cref{fig:challenges}) into a positive feedback loop.
After the line tripping, the undervoltages motivate the \glspl{LTC} to reduce
the tap ratios, which in turn causes the voltages to drop further, leading to
instability.
The asymmetry between the effects of $\myvector\tapratio$ on
$\myvector\voltage\attransmission$ and $\myvector\voltage\atdistribution$ also
explains why the horizontal drift in \cref{fig:collapse} is always towards the
west, while the vertical drift is towards the north and south directions.

The qualitative model of \cref{fig:challenges} captures another complication
that appears when \glspl{DER} come into the picture, especially if they are
controlled \secondaddition{without intersubstation coordination.}
The key observation is that changing the \gls{DER} injections
$\myvector\activepower\fromaggregatedDER$ and
$\myvector\reactivepower\fromaggregatedDER$ also affects the consumption of the
inflexible loads, and hence the overall effect on \glsin{VS} is not
straightforward.
For example, if the \gls{DER} control only increases the reactive injections
$\myvector\reactivepower\fromaggregatedDER$ to reduce the net
loads~$\myvector\reactivepower$, this will raise the
voltages~$\myvector\voltage\atdistribution$ and hence restore the
voltage-sensitive loads, leading to an increase in~$\myvector\activepower$.
Depending on the \gls{TN} physics, on the load sensitivities, and on the
precise \gls{DER} control~\cite{liemann2019}, this hidden effect might
precipitate \glsin{VS}~\cite{aristidou2017}.

The previous interactions become more complex when considering the system
dynamics, the simultaneous actions of both \glspl{LTC} and \glspl{DER}, and the
effects of other voltage regulators operating in the system. 
Despite these simplifications, the discussion highlights that instability might
be precipitated, or even driven, by the lack of \gls{COO}.

\subsection{Real-Time Detection of Voltage Instability}
\label{sub:monitoring}

\begin{figure}
    \centering
    \includegraphics{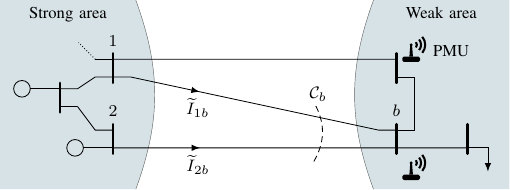}%
    \caption{Transmission corridor between a strong and a weak (receiving)
             area.}
    \label{fig:NLI}
\end{figure}

Still another degree of complexity is added by the detection of \glsin{VS}.
In general, this detection cannot rely on the voltages alone, and for high
\gls{DER} shares, it cannot rely only on the interplay of voltages and
\gls{LTC} movements.
To solve this problem, we resort in this paper to the \ac{NLI}, proposed
in~\cite{vournas2017} and later validated in~\cite{lambrou2021}.
Calculations performed on real historic measurements show that the \gls{NLI} is
effective, selective, and reliable~\cite{lambrou2021}.
An assumption for computing this indicator is that the boundary buses of the
weak area are equipped with \acp{PMU}, as shown in \cref{fig:NLI}.
We summarize the calculation procedure below.

At the boundary bus~$\boundarybus$, the \ac{PMU} measures the bus
voltage~$\phasor\voltage\atboundarybus$ and the total imported current
$
    \phasor\current\atboundarybus
    =
    - \sum_{\omega\,\in\,\nlicut\atboundarybus}
    \phasor\current_{\boundarybus\,\omega}
$,
where $\nlicut\atboundarybus$ is the set of sending buses that are adjacent to
bus~$\boundarybus$, and $\phasor\current_{\boundarybus\,\omega}$ is the total
current flowing from~$\boundarybus$ to bus~$\omega$.
After computing the derived quantities
$
    \activepower\atboundarybus
    =
    \Re (\phasor\voltage\atboundarybus\conj{\phasor\current}\atboundarybus)
$
and
$
    \conductance\atboundarybus
    =
    \Re (\phasor\current\atboundarybus/\phasor\voltage\atboundarybus)
$, 
the instantaneous \gls{NLI} at bus~$\boundarybus$ is defined as
\begin{equation}
    \label{eq:NLI}
    \nli\atboundarybus = 
    \frac{
        \Delta\activepower\atboundarybus
    }{
        \Delta\conductance\atboundarybus
    }\,,
\end{equation}
where $\Delta$ signifies the difference between two time instances.
This definition is motivated by the fact that when the corridor reaches its
power-transfer limit, an increase of $\conductance\atboundarybus$ does not lead
to an increase of $\activepower\atboundarybus$, and hence the indicator becomes
nonpositive. 
Fast transients can be suppressed by subjecting all measurements to moving
averages~\cite{vournas2017}, and the state of the \glspl{OEL} elsewhere in the
system can be used as a criterion to reset the \glspl{NLI}~\cite{escobar2022}.

\section{Proposed Control}
\label{sec:proposed-control}
{\newcommand\cost{
    \costfunction(
        \increment\myvector\controlsignal, 
        \myvector\slack
    )
}
\newcommand\emergencyconstant{z_\mathrm{em}}
\newcommand\restorationconstant{z_\mathrm{re}}
\newcommand\initialu{\myvector\controlsignal\attime{-1}}

We propose a hierarchical control scheme where the
centralized~\cite{antoniadou-plytaria2017} controller at the higher level has
the objective of maintaining long-term \gls{VS} in the \gls{TN}.
Because an adequate emergency control must foresee the effects of different
actions on the voltages and the \glspl{NLI}, a method that naturally lends
itself for the implementation of this higher-level controller is \gls{MPC}.
In cases where the constraints of the control problem do not exhibit high
time-variability, it might suffice to consider unitary control and prediction
horizons, but we consider arbitrary horizons for completeness.

\subsection{Control Architecture}
\label{sub:architecture}

The controller acts on the \gls{LTC} of step-down transformers in \glspl{WA}
(see \cref{fig:scheme}).
By sending remote commands, the controller lowers, maintains, or raises the tap
position of these \glspl{LTC}, whichever is more effective for preventing
\glsin{VS}. 
Unlike~\cite{valverde2013}, this direct action makes it unnecessary to modify
\gls{LTC} setpoints or anticipate tap movements.  
The controller also requests support from \glspl{DER} in the \gls{DN} that is
fed by the step-down transformer.
We assume that the share of \glspl{DER} is high enough to enable \glspl{AS}
that have an appreciable effect on the \gls{TN}. 
All information sent to the \glspl{DER} is preprocessed by an
\secondaddition{\gls{CO}}.

\begin{figure}
    \centering
    \includegraphics{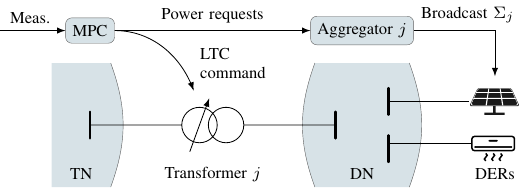}%
    \caption{Proposed control architecture.}
    \label{fig:scheme}
\end{figure}

\subsection{\gls{MPC} and \secondaddition{Aggregator} Interaction}

We assume that there is one \secondaddition{\gls{CO}} per step-down
transformer~$\substation$. 
This agent translates the power requests of the \gls{MPC} into a common,
simplified signal~$\signal\atsubstation$ that is broadcast to all downstream
devices. 
The \glspl{DER} receive and process the \secondaddition{\gls{CO}} signals but
they do not communicate back their decision to participate. 
Because of this, the exact number of downstream \glspl{DER} and their
availability to support the \gls{TN} remain uncertain to the
\secondaddition{\gls{CO}} and the \gls{MPC} controller.
The control loop is closed by measuring the joint contribution of \glspl{LTC}
and \glspl{DER} and feeding those measurements back into the \gls{MPC}
controller.  
However, since it may be impossible to measure the response of each \gls{DER}
individually, all measurements are taken at the \gls{TN} level {alone}. 

The \secondaddition{\gls{CO}} signal~$\signal\atsubstation$ has two channels
for independent active- and reactive-power requests, and we denote this as
$\signal\atsubstation = \channels{\signal[p]}{\signal[q]}$.
This signal is issued with a period~$\samplingperiod$ of a few seconds,
and~$\signal[p]$ and~$\signal[q]$ take independent values in the set of
integers $\{-\sizevalueset, \ldots, \sizevalueset\}$. 
Power injection is denoted by positive values, consumption by negative values,
and inaction by~$0$. 
Furthermore, the \secondaddition{\gls{CO}} sets an injection (resp.
consumption) limit at the \gls{T-D} boundary and broadcasts~$\sizevalueset$
(resp.\,$-\sizevalueset$) in the corresponding channel for all requests beyond
that limit.

Although~$\signal\atsubstation$ is received by a large number of \glspl{DER},
the \gls{MPC} controller sticks to the simplified model of \cref{fig:variables}
and pictures these devices as a fictitious, aggregate \gls{DER}~$\substation$.
This \gls{DER} is assumed to respond fast, and its initial output is zero for
both $\activepower\fromactualDER\atsubstation$
and~$\reactivepower\fromactualDER\atsubstation$.
Similarly, the controller lumps all loads fed by the \gls{DN} into an
aggregate, voltage-sensitive load, represented by the exponential load model
\begin{equation}
    \label{eq:exponential-model}
    {\estimated\activepower}\atsubstation\fromload
    = 
    \activepower\atsubstation[\initialsymbol]\fromload
    \left(
        \frac{
            \voltage
            \atsubstation
            \atdistribution
        }{
            \voltage
            \atsubstation[\initialsymbol]
            \atdistribution
        }
    \right) ^ {\!\pexponent\atsubstation}
    \quad
    \text{and}
    \quad
    {\estimated\reactivepower}\atsubstation\fromload
    = 
    \reactivepower\atsubstation[\initialsymbol]\fromload
    \left(
        \frac{
            \voltage
            \atsubstation
            \atdistribution
        }{
            \voltage
            \atsubstation[\initialsymbol]
            \atdistribution
        }
    \right) ^ {\!\qexponent\atsubstation},
\end{equation}
where the hat denotes estimated values, the subscript~$0$ denotes measurements
collected before the \gls{MPC} starts acting, and~$\pexponent\atsubstation$
and~$\qexponent\atsubstation$ are parameters that can be
estimated~\cite{milanovic2013}. 
In \glspl{DN} with a large share of induction motors, the model could be
replaced by a composite load model~\cite{arif2018}.
The inherent feedback of \gls{MPC} should compensate for the use of such
simplified models.

One set of measurements received by the \gls{MPC} controller contains the
manipulated variables~$\myvector\controlsignal$. 
This vector includes the ratio~$\tapratio\atsubstation$ and the \gls{DER}
contributions~${\estimated\activepower}\fromactualDER\atsubstation$
and~${\estimated\reactivepower}\fromactualDER\atsubstation$ at each step-down
transformer~$\substation$.
While the tap ratio can be measured directly, the net contribution of the
\gls{DER} population must be estimated, since the change in power measured at
the \gls{T-D} boundary includes both the power losses in the \gls{DN} and the
\gls{LR} due to the load sensitivity to voltages. 
This net contribution is estimated as
\begin{equation}
    \label{eq:estimated-DER-contribution}
    {\estimated\activepower}\fromactualDER\atsubstation
    =
    {\estimated\activepower}\fromload\atsubstation
    -
    \activepower\atsubstation
    \quad
    \text{and}
    \quad
    {\estimated\reactivepower}\fromactualDER\atsubstation
    =
    {\estimated\reactivepower}\fromload\atsubstation
    -
    \reactivepower\atsubstation\,,
\end{equation}
where $\activepower\atsubstation$ and~$\reactivepower\atsubstation$ are
directly measured.
The other set of measurements received by the \gls{MPC} contains the output
variables~$\myvector\outputsignal$. 
They include voltage magnitudes at both sides of the step-down transformers,
along with the \gls{NLI} as an indicator of \glsin{VS} in the~\gls{TN}.

\subsection{\gls{MPC} Formulation}
\label{sub:mpc-formulation}

To request the changes in the manipulated variables, the \gls{MPC} controller
solves an optimization problem at discrete time
instances~$\horizonindex\in\setk$ separated by the so-called sampling period.
The vector of \emph{hypothetical} manipulated variables at
time~$\horizonindex+\dummyindex$ thus takes the~form
\begin{equation}
    \label{eq:manipulated-variables}
    \myvector\controlsignal\attime{\horizonindex+\dummyindex}
    =
    \horarrayofvectors{\myvector\tapratio\attime{\horizonindex+\dummyindex};
                       \myvector\activepower\fromaggregatedDER\attime{\horizonindex+\dummyindex};
                       \myvector\reactivepower\fromaggregatedDER\attime{\horizonindex+\dummyindex}}
\end{equation}
for $\dummyindex=0, \ldots, \controlhorizon-1$, where $\controlhorizon$ denotes
the control horizon and\,\,$\transpose{}$\,denotes transposition.
Similarly, the vector of \emph{measured} outputs at time~$\horizonindex$~is
\begin{equation}
    \label{eq:outputs}
    \myvector\outputsignal\attime\horizonindex
    =
    \horarrayofvectors{
        \myvector\voltage\attransmission\attime\horizonindex;
        \myvector\voltage\atdistribution\attime\horizonindex;
        \myvector\nlivector\attime\horizonindex
    }\,,
\end{equation}
where the subvector $\myvector\nlivector\attime\horizonindex$ containts the
\glspl{NLI} of the $\boundarybuscount$ monitored \glspl{BB} in the \gls{WA}.  
For convenience, the decision variables of the optimization problem are
considered to be the \emph{changes} in the manipulated variables, which can be
defined in terms of the previous values as
\begin{equation}
    \increment\myvector\controlsignal\attime{\horizonindex+\dummyindex}
    =
    \myvector\controlsignal\attime{\horizonindex+\dummyindex}
    -
    \myvector\controlsignal\attime{\horizonindex+\dummyindex-1}
    \,.
\end{equation}

To
predict~$\myvector\outputsignal\attimegiven{\horizonindex+\dummyindex}\horizonindex$,
understood as the outputs at time~$\horizonindex+\dummyindex$ given their
values at time~$\horizonindex$, we employ the linearization
\begin{equation}
    \myvector\outputsignal\attimegiven{\horizonindex+\dummyindex}{\horizonindex}
    =
    \myvector\outputsignal\attimegiven{\horizonindex+\dummyindex-1}{\horizonindex}
    +
    \displaysensitivity{\myvector\outputsignal}{\myvector\controlsignal}
    \,
    \increment\myvector\controlsignal\attime{\horizonindex+\dummyindex-1}
\end{equation}
for $\dummyindex = 1, \ldots, \predictionhorizon$,
where~$\predictionhorizon\geq\controlhorizon$ is the prediction horizon and
$\inlinesensitivity{\myvector\outputsignal}{\myvector\controlsignal}$ contains
the sensitivities of voltages and \glspl{NLI} to changes in the manipulated
variables. 
The
matrix~$\inlinesensitivity{\myvector\outputsignal}{\myvector\controlsignal}$
must correspond to the post-\gls{DIST} operating point and the \gls{TN} topology
and can be updated very infrequently, if at
all~\cite{valverde2013,soleimanibidgoli2016,valverde2013a}. 
Furthermore,
$\inlinesensitivity{\myvector\outputsignal}{\myvector\controlsignal}$ can be
either computed from a power-flow model or estimated from
measurements~\cite{chen2016a} (see \cref{sub:nli-embedding} for details on the
\gls{NLI} sensitivities).
The inaccuracies derived from the infrequent update, the model mismatch, or the
estimation are, once again, compensated for by the closed-loop nature of
\gls{MPC}.

When the emergency unfolds, the \gls{MPC} has the objective of minimizing the
changes~$\increment\myvector\controlsignal\attime{\horizonindex+\dummyindex}$,
since abrupt changes could deteriorate the \gls{TN} dynamics and hence
precipitate instability.
Once the emergency is over, the objective shifts to restoring the manipulated
variables~$\myvector\controlsignal$, especially the \gls{DER} powers, to their
pre-\gls{DIST} value.
We thus consider the \objective function
\begin{equation}
    \label{eq:objective-emergency}
        \begin{split}
            &\emergencyconstant\!
            \sum_{\dummyindex = 0}^{\controlhorizon - 1}
            \weightednorm{\weightchanges}
                         {2}
                         {\increment\myvector\controlsignal(\horizonindex+\dummyindex)}
            \\
            &
        \quad
        \quad
        \quad
            + \restorationconstant\!
        \sum_{\dummyindex = 0}^{\controlhorizon - 1}
        \weightednorm{\weightu}{2}{\myvector\controlsignal(\horizonindex+\dummyindex) - \initialu}
        +
        \weightednorm{\weightslacks}{2}{\myvector\slack},
    \end{split}
\end{equation}
where $\emergencyconstant$ and $\restorationconstant$ are binary, mutually
exclusive parameters to select the objective (emergency or restoration),
$\weightednorm{\dummymatrix}2{\dummyvector}\! =\!
\transpose{\dummyvector}\dummymatrix\dummyvector$ denotes the norm of a
vector~$\dummyvector$ dictated by a diagonal, nonnegative weight
matrix~$\dummymatrix$, $\myvector\controlsignal(-1)$ denotes the pre-\gls{DIST}
value of the manipulated variables, \ie before the first \gls{MPC} iteration
($\horizonindex=0$), and~$\myvector\slack$ is a vector of positive slack
variables that render the constrained problem feasible when needed. 
The weight matrix~$\weightchanges$ discriminates between \incorrect{cheap} and
\incorrect{expensive} changes in the manipulated variables, $\weightu$
prioritizes the manipulated variables to be restored, and~$\weightslacks$
penalizes the slack variables to keep them at zero whenever possible.

\newcommand\totalinputs{3\substationcount}
\newcommand\totaloutputs{2\substationcount + \boundarybuscount}

The optimization problem is more succinctly formulated with all quantities in
matrix form. 
The sequences of manipulated variables and their changes over the control
horizon become
\begin{equation}
    \myvector\controlsignal
    =
    \bigverarray{
        \myvector\controlsignal\attime\horizonindex}
                {\myvector\controlsignal\attime{\horizonindex+1}}
                {\myvector\controlsignal\attime{\horizonindex+\controlhorizon\!-\!1}}\!
    \text{ and }
    \increment\myvector\controlsignal
    =
    \bigverarray{
        \increment\myvector\controlsignal\attime\horizonindex}
                {\increment\myvector\controlsignal\attime{\horizonindex+1}}
                {\increment\myvector\controlsignal\attime{\horizonindex+\controlhorizon\!-\!1}}\!.
\end{equation}
If~$\increment\myvector\controlsignal$ is known, then~$\myvector\controlsignal$
can be constructed directly from
$\myvector\controlsignal\attime{\horizonindex-1}$, \ie from the measurements
that are estimated right before running the optimization:
\begin{equation}
    \label{eq:affinity-of-u}
    \myvector\controlsignal
    =
    \measuredinputmatrix\myvector\controlsignal\attime{\horizonindex-1}
    +
    \inputincrementmatrix\increment\myvector\controlsignal\,,
\end{equation}
where $\measuredinputmatrix$ is a vertical stack of~$\controlhorizon$ identity
matrices~$\identity$ of dimension~$\totalinputs\times\totalinputs$ each,
and~$\inputincrementmatrix$ is a block lower triangular matrix
with~$\totalinputs\controlhorizon$ rows and columns, where the nonzero
submatrices are copies of~$\identity$~\cite{soleimanibidgoli2018}. 
Similarly, the sequence of predicted outputs over the prediction horizon
becomes
\begin{equation}
    \myvector\outputsignal
    =
    \bighorarray{\transpose{\myvector\outputsignal\attimegiven{\horizonindex+1}\horizonindex}}
                {\!\!\!\!\!\!\!}
                {\transpose{\myvector\outputsignal\attimegiven{\horizonindex+\predictionhorizon}\horizonindex}}\!.
\end{equation}
This sequence can be constructed from the measurements
$\myvector\outputsignal\attimegiven \horizonindex\horizonindex$ using
\begin{equation}
    \label{eq:affinity-of-y}
    \myvector\outputsignal
    =
    \measuredoutputmatrix\,
    \myvector\outputsignal\attimegiven \horizonindex\horizonindex
    +
    \sensitivitymatrix
    \increment\myvector\controlsignal\,,
\end{equation}
where $\measuredoutputmatrix$ is a vertical stack of~$\predictionhorizon$
identity matrices~$\identity$ with dimension
$(\totaloutputs)\times(\totaloutputs)$ each, and~$\sensitivitymatrix$ is a
block lower trapezoidal matrix with $(\totaloutputs)\predictionhorizon$~rows
and $\totalinputs\controlhorizon$~columns, where the nonzero submatrices are
copies
of~$\inlinesensitivity{\myvector\outputsignal}{\myvector\controlsignal}$~\cite{soleimanibidgoli2018}.
Note that the matrices $\measuredinputmatrix$, $\inputincrementmatrix$ and
$\measuredoutputmatrix$ are purely structural and their entries only depend on
the number of boundary buses and step-down transformers, while
$\sensitivitymatrix$ is the matrix that models the evolution of the controlled
variables.

The optimization problem at each time instance~$\horizonindex$ can now be
formulated as follows:
\begin{subequations}
    \begin{alignat}{2}
        & \!\minimize_{\increment\myvector\controlsignal,\,\myvector\slack}
            & \quad &
            \emergencyconstant
            \transpose{\increment\myvector\controlsignal}
            \weightchanges
            \increment\myvector\controlsignal
            \nonumber \\[-0.2cm]
        & & &
            +
            \restorationconstant
            \transpose{
                \left(
                    \myvector\controlsignal-\myvector\controlsignal^-
                \right)
            }
            \weightu
                \left(
                    \myvector\controlsignal-\myvector\controlsignal^-
                \right)
            \nonumber \\
        & & &
            +
            \transpose{\myvector\slack}
            \weightslacks\,
            \myvector\slack
            \label{eq:canonical-objective}\\
        & \text{subject to} & &
            \increment\lowerbound{\myvector\controlsignal}
            \vectorleq
            \increment\myvector\controlsignal
            \vectorleq
            \increment\upperbound{\myvector\controlsignal}\,,
            \label{eq:canonical-constraint-du} \\
        & & &
            \lowerbound{\myvector\controlsignal}
            \vectorleq
            \myvector\controlsignal
            \vectorleq
            \upperbound{\myvector\controlsignal}\,,
            \label{eq:canonical-constraint-u} \\
        & & &
            -\lowerslacks + \lowerbound{\myvector\outputsignal}
            \vectorleq
            \myvector\outputsignal
            \vectorleq
            \upperbound{\myvector\outputsignal} + \upperslacks\,,
            \label{eq:canonical-constraint-y}
    \end{alignat}
\end{subequations}
where~$\myvector\controlsignal^-$ is a vertical stack of $\controlhorizon$
copies of $\myvector\controlsignal(-1)$, $\lowerslacks$ and~$\upperslacks$ are
two vectors whose components are contained in the vector of slack
variables~$\myvector\slack$, and $\vectorleq$~denotes component-wise
inequality. 
Constraints~\eqref{eq:canonical-constraint-du}
and~\eqref{eq:canonical-constraint-u} account for the intrinsic physical
limitations of tap movements and for possible contractual limitations on the
power exchange at the \gls{T-D} boundary.
Constraints~\eqref{eq:canonical-constraint-y} try to keep the monitored
voltages and \glspl{NLI} within acceptable limits, but they are treated, in
contrast to~\eqref{eq:canonical-constraint-du}
and~\eqref{eq:canonical-constraint-u}, as soft constraints. 
This is done to prevent infeasibility when a large \gls{DIST} pushes the system
far from the desired state.  
In the general case, $\lowerslacks$ and~$\upperslacks$ can contain
$(\totaloutputs)\predictionhorizon$ independent slack variables each, so that
the vector $\horarrayofvectors{\lowerslacks;\upperslacks}$ can be used directly
as the decision variable~$\myvector\slack$.
Furthermore, $\lowerbound{\myvector\outputsignal}$
and~$\upperbound{\myvector\outputsignal}$ can be tightened progressively along
the prediction horizon to funnel the outputs into a narrow range of
operation~\cite{valverde2013a}. 

To keep the optimization problem free from integer variables despite the
discreteness of \gls{LTC} actions, tap ratios are treated as continuous
variables during the optimization. 
The tap-ratio changes that result from each solution of
\eqref{eq:canonical-objective}--\eqref{eq:canonical-constraint-y} are then
accumulated (\ie time-integrated with signs) and enacted every time that they exceed
the physical \gls{LTC} step, at which point the enacted action is subtracted
from the accumulated variable.

Since the \objective function in \eqref{eq:canonical-objective} is quadratic
and the constraints~\eqref{eq:canonical-constraint-u}
and~\eqref{eq:canonical-constraint-y} are affine
on~$\increment\myvector\controlsignal$ as per~\eqref{eq:affinity-of-u}
and~\eqref{eq:affinity-of-y}, the optimization problem can be cast as a
\gls{QP} through standard manipulations~\cite{soleimanibidgoli2018}.
Because the weight matrices~$\weightchanges$ and~$\weightslacks$ are diagonal
and nonnegative (and therefore positive-definite), the resulting \gls{QP} is
convex and can be easily handled by several solvers. 
Convexity and the lack of integer variables make this problem highly scalable. 
While the \gls{MPC} could easily handle decision variables from hundreds of
substations, it must be noted that the involved substations will likely be
limited to the weak area, as voltage problems are mainly local.  
Furthermore, the size of the optimization problem is not determined by the
number of \acp{DER}, but by the number of substations to be coordinated, as the
\ac{MPC} works with \ac{DER} aggregations. 
The corresponding \secondaddition{aggregator}, one per substation, is the agent
that is responsible for transmitting the required powers to the downstream
\acp{DER}.

The rows of~\eqref{eq:canonical-constraint-u} associated
to~$\myvector\activepower\fromaggregatedDER$
and~$\myvector\reactivepower\fromaggregatedDER$ can be augmented to consider
physical and operational constraints of the \glspl{DN}.
If each \gls{DSO} possesses and communicates enough information to the
\gls{TSO}, the box defined by the constant
bounds~$\lowerbound{\myvector\controlsignal}$
and~$\upperbound{\myvector\controlsignal}$ can be modified to consider a
feasible operating region~\cite{riaz2019}.
One approach is to still enforce box constraints but update the bounds
depending on the dynamic operating point and its position within the feasible
operating~region~\cite{escobar2025}.
Another approach is to enforce linear constraints that define a convex
approximation of said region~\cite{prionistis2022a}.
In this work, we keep constant~$\lowerbound{\myvector\controlsignal}$
and~$\upperbound{\myvector\controlsignal}$ as we assume limited knowledge about
the \glspl{DN} and do not prioritize their operational constraints during an
emergency.

As defined by IEEE Std.\,1547-2018~\cite{IEEEStd2018}, voltage regulation
capability through reactive power is mandatory for \acp{DER}, but the
utilization is at the discretion of the operator. 
The proposed scheme is activated in response to large disturbances only. 
Therefore, the participation of \acp{DER} is temporary and should not result in
stress for inverters or conventional units, as they contribute while respecting
their capacity limits.

For reactive-power requests only, the \acp{DER} could operate under
\textit{P-priority} mode, meaning that the maximum available reactive power is
a function of the nominal capacity of the \gls{DER} and its actual active-power
output. 
However, when the aggregate reactive power from available \acp{DER} is
insufficient, economic compensation could be offered to increase the \ac{DER}
reactive power injection or absorption.  
A signal from the \secondaddition{aggregator} could alert the \acp{DER} that
additional reactive-power capacity is required, prompting some units to switch
to \textit{Q-priority}~\cite{IEEEStd2018} while offering economic compensation
for the potential curtailment of active power.

In the case of active- and reactive-power availability from \acp{DER}, the
weight matrix\! $\weightchanges$ in~\eqref{eq:canonical-objective} could
penalize active-power requests more heavily to prioritize reactive-power use,
or the matrix could map the actual service cost computed offline and updated
based on weather, hour, day, and season. 
Active-power support from \acp{DER} could be compensated by the
\secondaddition{aggregator} to acknowledge its worth compared to reactive
power. 
However, compensation schemes for \ac{DER} participation are beyond the scope
of this paper and are left for future work.

The scalability of the proposed scheme, expressed in terms of the number of
\acp{DER} participating per substation, mainly depends on the communication
infrastructure required to transmit the signals. 
A cost-benefit analysis, as part of a planning exercise, would determine the
number of \acp{DER} per substation, possibly prioritizing the largest units to
maximize~impact.

\subsection{Sequence of Events}

The proposed control scheme is summarized in \cref{fig:flowchart}. 
\begin{figure}
    \centering
    \includegraphics{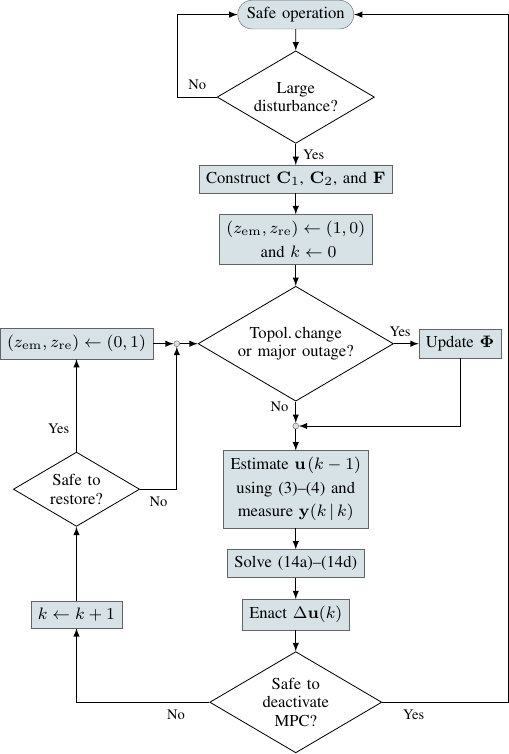}%
    \caption{Sequence of events in the proposed \acs{MPC} scheme.}
    \label{fig:flowchart}
\end{figure}
Once a large disturbance is detected, counter~$k$ is set to zero and the
control starts with the construction of matrices $\measuredinputmatrix$,
$\inputincrementmatrix$, and~$\measuredoutputmatrix$ for the chosen
$\controlhorizon$ and~$\predictionhorizon$ horizons. 
Then, the objective function \eqref{eq:canonical-objective} is configured to
operate in \textit{emergency mode} by setting $\emergencyconstant = 1$ and
$\restorationconstant = 0$. 
If there is a change in the network topology or a major outage, the sensitivity
matrix~$\sensitivitymatrix$ is updated.
Subsequently, the current manipulated variables~$\myvector\controlsignal$ are
estimated, and the latest measurements of $\myvector\outputsignal$ are
collected.  
The controller solves
\eqref{eq:canonical-objective}--\eqref{eq:canonical-constraint-y} but enacts
only the first change~$\increment\myvector\controlsignal\attime\horizonindex$
in the manipulated variables.  
If the monitored voltages and computed \gls{NLI} indicators are such that the
system condition is considered safe, the proposed \gls{MPC} can be deactivated. 
Otherwise, the controller must check if it is possible to restore the
\glspl{DER} to their pre-disturbance operating point because a)~other power
sources can now take over the support of the \glspl{DER}, or b)~the original
disturbance that triggered the activation of this scheme was resolved. 
If restoration is possible, the objective~\eqref{eq:canonical-objective} is
configured to \textit{restoration mode} by setting $\emergencyconstant = 0$ and
$\restorationconstant = 1$ to steer the \glspl{DER} and \glspl{LTC} towards the
pre-disturbance condition with the updated~$\sensitivitymatrix$,
$\myvector\controlsignal$, and~$\myvector\outputsignal$ in
\eqref{eq:canonical-objective}--\eqref{eq:canonical-constraint-y}. 
If safe restoration is not possible yet, the controller continues the
minimization of the control efforts. 
The \gls{MPC} is deactivated when safe operation is reached again.

\subsection{Computation of \gls{NLI} Sensitivities}
\label{sub:nli-embedding}

While voltage sensitivities can be found using standard power-flow
techniques~\cite{valverde2013}, computing the sensitivities of the \glspl{NLI}
requires further consideration. 
The \gls{NLI} is not a static property of the network; its definition only
makes sense when active power and conductance are changing. 
One way to turn this indicator into a static property is to trigger several
fictitious changes in the static model, one at a time, and define the most
pessimistic (\ie lowest) \gls{NLI} as the one associated with the current
operating point.
The changes considered in this work are perturbations of the transformer tap
ratios by a small amount~$\tapdifferential$.  
Perturbing the tap ratio~$\tapratio\atsubstation$ has the effect of changing
the active power at \gls{BB}~$\boundarybus$ by an amount
$
\increment\activepower_{\boundarybus}^{\substation}
=
\activepower\atboundarybus(\tapratio\atsubstation+\tapdifferential)
-
\activepower\atboundarybus(\tapratio\atsubstation)
$
and the conductance by an amount
$
\increment\conductance_{\boundarybus}^{\substation}
=
\conductance\atboundarybus(\tapratio\atsubstation+\tapdifferential)
-
\conductance\atboundarybus(\tapratio\atsubstation)
$.
It is then possible to define the static \gls{NLI} at \gls{BB}~$\boundarybus$
as
\begin{equation}
    \label{eq:static-NLI}
    \nli\atboundarybus
    =
    \min
    \left\{
        \increment\activepower_\boundarybus^\substation
        \,\middle/
        \increment\conductance_\boundarybus^\substation
        \ssep
        \substation=1,\ldots,\substationcount
    \right\}\,.
\end{equation}
The sensitivity of~$\nli\atboundarybus$ to a change in each manipulated
variable is approximated by running two successive power-flow~studies.

As discussed before, the load sensitivity to voltage plays a key role in the
unfolding of \glsin{VS} and should thus be considered when computing stability
indicators.
For improved accuracy, the computation in \eqref{eq:static-NLI} should retain
the dependence of~$\activepower\fromload\atsubstation$
and~$\reactivepower\fromload\atsubstation$
to~$\voltage\atdistribution\atsubstation$ (see \cref{fig:variables}).
This dependence, \eg in the form of an exponential model, can be integrated
into power-flow programs by adding terms to the power mismatch equations and to
the corresponding entries of the Jacobian matrix~\cite{el-hawary1987}.
The output of the fictitious, aggregate \gls{DER} can be assumed to be
independent of voltage as an approximation.
}

\section{Simulation Results}
\label{sec:simulation-results}
The proposed control is tested on the \nordic{} of \cref{sec:background}, whose
central area is shown in \cref{fig:central}.
Only the buses in this area are labeled, and they are linked through four
transformers between the \SI{400}{\kilo\volt} and \SI{130}{\kilo\volt} systems.
Six substations connect (sixth-order) synchronous machines, as shown for
bus~4051 (although not all~substations have a \gls{SCB}).
The step-down transformers are equipped with \glspl{LTC} to adjust the tap
ratios in the range \qtyrange{0.88}{1.20}{\pu} over 33 positions, \ie in steps
of \SI{0.01}{\pu}. 
When acting autonomously
\secondaddition{(\crefrange{sub:no-control}{sub:local-control})}, these
\glspl{LTC} try to keep the \gls{DN}-side voltages within a
$1\pm\SI{0.01}{\pu}$ deadband, acting with \secondaddition{(fixed)} initial and
subsequent delays \secondaddition{that are randomly distributed} in the ranges
\qtyrange{29}{31}{\second} and \qtyrange{8}{12}{\second},~respectively;
\secondaddition{in real life, these random distributions could be traced back to
construction differences that affect the total delay of the
\glspl{LTC}~\cite[\S4.4.2]{cutsem1998}.}

\newcommand\linelen{0.35cm}
\newcommand\mylegend{
    Central area of the \nordic, with \gls{HV} lines that are single-circuit 
    (\,\raisebox{0.7mm}{\tikz{\draw[nordic line] (0, 0) -- ++(0:\linelen);}}\,)
    or
    double-circuit
    (\,\raisebox{0.2mm}{\tikz{\draw[double circuit] (0, 0) -- ++(0:\linelen);}}\,),
    and a line that is tripped at 
    $t=\SI{1}{\second}$~(\,\raisebox{0.7mm}{\tikz{\draw[faulted] (0, 0) -- ++(0:\linelen);}}\,).%
}
\tikzset{nordic line/.style={400 kV lines}}
\begin{figure}
    \centering
    \newsavebox{\mybox}
\savebox{\mybox}{%
    \input{figures/central_expanded.tex}
}
\begin{tikzpicture}
    \coordinate (origin) at (0.1, 3.7);


    \path
        (origin) node[inner xsep=-15pt, inner ysep=-0pt]{%
            \usebox\mybox%
        };

    \draw
        (origin) ++(6.2, 2.7)
        node[below left]{%
            \begin{tabular}{S[table-format=4.0]S[table-format=2.0]c}
                {\gls{HV}} & {\gls{MV}} & Gen. \\
                    \midrule
                1041 & 1 & \\
                1042 & 2 & $\generator_{6}$ \\
                1043 & 3 & $\generator_{7}$ \\
                1044 & 4 & \\
                1045 & 5 & \\
                4041 & 41 & $\generator_{13}$ \\
                4042 & 42 & $\generator_{14}$ \\
                4043 & 43 & \\
                4044 &    & \\
                4045 &    & \\
                4046 & 46 & \\
                4047 & 47 & $\generator_{15}$ \\
                4051 & 51 & $\generator_{16}$ \\
            \end{tabular}
        }
    ;

\path
    (origin) ++(1, -3.9) coordinate (origin);

\def\bussep{1.1}

\def\schematicthickness{substation bus}
\drawbus{bus-four}{0.5*\bussep, 0.1}{vertical}{2}{5}
\drawbus{bus-three}{2.0*\bussep, 0.1}{vertical}{2}{2}
\drawbus{bus-two}{-0.5*\bussep, -1.1}{horizontal}{1}{1}

\path (bus-three -2) node[below]{$51$};
\path (bus-four -5) node[below]{$4051$};

\def\schematicthickness{hv}

\draw[\schematicthickness]
    (bus-four -1)
    --
    (bus-four -1 -| bus-two 0)
    --
    ++(-90:0.01)
    coordinate(temp);
\drawtransformer{trafo-2}{bus-two 0}{temp}
\path
    ($(bus-four 0)!1!(bus-three 0)$)
    coordinate (temp);
\drawtransformer[true]{step-down}{bus-four 0}{temp}

\drawgenerator{gen-two}{-0.5*\bussep,-1.6}
\drawtransmissionline{line-g2-two}{gen-two north}{bus-two 0}


\path (gen-two west) node[left]{$\generator_{16}$};

\def\surfrad{0.35}
\def\busvsep{1.2}
\def\bushsep{5}
\def\busprot{0.3}
\def\busheight{0.45}
\def\genrad{4pt}
\def\loadh{0.35}
\def\loadv{0.4}
\def\dnsep{1.33}
\def\tnsep{0.9}
\pgfmathsetmacro\DERAsep{2.8}
\path
    (bus-three 1) ++(\DERAsep, 0)
    node[aggregated der] (der) {\gls{DER}};
\draw
    (der) -- (bus-three 1)
    node[current left={0.5}]{}
    node[pos={0.45}, above, align=left]{
        $\activepower\fromactualDER$
        and
        $\reactivepower\fromactualDER$
    };

    \def\loadsprot{0.6}
    \def\capacitorplates{0.3}
    \def\capacitorsep{0.1}
    \draw[hv]          
        (bus-four -4)
        --
        ++(0:\loadsprot-\capacitorsep)
        coordinate (temp)
        ++(-90:\capacitorplates/2)
        --
        ++(90:\capacitorplates)
        (temp)
        ++(0:\capacitorsep)
        coordinate (temp)
        ++(-90:\capacitorplates/2)
        --
        ++(90:\capacitorplates)
        (temp)
        -- ++(0:0.8*\loadsprot)
        node[tlground,rotate=90]{}
        ;

    \draw[double circuit]
        (bus-four 1) -- ++(180:1.95)
        node[above right]{To $4045$};

    \draw[-latex, opacity=0.6]
        (origin) ++(-0.5, 1.7) to[out=20, in=160, relative] ++(-90:0.85);

    \pgfmathsetmacro\nlicutlen{0.9}
    \draw[densely dotted, semithick]
        (origin) ++(-1.4, 6.2)
        ++(180:\nlicutlen/2)
        node[left]{$\nlicut_{4041}$}
        --
        ++(0:\nlicutlen);

    \draw[densely dotted, semithick]
        (origin) ++(0.15, 6.2)
        ++(180:\nlicutlen/2)
        --
        ++(0:\nlicutlen)
        node[right]{$\nlicut_{4042}$};


    \path
        (bus-three -1)
        pic{load}
        (end load)
        node[right]{$\load$};
\end{tikzpicture}
    \vspace{-0.6cm}
    \caption{\protect\mylegend}
    \label{fig:central}
\end{figure}

Aggregate \glspl{DER} are connected to the \gls{MV} buses of the central area,
and they supply $20\%$ of the total \gls{DN} consumption as
in~\cite{pabonospina2021}. 
Specifically, they supply a total of \SI{1547.5}{\mega\watt} while the total
load amounts to \SI{7737.5}{\mega\watt}.
The latter was enlarged from its original value of \SI{6190}{\mega\watt} so
that the operating point~A from~\cite{vancutsem2020} remained intact.
The loads consume both active and reactive power, modeled as constant-current
and constant-impedance loads, respectively, while the \glspl{DER} initially
operate at unity power factor.
Once these \glspl{DER} receive requests, they start following reactive-power
setpoints chosen by their local controllers, without ever curtailing their
active power. 
They have a headroom of~$20\%$ of their nominal capacity (\ie are loaded to
$80\%$) to provide support.
In all other respects, they behave like the \dera of~\cite{2019}, with their
parameters set to category~\textsc{II} of \ieeestandard~\cite{IEEEStd2018}.

Voltage stability is monitored through the \glspl{NLI} at boundary buses~$4041$
and~$4042$, computed using the sets $\nlicut_{4041}$ and $\nlicut_{4042}$
(dotted lines in \cref{fig:central}) and expressed in \si{\pu\per\pu} in a
\SI{100}{\mega\voltampere} base.
Although the central area is also adjacent to the southern area through
buses~$4041$ and~$4045$, the associated \glspl{NLI} are neglected based on
prior knowledge of the power transfers.

The system was simulated in \ramses~\cite{aristidou2016}, a time-domain,
root-mean-square~\cite{lara2024} power system simulator, using a fixed
integration step of~\SI{1}{\milli\second} to accurately capture the timing of
discrete events~\cite{fabozzi2011}.
The simulations include detailed models of slow-acting devices such as
\glspl{LTC} and \glspl{OEL}, required for \gls{LTVS} assessment. 
Load restoration was considered through the effect of \glspl{LTC} on
voltage-sensitive loads. 
However, slow self-restoring loads, such as thermostatically controlled loads,
were not included in the simulation, consistent with the load characteristics
described in~\cite{vancutsem2020}. 
Additionally, potential interactions between the slow-acting devices and
secondary frequency control were not considered, as the simulated scenarios did
not involve generation/load tripping.

The logic of the \gls{MPC} and the \secondaddition{\glspl{CO}} was implemented
in \python{}, with the \cvxopt{} optimization package~\cite{andersen2022}.

The rest of this section presents several scenarios that compare the system
response to the tripping of line 4032--4044, a critical contingency in the
\nordic.
\Cref{sub:no-control} first presents the response without control, used as base
case.
\Cref{sub:LTC-blocking} then presents the response under a local
\gls{LTC}-blocking strategy, highlighting the need for
\secondaddition{intersubstation} coordination, and \cref{sub:local-control}
presents the response under the \gls{LC} scheme of~\cite{pabonospina2021}, used
as benchmark.
\Cref{sub:MPC-control} tests the proposed coordinated scheme on a transmission
(T)-only system and \cref{sub:MPC-control-TD} tests it on a more realistic
\gls{T-D} system.
The effect of different parameters is then studied in \cref{sub:comparison}
through a sensitivity analysis.
To make the benchmark \gls{LC} scheme and the proposed scheme comparable, all
simulations up to this point only consider reactive-power support.
Active-power support is finally considered in \cref{sub:active-power}.

\subsection{Response Without Control}
\label{sub:no-control}

\newcommand\importexternalplot[4]{%
    \begin{figure}[t]
        \centering
        \includegraphics[scale=1]{tikz/#4}%
        \vspace{-\myvsepunderplot+0.35cm}
        \mycaption{\nordic #1}{,}{%
        following the post-fault outage of
        \gls{HV} line $4032$--$4044$ at $\conttime=\SI{1}{\second}$%
        #3%
        }
        \label{#2}
    \end{figure}
}

\Cref{fig:no-control-case} shows the response without control, \ie operating
the \glspl{DER} as constant-power sources and allowing the \glspl{LTC} to act
autonomously.
While the simulation captures relevant dynamics, such as jumps due to \gls{LTC}
actions and the subsequent electromechanical oscillations, both shown in the
enlarged section of \cref{fig:no-control-case}(a) as $A$ and $B$, respectively,
we focus in the remainder of the paper on the long-term behavior.
As seen in \cref{fig:no-control-case}(b), the natural tendency of the
\glspl{LTC} is to raise the \gls{MV}-side voltages of
\cref{fig:no-control-case}(a) back into the $1\pm\SI{0.010}{\pu}$ deadband by
lowering the tap ratio.
They succeed in doing so for about $\SI{1}{\minute}$ at the expense of
excessive field currents in the synchronous machines in the central area, as
shown in \cref{fig:no-control-case}(c).
However, after some \glspl{OEL} withdraw this support, the feasible region of
the system shrinks and the physics of power flow forces~the voltages to
\emph{decline} with each \gls{LTC} action.
This leads to negative \glspl{NLI} in \cref{fig:no-control-case}(d).
After other \glspl{OEL} act, the conditions deteriorate further and the
evolution ends up in a \gls{VC} at the \gls{HV} and \gls{MV} buses~$1042$
and~$2$, respectively, at which point the simulation is stopped.
The \glspl{DER} maintain null reactive power output, missing the opportunity to
support the \gls{TN}.

\def\myvsepunderplot{0.7cm}
\importexternalplot%
    {under no control scheme}
    {fig:no-control-case}
    {. To facilitate comparison across scenarios, the time axis is
    extended to \SI{8}{\minute} and the $1\pm\SI{0.025}{\pu}$ voltage
    deadband, later used for the \gls{MPC}, is shown. \tapcaption}
    {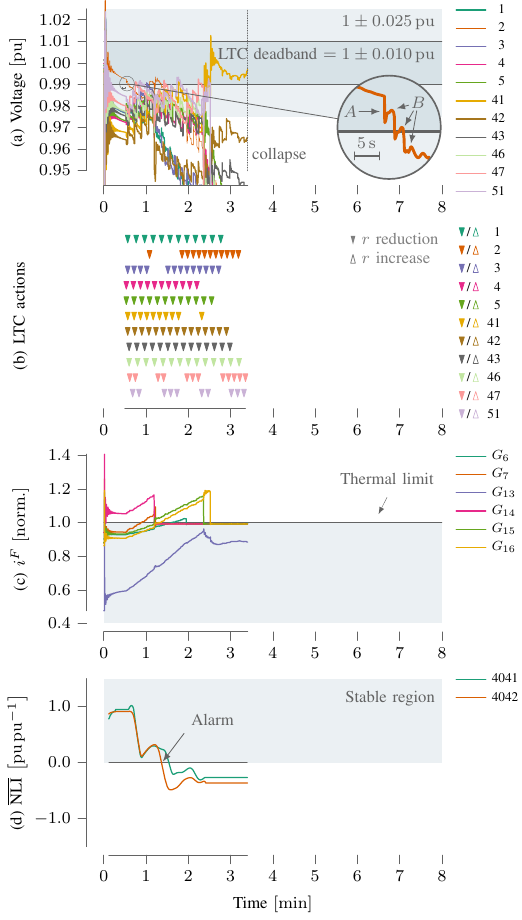}

Note that the \gls{VC} here is delayed with respect to the one in
\cite[Fig.\,3]{vancutsem2020} by about \SI{30}{\second}. 
This is because the loads were enlarged when introducing the \glspl{DER}, which
increases their sensitivity to voltages.%
    \def\pexponent{\alpha}%
    \footnote{%
        For example, a pre-disturbance active
        load~$\constantscalar\activepower_0$ that varies exponentially with
        voltage according to
        $
        \constantscalar\activepower
        =
        \constantscalar\activepower_0
        \left(
        \constantscalar\voltage
        /
        \constantscalar\voltage_0
        \right)^\pexponent
        $
        has a sensitivity
        $\partial\constantscalar\activepower
        /
        \partial\constantscalar\voltage
        =
        \pexponent
        \constantscalar\activepower_0
        \left(
            \constantscalar\voltage
            /
            \constantscalar\voltage_0
        \right)^{\pexponent-1}
        \!\!/\,
        \constantscalar\voltage_0
        $, which is itself proportional to~$\constantscalar\activepower_0$.%
    }
The declining voltages thus bring about an additional reduction in load powers,
slightly improving \gls{VS}.
In other hypothetical scenarios, \gls{VC} could be either delayed or
accelerated by the introduction of \glspl{DER} depending on their share and
their voltage-control mode~\cite{liemann2019}.

\begin{figure}
    \centering
    \includegraphics{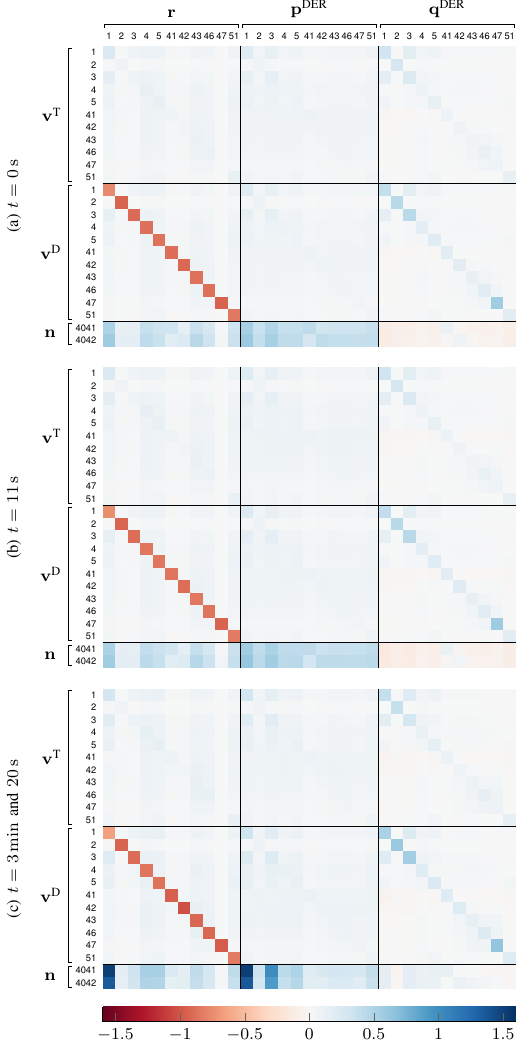}%
    \vspace{-0.5cm}
    \caption{Evolution
        of~$\inlinesensitivity{\myvector\outputsignal}{\myvector\controlsignal}$
        without control. 
        The vectors $\myvector\tapratio$, $\myvector\voltage\attransmission$
        and $\myvector\voltage\atdistribution$ are in \si{\pu} of the nominal
        tap ratios resp.\,voltages, while $\myvector\nlivector$ is in
        \si{\pu\per\pu} (\SI{100}{\mega\voltampere} base).  
        For readability, $\myvector\activepower\fromaggregatedDER$ and
        $\myvector\reactivepower\fromaggregatedDER$ have units of
        \SI{10}{\mega\watt} resp.\,\SI{10}{\mega\var}.}
    \vspace{-7pt}
    \label{fig:MPC-sensitivities}
\end{figure}

In the \SI{3}{\minute} and \SI{20}{\second} before the \gls{VC}, the
sensitivity matrix
$\inlinesensitivity{\myvector\outputsignal}{\myvector\controlsignal}$, later
used by the \gls{MPC}, evolves as shown in \cref{fig:MPC-sensitivities}. 
Columns correspond to the manipulated variables as defined
in~\eqref{eq:manipulated-variables}, while rows correspond to the outputs as
defined in~\eqref{eq:outputs}.
The matrix was computed using a perfect \gls{TN} model, reading the necessary
inputs (voltage magnitudes and active powers at generation buses) from the
snapshots of the dynamic simulation, and running successive power-flow studies
as explained in \cref{sec:proposed-control}.
\Cref{fig:MPC-sensitivities}(a) shows the pre-disturbance sensitivities,
\cref{fig:MPC-sensitivities}(b) shows them after the disturbance and when the
first iteration of the \gls{MPC} would occur, and
\cref{fig:MPC-sensitivities}(c) shows them shortly before the \gls{VC}.
As expected, $\myvector\tapratio$ has opposite effects on
$\myvector\voltage\attransmission$ and $\myvector\voltage\atdistribution$
(see negative-valued diagonal).
Furthermore, the sensitivities to $\myvector\activepower\fromaggregatedDER$ are
always positive, while the sensitivities to
$\myvector\reactivepower\fromaggregatedDER$ are only unambiguously positive in
the diagonals associated to $\myvector\voltage\attransmission$
and~$\myvector\voltage\atdistribution$; these diagonals are higher than for
$\myvector\activepower\fromaggregatedDER$ due to the inductive nature of the
\gls{TN}.
For this particular scenario, it can be seen that the signs and magnitudes in
$\inlinesensitivity{\myvector\outputsignal}{\myvector\controlsignal}$ remain
fairly constant (even after the outage, since the tripped line is not
\emph{within} the central area), supporting the claim that the \gls{MPC} can
update this matrix infrequently, if at all. 
However, this may not be the case under stressed conditions, as in
\cref{fig:MPC-sensitivities}(c).

\subsection{Response Under Local \gls{LTC} Blocking}
\label{sub:LTC-blocking}

We first counteract the \gls{VC} by blocking the \glspl{LTC} in the central
area, a strategy that has been in place in control rooms for decades;
see~\cite[\S\,3.1]{zotero-2267}.
Essentially, blocking the \glspl{LTC} has the effect of delaying the recovery
of the loads, which are sensitive to voltage.
Although there are centralized methods to determine which \glspl{LTC} to block,
\eg based on the sensitivities of voltages to tap ratios~\cite{vournas2001}, we
simulate a local method to highlight the importance of
\secondaddition{intersubstation} coordination.
Specifically, we implement the one-shot strategy from~\cite{capitanescu2009},
where an \gls{LTC} is blocked once the \gls{HV}-side voltage of the transformer
drops below a threshold (here \SI{0.9}{\pu}) for some predefined duration (here
\SI{3}\second).
In contrast to~\cite{capitanescu2009}, we do not optimize these parameters nor
do we complement the \gls{LTC} blocking with undervoltage \gls{LS}.

The system response is shown in \cref{fig:LTC-blocking-case}. 
Although the voltages do not collapse in the simulation, they reach the low
values of \cref{fig:LTC-blocking-case}(a).
One critical case is bus~$4$, which settles at a voltage of \SI{0.91}{\pu} (not
visible in the plot).
Its \gls{LTC} was the first one to block in \cref{fig:LTC-blocking-case}(b),
but its voltage kept being pulled down by the \glspl{LTC} that remained active.
The only voltages that successfully reenter the \gls{LTC} deadband are those of
buses~$41$, $47$, and $51$, but they do so because of their privileged
electrical location in the \gls{TN} and at the expense of their neighbors.
Besides the likelihood that the final voltages are unstable, \eg if generation
further down the \gls{DN} trips due to undervoltages, the system is severely
weakened, with several machines operating under field-current limitation in
\cref{fig:LTC-blocking-case}(c) and the \glspl{NLI} being negative in
\cref{fig:LTC-blocking-case}(d).
This highlights the need for \secondaddition{intersubstation} coordination.

\def\myvsepunderplot{0.7cm}
\importexternalplot%
    {under \gls{LTC}-blocking strategy}
    {fig:LTC-blocking-case}
    {}
    {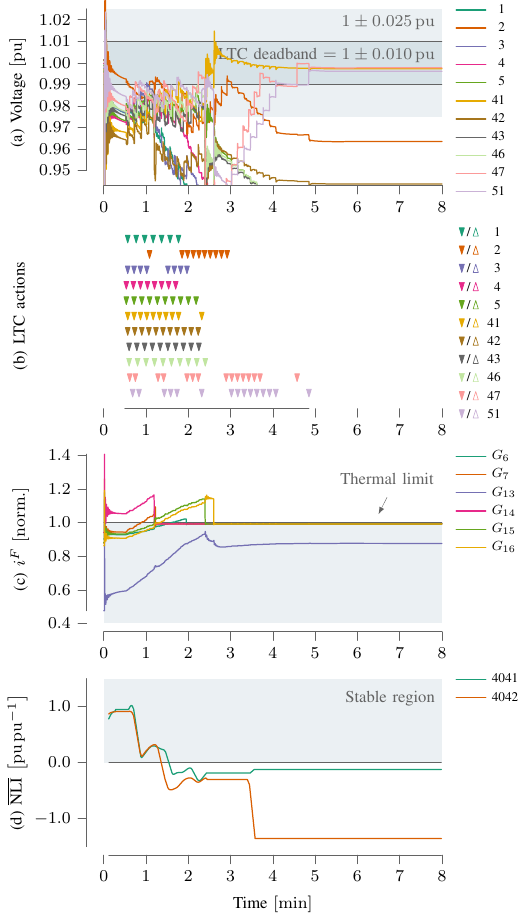}

\subsection{Response Under an \gls{LC} scheme}
\label{sub:local-control}

\begin{figure}
	\newcommand\drawarrow[1]{%
        \raisebox{-1pt}{\tikz{\draw[-latex](0,0) -- ++(#1:0.3cm);}}
	}
	\centering
    \includegraphics[scale=1]{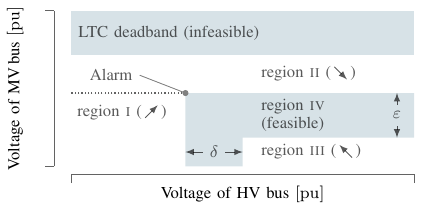}%
    \vspace{-10pt}
	\mycaption
        {Summary of the \acs{LC} scheme from~\cite{pabonospina2021}}
        {.}{The slanted arrows 
            denote the expected \emph{overall} effect of the
            control actions of each region%
        }
    \vspace{-2pt}
	\label{fig:pabon-scheme}
\end{figure}

The \nordic is now controlled with the \gls{LC} scheme
from~\cite{pabonospina2021}, a more sophisticated strategy that is schematized
in \cref{fig:pabon-scheme}.
Inside each step-down transformer of the \gls{WA}, this \gls{LC} scheme
continuously monitors the voltages of the \gls{HV} and \gls{MV} sides, as well
as a stability indicator.
Upon sensing an alarm, it takes a snapshot of both voltages and uses them as an
anchor to draw an L-shaped region~(\textsc{iv}) in the voltage space.
In contrast to the original \gls{LTC} deadband, the \gls{MV}-side voltages in
this region are, by design, lower than the pre-alarm voltages up to a
predefined tolerance~$\varepsilon$. 
Due to other events that occur elsewhere in the system, the \gls{LC} scheme
must make an effort to steer the operating point across the voltage space,
possibly passing through the non-permanent regions~\textsc{i} to~\textsc{iii},
and landing back on region~\textsc{iv}.
Motion along the northwest-southeast diagonal is achieved, as explained for
\cref{fig:collapse}, through changes in the tap ratio~$\tapratio\atsubstation$
of the controlled transformer~$\substation$ whereas motion towards the
northeast is achieved by requesting a common, positive rate of change of
per-unit reactive power,
$\derivative{\reactivepower}\fromactualDER\atsubstation$, from the \glspl{DER}
connected downstream.

Some implementation details are key to understanding the response under the
\gls{LC} scheme. 
Contrary to~\cite{pabonospina2021}, the alarm here is raised globally by the
first \gls{NLI} that changes sign and not locally by the \acs{LIVES} method
of~\cite{vournas2008}, meaning that all \gls{LC} schemes are triggered
simultaneously.
The recorded voltages are not filtered, and they anchor a region~\textsc{iv} of
dimensions $\delta=\SI{0.01}{\pu}$ and $\varepsilon=\SI{0.02}{\pu}$.
Changes in the tap ratio~$\tapratio\atsubstation$ of transformer~$j$ are
induced by moving the setpoint~$\setpoint\voltage\atsubstation$ of the
\gls{LTC} and resizing its tolerance~$\halfdeadband\atsubstation$.
It is important to recall that an autonomous \gls{LTC} is a \gls{FSA} that
jumps between the idle, waiting, or active states according to the controlled
voltage~$\voltage\atsubstation$ and an internal timer. 
In the active state, it tries to perform the action
\begin{equation}
    \tapratio\atsubstation
    \leftarrow
    \begin{cases}
        \tapratio\atsubstation - \abs{\increment\tapratio}\atsubstation
            & \text{if $\voltage\atsubstation < \setpoint\voltage\atsubstation - \halfdeadband\atsubstation$}\,, \\
        \tapratio\atsubstation
            & \text{if $\setpoint\voltage\atsubstation-\halfdeadband\atsubstation
                        \leq
                        \voltage\atsubstation
                        \leq
                        \setpoint\voltage\atsubstation+\halfdeadband\atsubstation$}\,, \\
        \tapratio\atsubstation + \abs{\increment\tapratio}\atsubstation
            & \text{if $\voltage\atsubstation > \setpoint\voltage\atsubstation + \halfdeadband\atsubstation$}\,,
    \end{cases}
\end{equation}
where $\abs{\increment\tapratio}\atsubstation$ is the physical tap-ratio step.
In our implementation, we raise (resp.\,lower) $\setpoint\voltage\atsubstation$
to a very large (resp.\,low) value when trying to decrease (resp.\,increase)
$\tapratio\atsubstation$, and enlarge~$\halfdeadband\atsubstation$ when trying
to freeze~$\tapratio\atsubstation$.
This logic does not interfere with the timing of the automaton and thus
prevents the \glspl{LTC} from synchronizing.
Additionally, we employ, as in~\cite{pabonospina2021},
$\derivative{\reactivepower}\fromactualDER\atsubstation=\SI{0.01}{\pu\per\second}$
(in the per-unit base of each \gls{DER}).

These considerations are seen in action in \cref{fig:LC-case}.
During the first \SI{1}{\minute} and \SI{20}{\second}, the voltage responses of
\cref{fig:LC-case}(a), as well as the \gls{LTC} actions of
\cref{fig:LC-case}(b), are the same as in the case without control.
However, as soon as the first \gls{NLI} becomes negative in
\cref{fig:LC-case}(d), the \gls{LC} schemes of the~$11$ transformers are
triggered.
Right after the activation, the voltages touch the upper-left corner of
region~\textsc{iv}, but their declining tendency at both the \gls{HV} and
\gls{MV} levels causes them to land immediately on region~\textsc{i}.
Each \gls{LC} scheme then commands reactive-power support from the \glspl{DER}
downstream, which ramp up as shown in \cref{fig:LC-case}(e) until the voltages
land on regions~\textsc{ii} to~\textsc{iv}.
At this point, the system is driven by the \gls{LTC} actions of
\cref{fig:LC-case}(b).

\def\myvsepunderplot{0.7cm}
\importexternalplot%
    {under the \acs{LC} from~\cite{pabonospina2021}} 
    {fig:LC-case}
    {}
    {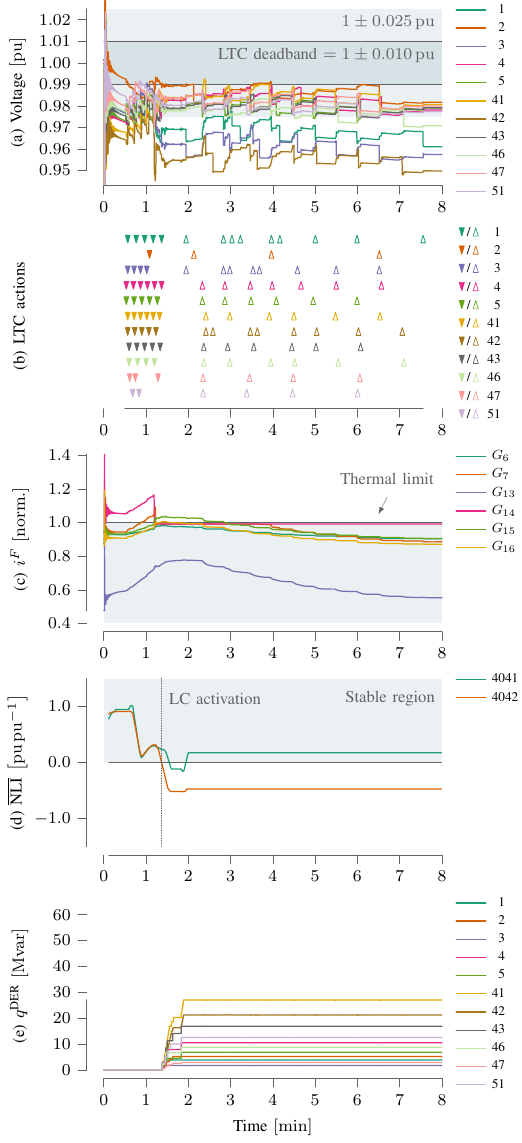}

The resulting dynamics are, as seen in \cref{fig:LC-case}(a), oscillatory.
This is because the \gls{LC} schemes act selfishly to achieve their own goals.
Indeed, in their attempt to decrease their own \gls{MV}-side voltage by
increasing the tap ratio, the \glspl{LTC} inadvertently raise neighboring
voltages.
A good example is the \gls{LC} controlling the voltage at~bus~$42$, the bottom
trajectory in \cref{fig:LC-case}(a).
After about $\conttime=\SI{2}{\minute}$, this voltage increases at the same
time that other \glspl{LTC} strive to reduce their own \gls{MV}-side voltages,
given that those reductions relieve the total consumption in the central area.
When this voltage exceeds the value it had when the snapshot was taken and thus
enters region~\textsc{ii}, its \gls{LTC} acts a first time after its initial
delay of \SI{31}{\second}, fails to return to region~\textsc{iv}, and then acts
successfully a second time after its subsequent delay of~\SI{10}{\second}.
These two successive actions (between \SI{2}{\minute} and \SI{3}{\minute}),
which produce sharp voltage drops at bus~$42$, are felt by the remaining
voltages as milder increases, \eg at buses~$1$ and~$3$.
It becomes clear that the controlled voltages are subject to antagonistic
effects of their own and other \glspl{LTC}, as explained for
\cref{fig:collapse}.
The response obtained here is less smooth than the one
from~\cite{pabonospina2021}, mainly because that reference splits the central
load among~$144$ transformers, instead of only~$11$. 
However, the comparison is fair because the \gls{MPC} faces the same severe
conditions.

\subsection{Response Under \acs{MPC}}
\label{sub:MPC-control}

The \gls{MPC} runs with a sampling period (\ie control step size)
of~\SI{10}{\second}, which accommodates the typical \SI{5}{\second} operating
time of \glspl{LTC}~\cite{IEEEStd2012}, but looking
only~$\controlhorizon=\predictionhorizon=3$ sampling periods into the future.
The \gls{MPC} employs a sensitivity
matrix~$\inlinesensitivity{\myvector\outputsignal}{\myvector\controlsignal}$
that is computed only once after the disturbance, given the slight variations
from \cref{fig:MPC-sensitivities}.
To compute these sensitivities, an inaccurate model of the \gls{TN} is used, in
which a normal error is added to the branch parameters and load exponents with
a zero mean and a standard deviation of~$10\%$ of the actual~values.

To make the \gls{MPC} and \gls{LC} comparable, the active-power changes from
the \glspl{DER} are suppressed.
Tap ratios have a range of~\SIrange{0.88}{1.20}{\pu} without changing more than
\SI{0.01}{\pu} per tap movement (physical limits) whereas reactive powers are
bounded by an estimate of the reactive-support potential of the \gls{DN} and
cannot change more than \SI{100}{\mega\var} per sampling period.
Regarding the outputs, the \glspl{NLI} must remain positive while \gls{HV}- and
\gls{MV}-side voltages are (softly) confined to $1\pm\SI{0.100}\pu$ and
$1\pm\SI{0.025}\pu$ deadbands, respectively.
Inside the diagonal weight matrices~$\weightchanges$ and~$\weightslacks$,
changes in per-unit tap ratios and reactive powers cost resp.\,$1$ and~$4$
while slack variables for voltages and \glspl{NLI} cost resp.\,$\num{e3}$
and~$\num{e4}$.
In a system like the \nordic, the quadratic
form~$\transpose{\increment\myvector\controlsignal}
\weightchanges\increment\myvector\controlsignal$ spreads control actions not
only across time, but also across substations.

Preliminary experiments demonstrated a poor performance under active-power
suppression with the computed sensitivity
matrix~$\inlinesensitivity{\myvector\outputsignal}{\myvector\controlsignal}$.
Essentially, this is because the two remaining manipulated variables
($\myvector\tapratio$ and $\myvector\reactivepower\fromaggregatedDER$) have
opposite effects on $\myvector\voltage\atdistribution$ and on the \glspl{NLI}
in $\myvector\nlivector$ (see \cref{sub:no-control}). 
If the system condition thus deteriorates to the point that the \gls{MPC} must
resort to $\myvector\reactivepower\fromaggregatedDER$\!, it may request a
\emph{reduction} in this variable (\ie an increase in reactive load) while also
requesting an increase in $\myvector\tapratio$, expecting the overall effect to
be an increase in both $\myvector\voltage\atdistribution$ and
$\myvector\nlivector$.
However, whether the \gls{MPC} manages to strike this fine balance is highly
dependent on the values of the sensitivities (not only on their sign), on the
costs in~$\weightchanges$, and on the actual fulfillment of the requests by the
\glspl{DER}.
We thus opt to overwrite the sensitivities of~$\myvector\nlivector$
to~$\myvector\reactivepower\fromaggregatedDER$ to~$0$.
This effectively decouples the control of voltages and \glspl{NLI}: If an
\gls{NLI} becomes negative, the \gls{MPC} can only resort to the tap ratios and
will command them to increase, at which point positive reactive-power
injections take over the voltage control.
We note, however, that a partial overwriting may suffice to
render~$\inlinesensitivity{\myvector\outputsignal}{\myvector\controlsignal}$
well-conditioned, as the \gls{NLI} sensitivities to some entries of
$\myvector\reactivepower\fromaggregatedDER$ may be positive; see for example
buses~$41$ and~$42$ in~\cref{fig:MPC-sensitivities}(a).
Furthermore, an overwriting may not be needed at all if enough active power,
which has the same effect on voltages and \glspl{NLI}, were available.

As soon as the reactive-power requests have been computed by the \gls{MPC},
they are sent to the~$11$ \secondaddition{\glspl{CO}} next to the transformers.
Each \secondaddition{\gls{CO}} has an estimate of the downstream \gls{DER}
installed capacity. 
The estimation error is modeled as a normal distribution with zero mean and a
standard deviation equal to~$10\%$ of the actual installed capacity.
Notice that this error could also model severe communication delays with some
\glspl{DER}, whose missing reaction before the next \gls{MPC} iteration could
be interpreted as an overestimation of the \gls{DER} availability.
In general, upon receiving a request, the \secondaddition{\gls{CO}}
$\substation$ translates it into a signal~$(\signal[q])\atsubstation$ that
saturates at~$\pm\upperbound{\abs{\signal}} = \pm5$ whenever the request
exceeds resp.\,$\pm \apparentpower^\mathrm{ins}_\substation$, and that is
linearly interpolated and discretized in such a way that a~$0$ is issued when
no power is requested.
This signal is then issued every $\samplingperiod=\SI{1}\second$, received by
the \gls{DER}~$\dummyindex$ downstream, and translated, in turn, into a
reactive-power
setpoint~$\setpoint{\left(\reactivepower\fromactualDER_\dummyindex\right)}$ via
a similar logic:
\begin{equation}
    \label{eq:LC-DER-translation}
    \setpoint{\left(\reactivepower\fromactualDER_\dummyindex\right)}
    \leftarrow
    \reactivepower\fromactualDER_{\dummyindex0}
    +
    \frac{(\signal[q])\atsubstation}5
    \left(
        \apparentpower^\mathrm{nom}_\dummyindex
        -
        \reactivepower\fromactualDER_{\dummyindex0}
        \cdot
        \sgn{
            (\signal[q])\atsubstation
        }
    \right)\,,
\end{equation}
where $\reactivepower\fromactualDER_{\dummyindex0}$~is the pre-disturbance
reactive power of the \gls{DER} and
$\apparentpower^\mathrm{nom}_\dummyindex$~is its rated capacity.
Essentially, the expression on the right
hits~$\pm\apparentpower^\mathrm{nom}_\dummyindex$ when
$(\signal[q])\atsubstation$ hits $\pm5 $, stays at
$\reactivepower\fromactualDER_{\dummyindex0}$ when $(\signal[q])\atsubstation =
0$, and forms~$5$ equal steps in between.

\Cref{fig:MPC-T-case} shows the system response in this scenario.
Already~\SI{10}{\second} after the line outage, the \gls{MPC} starts marching
the \glspl{LTC} in unison to stop the declining voltages.
However, it refrains from trying to restore them to $1\pm\SI{0.010}\pu$, and
instead keeps them fairly constant and close to the prespecified
$1\pm\SI{0.025}\pu$ deadband for about~\SI{1}{\minute} and~\SI{20}{\second}.
When the \gls{OEL} of~$\generator_{14}$ acts, in \cref{fig:MPC-T-case}(c), the
sudden drop in voltages and, more importantly, the first violation of an
\gls{NLI} constraint in \cref{fig:MPC-T-case}(d) give the \gls{MPC} a strong
incentive to take aggressive actions.
Before reaching~\SI{2}{\minute}, it moves \glspl{LTC}, \eg the one controlling
the voltage at bus~$41$, in their non-natural direction to restore
the~\gls{NLI}.
Furthermore, especially after~\SI{2}{\minute}, it requests reactive power from
the \glspl{DER} depending on their size and location, as shown in
\cref{fig:MPC-T-case}(e).
These joint actions improve \gls{VS} to the point that the \gls{NLI} can be
reset to~$0.1$ (chosen arbitrarily), enabled in part by the \gls{OEL} reset
of~$\generator_{14}$ after~$\normfieldcurrent$ returned below its thermal
limit.
Once the \gls{DER} powers have been deployed after $\conttime=\SI{3}{\minute}$,
the voltages reenter the $1\pm\SI{0.025}\pu$ deadband and eventually reach a
close-to-steady state, with sparse \gls{LTC} actions taking place as the
\gls{MPC} performs fine-grained voltage corrections.
Although~$\generator_{14}$ exceeds the thermal limit again, the overshoot is
small, so that the \gls{OEL} action will take place far into the future
(inverse-time characteristic) with a small impact on the system.

\def\myvsepunderplot{0.7cm}
\importexternalplot%
    {under \acs{MPC} (T-only system)} 
    {fig:MPC-T-case} 
    {}
    {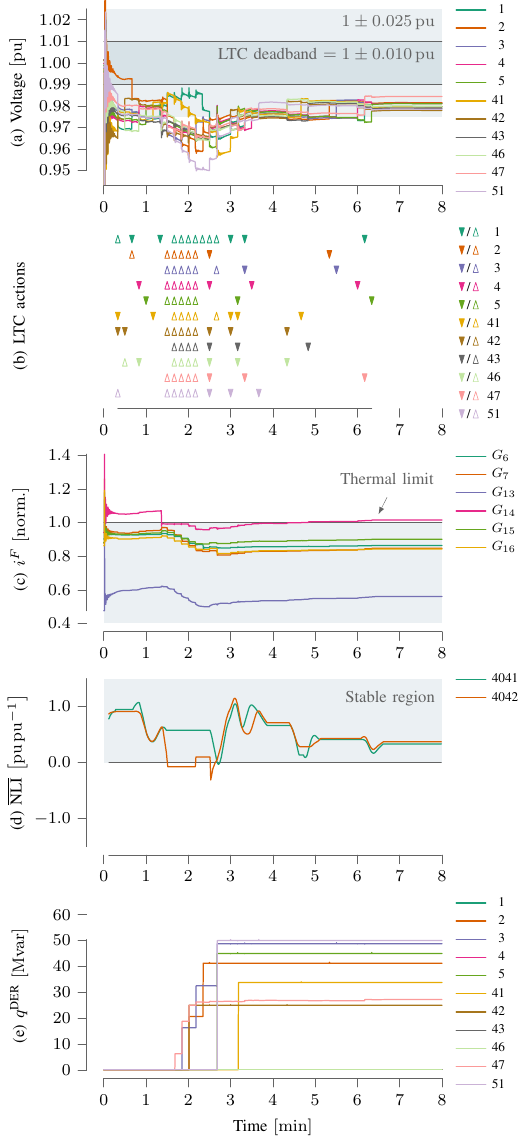}

\subsection{Response Under \acs{MPC} in a \acs{T-D} system}
\label{sub:MPC-control-TD}

The response is now simulated on a \gls{T-D} system that spans the \gls{HV} and
\gls{MV} levels.
This system is built by disaggregating the~$11$ \gls{MV} loads of the \nordic
with replicas of a \gls{DN}, a common
technique~\cite{pilatte2019,pabonospina2021} to avoid the difficulties of
network synthesis.
Specifically, this \gls{DN} consists of $76$ buses operating at \gls{MV} and a radial
topology~\cite{valverde2013}.
To disaggregate the central loads (all above \SI{100}{\mega\watt}) with as few
replicas as possible, the \gls{DN} is populated with loads (retaining the
exponents of the aggregate model) that maximize the \gls{DN} consumption with
acceptable voltages.
Finally, \glspl{DER} are placed with the same procedure, headroom, and
parameters of the T-only scenario, but spreading them over~$15\%$ of the load
buses and assigning them normally distributed outputs that add up to the
original~\gls{DER} share.

\newcommand\knittingvector{
    \transpose{
        \begin{bmatrix}
            \dummyvectorletter_1 &
            \dummyvectorletter_2
        \end{bmatrix}
    }
}
To disaggregate each central load, a subset of the replicas is chosen by
solving the subset-sum problem in~\cite{escobar2020}.
Then, the selected replicas are knitted together with the \gls{TN} by scaling
all the active (resp.\,reactive) powers of the dispersed \gls{DN} loads by a
common factor~$\dummyvectorletter_1$ (resp.\,$\dummyvectorletter_2$) that
ensures power continuity at the \gls{T-D} boundary.
For simplicity, the replicas are connected to the \gls{TN} through the
existing~$11$ transformers.
This process results in a \gls{T-D} system with $2624$~buses, $339$~instances
of the~\dera, and $2237$~exponential loads.  
Of these, only $50$~buses and $50$~\glspl{DER} are monitored to reduce
computational burden.
The \gls{MPC} uses the same model as the previous scenario.

\Cref{fig:MPC-TD-case} shows the simulated response.
As expected, the bus voltages of \cref{fig:MPC-TD-case}(a) and the \glspl{NLI}
of \cref{fig:MPC-TD-case}(b) have essentially the same behavior as the previous
scenario since the \gls{MPC} parameters were kept intact.
Minor differences are due to the smaller size of the \glspl{DER}, which respond
in \cref{fig:MPC-TD-case}(c) with no more than~\SI{2.5}{\mega\var}.

\def\myvsepunderplot{0.7cm}
\importexternalplot%
    {under \acs{MPC} (\acs{T-D} system)}
    {fig:MPC-TD-case}                    
    {. The bottom plot shows~$\reactivepower\fromactualDER$ for~$50$
    \acsp{DER} chosen at random. The vertical axis was adjusted
    accordingly}
    {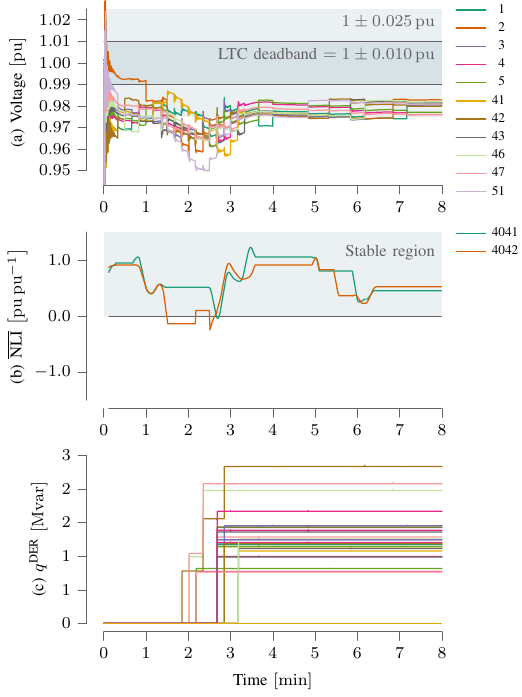}

\subsection{Scenario Comparison and Sensitivity Analysis}
\label{sub:comparison}

\newcommand\averageword{average }
\newcommand\VoltageIntegral{\averageword voltage dev.}
\newcommand\NLI{\averageword\gls{NLI}}
\newcommand\TapMovements{remaining taps}
\newcommand\TapDown{$\tapratio$ reductions}
\newcommand\TapUp{$\tapratio$ increases}
\newcommand\ControlEffortP{\averageword$\activepower\fromactualDER$ effort}
\newcommand\ControlEffortQ{\averageword$\reactivepower\fromactualDER$ effort}
\newcommand\ReactiveMargin{\averageword$1 - \normfieldcurrent$}
\newcommand\ActivatedOELs{activated \glspl{OEL}}
\newcommand\PowerReserve{\averageword$\apparentpower\fromactualDER$ reserve}
\newcommand\decisionvars{no.\,of decision vars.}
\newcommand\MPCiters{\gls{MPC} iterations}
\newcommand\mintime{min.\,optimization time~(\si\second)\!}
\newcommand\avgtime{avg.\,optimization time (\si\second)}
\newcommand\maxtime{max.\,optimization time~(\si\second)\!\!}

The previous scenarios are best compared based on average behaviors.
On the one hand, given a signal~$\dummyonevariablefunction\!$, its variation
across time is captured by the time average
$
    \timeaverage{\dummyonevariablefunction(t)}
    =
    \frac 1\simhorizon \int_0^{\simhorizon} \!\!\dummyonevariablefunction(t) \, \mathrm dt\,,
$
where~$\simhorizon$ is the simulation horizon (\SI{8}{\minute} in this
section). 
On the other hand, given a collection of values $x_1, \ldots, x_N$ associated
to~$N$ elements, such as voltages associated to~$N$ buses, the variation across
those elements is summarized by the usual average
$\avg_\dummyindex\,x_\dummyindex = \frac 1N \sum_{\dummyindex=1}^N
x_\dummyindex$.
We combine these two operations to compute the performance measures of
\cref{tab:metrics-meaning}.

\begin{table}
    \centering
    \caption{Considered performance measures}
    \label{tab:metrics-meaning}
    \newcommand\interlinespace{0.02cm}
\begin{tabularx}\columnwidth{Xlll}
    \toprule
        Measure & Computation & Units & Ideal \\
    \midrule
        \VoltageIntegral & $\VoltageIntegralFormula$ & \si\pu & 0                 \\[\interlinespace]
        \NLI & $\NLIFormula$  & \si\pu/\si\pu & high                                \\
        \TapDown & simple counting & none & low                                 \\
        \TapUp & simple counting & none & low                                 \\
        \TapMovements & simple counting  & none & high                       \\
        \ControlEffortP & $\ControlEffortPFormula$ & \si{\mega\watt} & 0         \\[\interlinespace]
        \ControlEffortQ & $\ControlEffortQFormula$ & \si{\mega\var} & 0          \\[\interlinespace]
        \PowerReserve & $\PowerReserveFormula$ & \si{\mega\voltampere} & high  \\
        \ActivatedOELs & simple counting & none & 0                         \\
        \ReactiveMargin & $\ReactiveMarginFormula$ & norm. & high                \\
    \bottomrule
\end{tabularx}

\end{table}

\cref{tab:metrics} compares the four scenarios.
Firstly, the scenario without control confirms the instability mechanism of the
\nordic: If left to act autonomously without support from the \glspl{DER}, the
\glspl{LTC} almost exhaust their mechanical positions at the expense of the
synchronous machines under high field currents.
Secondly, under the \gls{LC} scheme, the system recognizes that it should
refrain from making tap changes, but it does not receive enough support from
the \glspl{DER}.
This is not because the \glspl{DER} lack power reserves, but rather because the
scheme, being local and without prediction capabilities, is unable to send the
right requests.
Thirdly, in the scenario under \gls{MPC} (T), the system has a similar
\emph{average} voltage deviation and tap usage as under the \gls{LC} scheme,
but neither the voltages nor the tap ratios oscillate.
The scheme asks for enough reactive power, without significantly reducing the
\gls{DER} reserve, and thus relies less on the synchronous machines.
Finally, in the \gls{T-D} system, the behavior points to a slight improvement
in stability, due to the constant-impedance component added to the load mix
with the explicit modeling of the \gls{DN} lines.

\begin{table}
    \centering
    \caption{Performance measures for the simulated scenarios}
    \label{tab:metrics}
    \begin{tabularx}{\columnwidth}{XS[table-format=3.3]S[table-format=3.3]S[table-format=3.3]S[table-format=3.3]}
\toprule
& \multicolumn{4}{c}{Type of control}\\
\cmidrule(lr){2-5}
 Measure & {None} & {\acs{LC}~\cite{pabonospina2021}} & {\gls{MPC} (T)} & {\gls{MPC} (\gls{T-D})}\\
\midrule
\VoltageIntegral &  0.035   &  0.022   &  0.022   &  0.022   \\
\NLI &  0.053   &  -0.038  &  0.493   &  0.622   \\
\TapDown &  135     &  46      &  39      &  34      \\
\TapUp &  0       &  67      &  60      &  61      \\
\TapMovements &  17      &  173     &  173     &  179     \\
\ControlEffortP &  0.000   &  0.000   &  0.000   &  -0.000  \\
\ControlEffortQ &  -0.000  &  8.566   &  17.072  &  0.446   \\
\PowerReserve &  35.170  &  34.541  &  30.005  &  1.093   \\
\ActivatedOELs &  7       &  2       &  0       &  0       \\
\ReactiveMargin &  0.038   &  0.098   &  0.158   &  0.165   \\
\bottomrule
\end{tabularx}
\end{table}

\Cref{tab:times} compares performance measures, but now for different horizons,
applied to the T-only system.
The last four rows refer to the \gls{QP}
\eqref{eq:canonical-objective}--\eqref{eq:canonical-constraint-y}, showing its
number of decision variables and the time required to solve it; this time is
averaged over~$47$ iterations, \ie once every~\SI{10}{\second} until
$\SI{8}{\minute}= \SI{480}{\second}$.
(The hardware ran Windows~11 on~4 cores of an \textsc{amd epyc 7763} processor
and had 16\,GB of RAM.) 
The choice of~$4$ sampling periods has the poorest performance (see
avg.\,voltage dev.\,and avg.\,\gls{NLI}), since a faster control is needed to
force the \gls{MPC} to act more aggressively (see
avg.\,$\reactivepower\fromaggregatedDER$), as noted in~\cite{glavic2011}.
The choices of $1$ and~$3$ sampling periods perform similarly to each other
without overusing the \gls{DER} reserves (see
avg.\,$\apparentpower\fromaggregatedDER$).
The advantage of non-unitary horizons is expected to increase when
\eqref{eq:canonical-constraint-du} becomes binding and when
\eqref{eq:canonical-constraint-u} varies along the horizon, \eg when
forecasting \gls{DER} availability.
It is noted that the running time of the optimization easily fits within
\SI{10}{\second}. 
Even lower times could be achieved with more efficient commercial solvers. 
Therefore, many more substations could be considered and solved, if required,
in under \SI{1}{\second}, which is suitable for real-time implementation.

\newcommand\basecasesymbol{\dagger}
\newcommand\basecase[1]{#1\,$^\basecasesymbol$}
\begin{table}
    \centering
    \caption{Performance measures for different horizons,
    with~$\controlhorizon=\predictionhorizon$}
    \label{tab:times}
    \begin{threeparttable}
        \begin{tabularx}{\columnwidth}{XS[table-format=3.3]S[table-format=3.3]S[table-format=3.3]S[table-format=3.3]}
\toprule
& \multicolumn{4}{c}{$\controlhorizon$ and $\predictionhorizon$ in sampling periods (\SI{10}{\second})}\\
\cmidrule(lr){2-5}
 Measure & {1} & {2} & {\basecase3} & {4}\\
\midrule
\VoltageIntegral  &  0.022   &  0.021   &  0.022   &  0.024   \\
\NLI              &  0.450   &  0.542   &  0.493   &  0.439   \\
\TapDown          &  39      &  29      &  39      &  45      \\
\TapUp            &  60      &  70      &  60      &  158     \\
\TapMovements     &  173     &  193     &  173     &  265     \\
\ControlEffortP   &  0.000   &  0.000   &  0.000   &  0.000   \\
\ControlEffortQ   &  17.042  &  23.702  &  17.072  &  17.743  \\
\PowerReserve     &  30.005  &  27.106  &  30.005  &  22.582  \\
\ActivatedOELs    &  0       &  0       &  0       &  0       \\
\ReactiveMargin   &  0.157   &  0.181   &  0.158   &  0.166   \\\midrule
\decisionvars     &  81      &  114     &  147     &  180     \\
\mintime          &  0.016   &  0.078   &  0.188   &  0.358   \\
\avgtime          &  0.044   &  0.110   &  0.219   &  0.391   \\
\maxtime          &  0.075   &  0.134   &  0.241   &  0.444   \\
\bottomrule
\end{tabularx}
        \begin{tablenotes}
            \item[$\basecasesymbol$] Base case from \cref{sub:MPC-control}
        \end{tablenotes}
    \end{threeparttable}
\end{table}

To test the sensitivity of the proposed scheme to reduced and 
spatially-heterogeneous \gls{DER} availability, we consider a scenario where
substations~$5$, $43$, and~$46$ (chosen randomly) contribute only through
\gls{LTC} actions, but not \gls{DER} powers.
The corresponding response is shown in \cref{fig:MPC-8-case}.
As seen in the voltage and \gls{NLI} responses from \cref{fig:MPC-8-case}(a)
and \cref{fig:MPC-8-case}(d), respectively, the \gls{MPC} achieves a similar
performance to the base case.
It does so, in part, by requesting the reactive powers in
\cref{fig:MPC-8-case}(e) from all substations but four.
Besides the unavailable substations ($5$, $43$, and~$46$), the \gls{MPC}
refrains from requesting powers from substation~$1$. 
On the contrary, substation~$4$, whose \glspl{DER} remained inactive in the
base case, injects a total of \SI{52.5}{\mega\var}, becoming the most important
reactive-power contributor.
These decisions from the \gls{MPC} highlight its ability to adapt to different
\gls{DER} availabilities.

\def\myvsepunderplot{0.7cm}
\importexternalplot%
    {under \acs{MPC} with \gls{DER} support from only 8 substations}
    {fig:MPC-8-case}
    {}
    {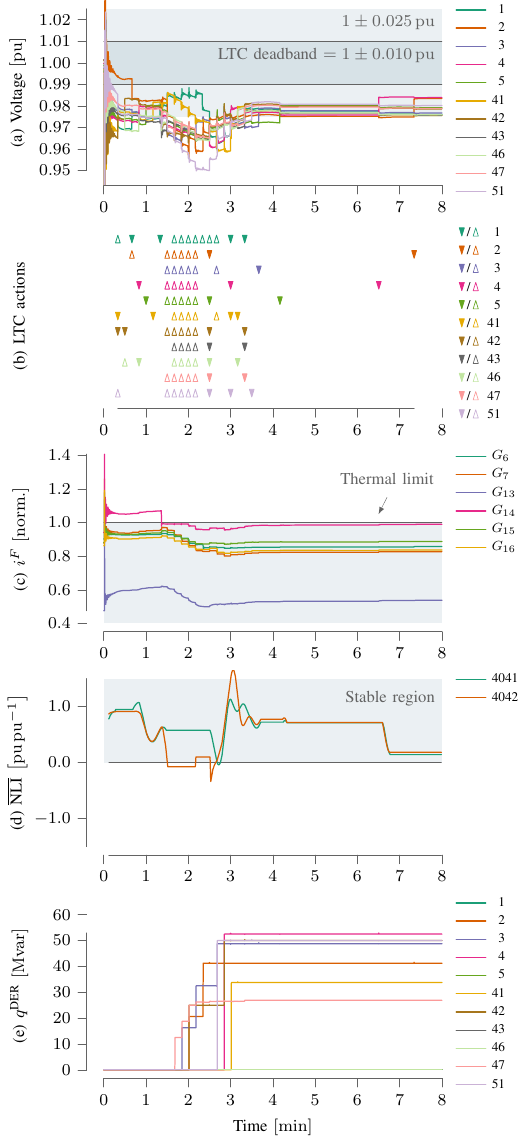}

Another series of experiments, reported in \cref{tab:deadbands}, illustrates
the effect of the \gls{MV}-side voltage half-deadband.
Note that for \SI{0.015}{\pu} and \SI{0.020}{\pu} the average voltage deviation
is larger than the half-deadband width, meaning that the \gls{MPC} fails to
satisfy all voltage constraints and must incur in expensive slacks.
Furthermore, the average \gls{NLI} is lower in those two cases as the \gls{MPC}
focuses on voltage control.
This voltage control is performed in large part by the reactive powers of the
\gls{DER}, highly incentivized by the expensive slacks. 
Halving the half-deadband width from~\SI{0.030}{\pu} to~\SI{0.015} thus results
in a tenfold increase in reactive-power effort.
In turn, this alleviates the synchronous machines, as indicated by the
\ReactiveMargin.

\begin{table}
    \centering
    \caption{Performance measures for different \gls{MV}-side deadbands}
    \label{tab:deadbands}
    \begin{threeparttable}
        \begin{tabularx}{0.89\columnwidth}{XS[table-format=2.3]
								   S[table-format=2.3]
								   S[table-format=2.3]
								   S[table-format=1.3]}
\toprule
& \multicolumn{4}{c}{\gls{MV}-side half-deadband (\si\pu)}\\
\cmidrule(lr){2-5}
    Measure & 0.015 & 0.020 & {\basecase{0.025}} & 0.030 \\
\midrule
\VoltageIntegral & 0.0153  	&  0.027 &  0.022  &  0.026 \\
\NLI             & 0.3912	&  0.365  &  0.493  &  0.491\\
\ControlEffortP  & 0.000	&  0.000   &  0.000   &  0.000 \\
\ControlEffortQ  & 54.416	&  43.896 &  17.072 &  5.222 \\
\ActivatedOELs   & 0		&  0       &  0       &  0   \\
\ReactiveMargin  & 0.212	&  0.229  &  0.158  &  0.141 \\
\bottomrule
\end{tabularx}
        \begin{tablenotes}
            \item[$\basecasesymbol$] Base case from \cref{sub:MPC-control}
        \end{tablenotes}
    \end{threeparttable}
\end{table}

To conclude the sensitivity analysis, we consider the \gls{DER} penetrations of
\cref{tab:penetrations}, expressed as the percentage of the substation load
that is supplied by the \glspl{DER}.
In all cases, the net load at each substation was left intact by increasing (or
decreasing) the \gls{DER} active power and the active load by the same amount.
The most affected performance measure is the reactive-power effort.
This is because the load increase (or reduction) affects, as explained in
\cref{sub:no-control}, the load sensitivity to voltages and hence the voltage
stability.
The \gls{MPC} then devotes the tap ratios to improve stability, while the
voltages are controlled exclusively by the reactive powers.

\begin{table}
    \centering
    \caption{Performance measures for different \gls{DER} penetrations}
    \label{tab:penetrations}
    \begin{threeparttable}
        \begin{tabularx}{0.75\columnwidth}{XS[table-format=2.3]S[table-format=2.3]S[table-format=1.3]}
\toprule
& \multicolumn{3}{c}{\gls{DER} penetration (\%)}\\
\cmidrule(lr){2-4}
    Measure & {10} & {\basecase{20}} & {30}\\
\midrule
\VoltageIntegral  &  0.023  &  0.022  &  0.023	\\
\NLI              &  0.319  &  0.493  &  0.382	\\
\ControlEffortP   &  0.000   &  0.000   &  0.000  	\\
\ControlEffortQ   &  30.629 &  17.072 &  8.573 	\\
\ActivatedOELs    &  0       &  0       &  0     	\\
\ReactiveMargin   &  0.190  &  0.158  &  0.152  	\\
\bottomrule
\end{tabularx}
        \begin{tablenotes}
            \item[$\basecasesymbol$] Base case from \cref{sub:MPC-control}
        \end{tablenotes}
    \end{threeparttable}
\end{table}

\subsection{Response Under \gls{MPC} Including Active-Power Support}
\label{sub:active-power}

Finally, we consider a scenario with extended \gls{DER} capabilities,
specifically active-power support.
While active power plays only a secondary role during normal (\eg secondary)
voltage control, it can have a significant impact during emergencies, given its
positive effect on both voltages and stability (see
\cref{fig:MPC-sensitivities}).
By simulating the \dera model, which represents inverter-based resources, we
implicitly assume that the active power stems from available headroom and that
the P-priority mode follows a circular capability curve.
In practice, however, the active power may be provided also by flexible loads,
and the power priority may conform to more complex control modes.
Changes in active power are bounded by $\pm\SI{5}{\mega\watt}$ per sampling
period, and they cost 10 times more than for reactive~power.

The system response is shown in \cref{fig:MPC-PQ-case}. 
During the first \SI{1}{\minute} and \SI{30}{\second}, the \gls{MPC} has no
incentive to manipulate variables other than the tap ratios, and hence the
response is the same as in the previous scenarios.
However, once the \gls{NLI} at bus~$4042$ becomes negative in
\cref{fig:MPC-PQ-case}(d) and especially after its more pronounced negative
peak at \SI{2}{\minute} and \SI{30}{\second} (also visible in previous plots),
active power is requested. 
The active-power injections do not increase in sharp steps due to the ramp
limiters in the \dera model, and they may cause the reactive power to recede;
this is the case of the \gls{DER} at bus~$47$, whose initial reactive-power
contribution of about \SI{25}{\mega\var} had brought it to a 120\% of its capacity
(\SI{31.25}{\mega\voltampere}).
Once the system receives the additional support from the active powers, the
\gls{NLI} at bus~$4041$ evolves in such a way that it remains negative slightly
longer. 
This motivates the \gls{MPC} to further request active power and increase the
tap ratios in \cref{fig:MPC-PQ-case}(b), which in turn alleviates the
synchronous machines from \cref{fig:MPC-PQ-case}(c).

\def\myvsepunderplot{0.7cm}
\importexternalplot%
    {under \acs{MPC} (T-only system) with active- and reactive-power support} 
    {fig:MPC-PQ-case}
    {}
    {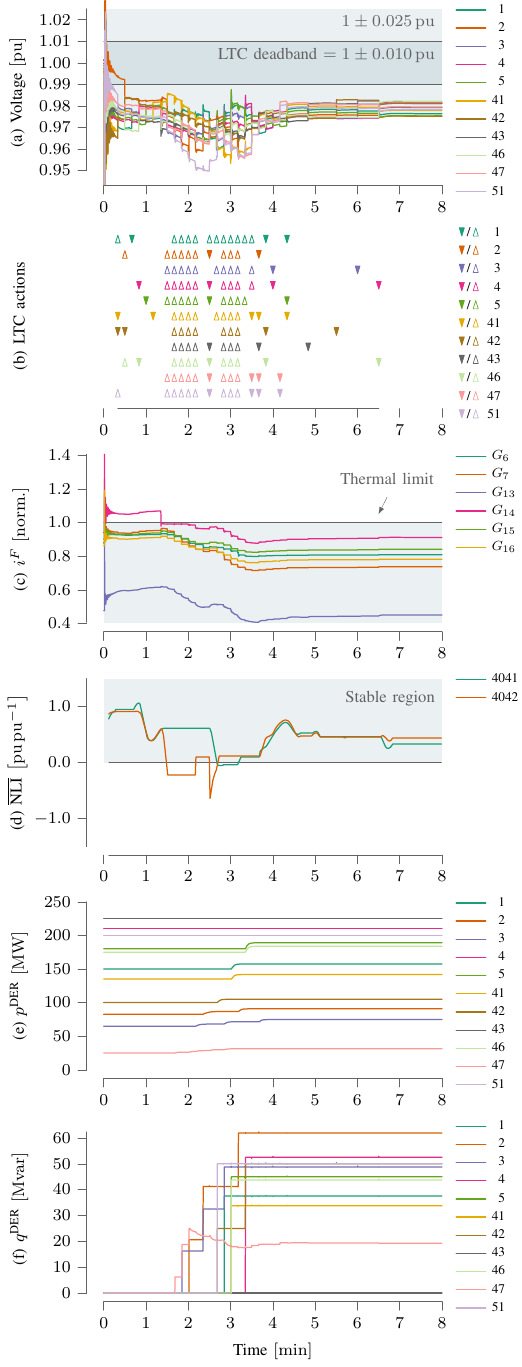}

While resorting to active power may cause other detrimental effects to appear,
such as depletion of \glspl{BESS} and synchronization of \glspl{TCL}, we
emphasize that the proposed scheme can be used as a first line of defense
during emergencies. 
By the time that those effects come into play, the network operator is expected
to have reacted with more sustainable countermeasures.

\section{Conclusion}
\label{sec:conclusion}
This work proposed a control scheme that coordinates the \glspl{DER} and
\glspl{LTC} from multiple substations.
Simulations on a realistic \gls{TN} showed that this scheme can successfully
coordinate \glspl{DER} and \glspl{LTC} while maintaining \gls{LTVS}. 
The proposed scheme is robust against inaccuracies in the predictor and it
drives the system smoothly to a new steady state after a \gls{DIST}. 
It showed better performance when compared to a model-free \gls{LC}.

There are various opportunities to extend the topics addressed in this work. 
The \gls{MPC} could be augmented with historic measurements, so that it could
adjust its weights and horizons based on past events, while the
\secondaddition{\glspl{CO}} could be reinforced by increasing the amount of
knowledge they have about the \glspl{DER}. 
Additionally, the way how the \secondaddition{\glspl{CO}} disaggregate the
power requests among \glspl{DER} from the same substation while respecting
local \gls{DN} constraints, \secondaddition{\ie \textit{intra}substation
coordination,} must be explored further. 
Finally, strategies to incentivize or compensate \acp{DER} that participate in
emergency control will be proposed in the future.

\bibliographystyle{IEEEtran}
\balance
\bibliography{bib/bstcontrol,bib/IEEEabrv,bib/library}

@IEEEtranBSTCTL{BSTcontrol, 
    CTLuse_forced_etal = "yes", 
    CTLmax_names_forced_etal = "6", 
    CTLnames_show_etal = "1", 
}

@article{pabonospina2021,
  title = {Emergency Support of Transmission Voltages by Active Distribution Networks: {{A}} Non-Intrusive Scheme},
  author = {Pab{\'o}n Ospina, Luis David and Van Cutsem, Thierry},
  year = {2021},
  month = sep,
  journal = {IEEE Trans. Power Syst.},
  volume = {36},
  number = {5},
  pages = {3887--3896},
  publisher = {IEEE},
  issn = {0885-8950},
  doi = {10.1109/TPWRS.2020.3027949},
}

@article{wu2001,
  title = {Voltage Security Enhancement via Coordinated Control},
  author = {Wu, Qiang and Popovi{\'c}, Dragana H. and Hill, David J. and Parker, Colin J.},
  year = {2001},
  month = feb,
  journal = {IEEE Trans. Power Syst.},
  volume = {16},
  number = {1},
  pages = {127--135},
  publisher = {IEEE},
  issn = {08858950},
  doi = {10.1109/59.910790},
}

@article{chen2016a,
  title = {Measurement-Based Estimation of the Power Flow {{Jacobian}} Matrix},
  author = {Chen, Yu Christine and Wang, Jianhui and {Dom{\'i}nguez-Garc{\'i}a}, Alejandro D. and Sauer, Peter W.},
  year = 2016,
  month = sep,
  journal = {IEEE Trans. Smart Grid},
  volume = {7},
  number = {5},
  pages = {2507--2515},
  issn = {1949-3061},
  doi = {10.1109/TSG.2015.2502484},
  urldate = {2024-09-21}
}

@article{glavic2011,
  title = {Receding-Horizon Multi-Step Optimization to Correct Nonviable or Unstable Transmission Voltages},
  author = {Glavic, Mevludin and Hajian, Mahdi and Rosehart, William and Van Cutsem, Thierry},
  year = {2011},
  month = aug,
  journal = {IEEE Trans. Power Syst.},
  volume = {26},
  number = {3},
  pages = {1641--1650},
  publisher = {IEEE},
  issn = {0885-8950},
  doi = {10.1109/TPWRS.2011.2105286},
}

@techreport{zotero-5591,
  title = {Final Report of the Investigation Committee on the 28 {{September}} 2003 Blackout in {{Italy}}},
  year = {2004},
  month = apr,
  pages = {1--120},
  institution = {UCTE},
  langid = {english},
}

@article{mandoulidis2022a,
  title = {Overview, Comparison, and Extension of Emergency Controls against Voltage Instability Using Inverter-Based Generators},
  author = {Mandoulidis, P. and Chaspierre, G. and Vournas, C. and Van Cutsem, T.},
  year = {2022},
  month = sep,
  journal = {Sustainable Energy, Grids and Networks},
  volume = {31},
  pages = {100710},
  issn = {23524677},
  doi = {10.1016/j.segan.2022.100710},
  urldate = {2023-01-04},
  langid = {english},
}

@book{cutsem1998,
  title = {Voltage Stability of Electric Power Systems},
  author = {Van Cutsem, Thierry and Vournas, Costas},
  year = {1998},
  publisher = {Springer},
  address = {Boston, MA, USA},
  doi = {10.1007/978-0-387-75536-6},
  urldate = {2024-02-06},
  langid = {english},
}

@inproceedings{valverde2013a,
  title = {Control of Dispersed Generation to Regulate Distribution and Support Transmission Voltages},
  booktitle = {2013 {{IEEE Grenoble Conference}}},
  author = {Valverde, Gustavo and Van Cutsem, Thierry},
  year = {2013},
  month = jun,
  pages = {1--6},
  publisher = {IEEE},
  address = {Grenoble, France},
  doi = {10.1109/PTC.2013.6652119},
}

@article{el-hawary1987,
  title = {Incorporation of Load Models in Load-Flow Studies: {{Form}} of Model Effects},
  author = {{El-Hawary}, M. E. and Dias, L. G.},
  year = 1987,
  month = jan,
  journal = {IEE Proc. C Gener. Transm. Distrib.},
  volume = {134},
  number = {1},
  pages = {27--30},
  issn = {01437046},
  doi = {10.1049/ip-c.1987.0004},
  urldate = {2025-04-05},
  langid = {english}
}

@article{ma2014a,
  title = {Adaptive Coordinated Voltage Control---{{Part II}}: {{Use}} of Learning for Rapid Response},
  shorttitle = {Adaptive Coordinated Voltage Control---Part Ii},
  author = {Ma, Haomin and Hill, David J.},
  year = {2014},
  month = jul,
  journal = {IEEE Trans. Power Syst.},
  volume = {29},
  number = {4},
  pages = {1554--1561},
  issn = {1558-0679},
  doi = {10.1109/TPWRS.2013.2293572},
}

@article{vournas2017,
  title = {Voltage Stability Monitoring from a Transmission Bus {{PMU}}},
  author = {Vournas, Costas D. and Lambrou, Charalambos and Mandoulidis, Panagiotis},
  year = {2017},
  month = jul,
  journal = {IEEE Trans. Power Syst.},
  volume = {32},
  number = {4},
  pages = {3266--3274},
  publisher = {IEEE},
  issn = {08858950},
  doi = {10.1109/TPWRS.2016.2629495},
}

@article{kraiczy2018,
  title = {Parallel Operation of Transformers with on Load Tap Changer and Photovoltaic Systems with Reactive Power Control},
  author = {Kraiczy, Markus and Stetz, Thomas and Braun, Martin},
  year = {2018},
  month = nov,
  journal = {IEEE Trans. Smart Grid},
  volume = {9},
  number = {6},
  pages = {6419--6428},
  issn = {1949-3061},
  doi = {10.1109/TSG.2017.2712633},
}

@techreport{zotero-1137,
  title = {Final Report: {{System}} Disturbance on 4 {{November}} 2006},
  year = {2007},
  month = jan,
  pages = {1--85},
  address = {Brussels, Belgium},
  institution = {UCTE},
  langid = {english},
}

@standard{IEEEStd2018,
  title         = "IEEE standard for interconnection and interoperability of distributed energy resources with associated electric power systems interface",
  organization  = "IEEE",
  address       = "Piscataway, NJ, USA",
  number        = "1547-2018",
  year          = "2018",
  month         = Feb,
}

@inproceedings{vournas2001,
  title = {Emergency Tap-Blocking to Prevent Voltage Collapse},
  booktitle = {2001 {{IEEE Porto PowerTech}}},
  author = {Vournas, C.D. and Manos, G.A.},
  year = 2001,
  month = sep,
  volume = {2},
  pages = {1--5},
  publisher = {IEEE},
  address = {Porto, Portugal},
  doi = {10.1109/PTC.2001.964741},
  urldate = {2025-12-10},
}

@misc{andersen2022,
  title = {{{CVXOPT}}: {{Python}} Software for Convex Optimization},
  author = {Andersen, Martin and Dahl, Joachim and Vandenberghe, Lieven},
  year = {2022},
  month = mar
}

@techreport{2019,
  title = {Reliability Guideline: {{Parameterization}} of the {{DER}}\_{{A}} Model},
  year = {2019},
  month = sep,
  pages = {32 pp.},
  address = {Atlanta, GA, USA},
  institution = {North American Electric Reliability Corporation (NERC)},
  langid = {english},
}

@inproceedings{fabozzi2011,
  title = {On Simplified Handling of State Events in Time-Domain Simulation},
  booktitle = {17th {{Power Systems Computation Conference}}},
  author = {Fabozzi, Davide and Chieh, Angela S. and Panciatici, Patrick and Van Cutsem, Thierry},
  year = {2011},
  month = aug,
  pages = {1--9},
  address = {Stockholm, Sweden},
}

@standard{IEEEStd2012,
  title         = "IEEE standard requirements for tap changers",
  organization  = "IEEE",
  address       = "Piscataway, NJ, USA",
  number        = "C57.131",
  year          = "2012",
  month         = May,
}

@article{antoniadou-plytaria2017,
  title = {Distributed and Decentralized Voltage Control of Smart Distribution Networks: {{Models}}, Methods, and Future Research},
  author = {{Antoniadou-Plytaria}, Kyriaki E. and {Kouveliotis-Lysikatos}, Iasonas N. and Georgilakis, Pavlos S. and Hatziargyriou, Nikos D.},
  year = 2017,
  month = nov,
  journal = {IEEE Trans. Smart Grid},
  volume = {8},
  number = {6},
  pages = {2999--3008},
  publisher = {IEEE},
  issn = {1949-3053},
  doi = {10.1109/TSG.2017.2679238}
}

@inproceedings{soleimanibidgoli2016,
  title = {Receding-Horizon Control of Distributed Generation to Correct Voltage or Thermal Violations and Track Desired Schedules},
  booktitle = {2016 {{Power Systems Computation Conference}} ({{PSCC}})},
  author = {Soleimani Bidgoli, Hamid and Glavic, Mevludin and Van Cutsem, Thierry},
  year = {2016},
  month = jun,
  pages = {1--8},
  publisher = {IEEE},
  address = {Genoa, Italy},
  doi = {10.1109/PSCC.2016.7540818},
}

@article{arif2018,
  title = {Load Modeling---{{A}} Review},
  author = {Arif, Anmar and Wang, Zhaoyu and Wang, Jianhui and Mather, Barry and Bashualdo, Hugo and Zhao, Dongbo},
  year = 2018,
  month = nov,
  journal = {IEEE Trans. Smart Grid},
  volume = {9},
  number = {6},
  pages = {5986--5999},
  issn = {1949-3061},
  doi = {10.1109/TSG.2017.2700436},
  urldate = {2025-04-05}
}

@article{escobar2025,
  title = {Data-Driven Participation of Active Distribution Networks in Transmission Voltage Control},
  author = {Escobar, Francisco and Pierrou, Georgia and Valverde, Gustavo and Hug, Gabriela},
  year = 2025,
  month = sep,
  journal = {Sustainable Energy, Grids and Networks},
  volume = {43},
  pages = {101906},
  issn = {23524677},
  doi = {10.1016/j.segan.2025.101906},
  urldate = {2025-12-29},
  langid = {english}
}

@article{hatziargyriou2021,
  title = {Definition and Classification of Power System Stability -- {{Revisited}} \& Extended},
  author = {Hatziargyriou, Nikos and Milanovic, Jovica and Rahmann, Claudia and Ajjarapu, Venkataramana and Canizares, Claudio and Erlich, Istvan and Hill, David and Hiskens, Ian and Kamwa, Innocent and Pal, Bikash and Pourbeik, Pouyan and {Sanchez-Gasca}, Juan and Stankovic, Aleksandar and Van Cutsem, Thierry and Vittal, Vijay and Vournas, Costas},
  year = {2021},
  month = jul,
  journal = {IEEE Trans. Power Syst.},
  volume = {36},
  number = {4},
  pages = {3271--3281},
  publisher = {IEEE},
  issn = {0885-8950},
  doi = {10.1109/TPWRS.2020.3041774},
}

@article{valverde2013,
  title = {Model Predictive Control of Voltages in Active Distribution Networks},
  author = {Valverde, Gustavo and Van Cutsem, Thierry},
  year = {2013},
  month = dec,
  journal = {IEEE Trans. Smart Grid},
  volume = {4},
  number = {4},
  pages = {2152--2161},
  publisher = {IEEE},
  issn = {19493053},
  doi = {10.1109/TSG.2013.2246199},
}

@techreport{zotero-2267,
  title = {Voltage Stability of Power Systems: {{Concepts}}, Analytical Tools, and Industry Experience},
  year = {1990},
  month = aug,
  number = {90TH0358-2-PWR},
  pages = {1--189},
  address = {Piscataway, NJ, USA},
  institution = {{IEEE Power and Energy Society}},
  urldate = {2022-12-22},
}

@article{cai2020,
  title = {A Data-Based Learning and Control Method for Long-Term Voltage Stability},
  author = {Cai, Huaxiang and Ma, Haomin and Hill, David J.},
  year = {2020},
  month = jul,
  journal = {IEEE Trans. Power Syst.},
  volume = {35},
  number = {4},
  pages = {3203--3212},
  issn = {1558-0679},
  doi = {10.1109/TPWRS.2020.2967434},
}

@article{escobar2020,
  title = {A Combined High-, Medium-, and Low-Voltage Test System for Stability Studies with {{DERs}}},
  author = {Escobar, Francisco and Garc{\'i}a, Jorge and V{\'i}quez, Juan M. and Valverde, Gustavo and Aristidou, Petros},
  year = {2020},
  month = dec,
  journal = {Electric Power Systems Research},
  volume = {189},
  pages = {106671},
  publisher = {Elsevier},
  issn = {03787796},
  doi = {10.1016/j.epsr.2020.106671},
}

@article{vournas2008,
  title = {Local Identification of Voltage Emergency Situations},
  author = {Vournas, C.D. and Van Cutsem, Thierry},
  year = {2008},
  month = aug,
  journal = {IEEE Trans. Power Syst.},
  volume = {23},
  number = {3},
  pages = {1239--1248},
  publisher = {IEEE},
  issn = {0885-8950},
  doi = {10.1109/TPWRS.2008.926425},
}

@inbook{soleimanibidgoli2018,
  title = {Operation of Distribution Systems within Secure Limits Using Real-Time Model Predictive Control},
  booktitle = {Dynamic {{Vulnerability Assessment}} and {{Intelligent Control}} for {{Sustainable Power Systems}}},
  author = {Soleimani Bidgoli, Hamid and Valverde, Gustavo and Aristidou, Petros and Glavic, Mevludin and Van Cutsem, Thierry},
  year = {2018},
  month = jan,
  pages = {283--309},
  publisher = {John Wiley \& Sons},
  address = {Chichester, UK},
  doi = {10.1002/9781119214984.ch14},
  urldate = {2022-08-28},
  collaborator = {{Rueda-Torres}, Jos{\'e} Luis and {Gonz{\'a}lez-Longatt}, Francisco},
  langid = {english},
}

@inproceedings{jaramillo2023,
  title = {Coordinated Control of Load Tap Changer Transformers for Voltage Regulation and Voltage Hunting Prevention: {{A}} Switched Systems Approach},
  shorttitle = {Coordinated Control of Load Tap Changer Transformers for Voltage Regulation and Voltage Hunting Prevention},
  booktitle = {2023 62nd {{IEEE Conf}}. {{Decis}}. {{Control CDC}}},
  author = {Jaramillo, Ismael and {Mercado-Uribe}, {\'A}ngel and Schiffer, Johannes},
  year = 2023,
  month = dec,
  pages = {8553--8558},
  publisher = {IEEE},
  address = {Singapore, Singapore},
  doi = {10.1109/CDC49753.2023.10383763},
  urldate = {2024-06-03},
  langid = {english},
}

@techreport{2025,
  title = {Grid Incident in {{South-East Europe}} on 21 {{June}} 2024: {{Final}} Report},
  shorttitle = {Grid Incident in South-East Europe on 21 June 2024},
  year = 2025,
  month = feb,
  institution = {ICS Investigation Expert Panel},
  urldate = {2025-04-22},
  langid = {english}
}

@article{prionistis2021,
  title = {Voltage Stability Support Offered by Active Distribution Networks},
  author = {Prionistis, Giorgos and Souxes, Theodoros and Vournas, Costas},
  year = {2021},
  month = jan,
  journal = {Electric Power Systems Research},
  volume = {190},
  pages = {106728},
  publisher = {Elsevier},
  issn = {03787796},
  doi = {10.1016/j.epsr.2020.106728},
}

@article{vancutsem2020,
  title = {Test Systems for Voltage Stability Studies},
  author = {Van Cutsem, Thierry and Glavic, Mevludin and Rosehart, William and Canizares, Claudio and Kanatas, Marios and Lima, Leonardo and Milano, Federico and Papangelis, Lampros and Ramos, Rodrigo Andrade and Santos, Jhonatan Andrade Dos and Tamimi, Behnam and Taranto, Glauco and Vournas, Costas},
  year = {2020},
  month = sep,
  journal = {IEEE Trans. Power Syst.},
  volume = {35},
  number = {5},
  pages = {4078--4087},
  publisher = {IEEE},
  issn = {0885-8950},
  doi = {10.1109/TPWRS.2020.2976834},
}

@article{milanovic2013,
  title = {International Industry Practice on Power System Load Modeling},
  author = {Milanovic, Jovica V. and Yamashita, Koji and Mart{\'i}nez Villanueva, Sergio and Djokic, Sasa {\v Z}. and Korunovi{\'c}, Lidija M.},
  year = {2013},
  month = aug,
  journal = {IEEE Trans. Power Syst.},
  volume = {28},
  number = {3},
  pages = {3038--3046},
  issn = {1558-0679},
  doi = {10.1109/TPWRS.2012.2231969},
}

@article{aristidou2016,
  title = {Power System Dynamic Simulations Using a Parallel Two-Level {{Schur-complement}} Decomposition},
  author = {Aristidou, Petros and Lebeau, Simon and Van Cutsem, Thierry},
  year = {2016},
  month = sep,
  journal = {IEEE Trans. Power Syst.},
  volume = {31},
  number = {5},
  pages = {3984--3995},
  publisher = {IEEE},
  issn = {08858950},
  doi = {10.1109/TPWRS.2015.2509023},
}

@article{lara2024,
  title = {Revisiting Power Systems Time-Domain Simulation Methods and Models},
  author = {Lara, Jose Daniel and {Henriquez-Auba}, Rodrigo and Ramasubramanian, Deepak and Dhople, Sairaj and Callaway, Duncan S. and Sanders, Seth},
  year = 2024,
  month = mar,
  journal = {IEEE Trans. Power Syst.},
  volume = {39},
  number = {2},
  pages = {2421--2437},
  issn = {0885-8950, 1558-0679},
  doi = {10.1109/TPWRS.2023.3303291},
  urldate = {2024-03-22},
  langid = {english}
}

@article{sun2019,
  title = {Review of Challenges and Research Opportunities for Voltage Control in Smart Grids},
  author = {Sun, Hongbin and Guo, Qinglai and Qi, Junjian and Ajjarapu, Venkataramana and Bravo, Richard and Chow, Joe and Li, Zhengshuo and Moghe, Rohit and {Nasr-Azadani}, Ehsan and Tamrakar, Ujjwol and Taranto, Glauco N. and Tonkoski, Reinaldo and Valverde, Gustavo and Wu, Qiuwei and Yang, Guangya},
  year = 2019,
  month = jul,
  journal = {IEEE Trans. Power Syst.},
  volume = {34},
  number = {4},
  pages = {2790--2801},
  publisher = {IEEE},
  issn = {0885-8950},
  doi = {10.1109/TPWRS.2019.2897948}
}

@inproceedings{prionistis2022a,
  title = {Using Active Distribution Network Flexibility to Increase Transmission System Voltage Stability Margins},
  booktitle = {11th {{Bulk Power Systems Dynamics}} and {{Control Symposium}} ({{IREP}} 2022)},
  author = {Prionistis, Giorgos and Vournas, Costas},
  year = {2022},
  month = jul,
  eprint = {2208.08920},
  primaryclass = {cs, eess},
  pages = {1--12},
  address = {Banff, Canada},
  doi = {10.48550/arXiv.2208.08920},
  urldate = {2022-09-10},
  archiveprefix = {arXiv},
  langid = {english},
}

@inproceedings{liemann2019,
  title = {Impact of Varying Shares of Distributed Energy Resources on Voltage Stability in Electric Power Systems},
  booktitle = {2019 {{IEEE Milan PowerTech}}},
  author = {Liemann, Sebastian and Robitzky, Lena and Rehtanz, Christian},
  year = {2019},
  month = jun,
  pages = {1--6},
  publisher = {IEEE},
  address = {Milan, Italy},
  doi = {10.1109/PTC.2019.8810761},
}

@inproceedings{escobar2023a,
  title = {Predictive Control of {{TN-DN}} Boundary Bus Voltages with Long-Term Stability Constraints},
  booktitle = {2023 {{IEEE Power}} \& {{Energy Society Innovative Smart Grid Technologies Conference}} ({{ISGT}})},
  author = {Escobar, Francisco and Valverde, Gustavo},
  year = {2023},
  month = jan,
  pages = {1--5},
  address = {Washington, D.C., USA},
  issn = {2472-8152},
  doi = {10.1109/ISGT51731.2023.10066454},
  langid = {english},
}

@article{capitanescu2009,
  title = {Decentralized Tap Changer Blocking and Load Shedding against Voltage Instability: {{Prospective}} Tests on the {{RTE}} System},
  author = {Capitanescu, F. and Otomega, B. and Lefebvre, H. and Sermanson, V. and Van Cutsem, T.},
  year = 2009,
  month = oct,
  journal = {Int. J. Electr. Power Energy Syst.},
  volume = {31},
  number = {9},
  pages = {570--576},
  publisher = {Elsevier},
  issn = {01420615},
  doi = {10.1016/j.ijepes.2009.03.025},
}

@inproceedings{riaz2019,
  title = {On Feasibility and Flexibility Operating Regions of Virtual Power Plants and {{TSO}}/{{DSO}} Interfaces},
  booktitle = {2019 {{IEEE Milan PowerTech}}},
  author = {Riaz, Shariq and Mancarella, Pierluigi},
  year = 2019,
  month = jun,
  pages = {1--6},
  publisher = {IEEE},
  address = {Milan, Italy},
  doi = {10.1109/PTC.2019.8810638},
  urldate = {2024-08-28}
}

@techreport{wirth2021,
  title = {Recent Facts about Photovoltaics in {{Germany}}},
  author = {Wirth, Harry},
  year = {2021},
  month = may,
  address = {Freiburg, Germany},
  institution = {Fraunhofer Institute for Solar Energy Systems},
  langid = {english},
}

@article{escobar2022,
  title = {Coordination of {{DERs}} and Flexible Loads to Support Transmission Voltages in Emergency Conditions},
  author = {Escobar, Francisco and V{\'i}quez, Juan Manuel and Garc{\'i}a, Jorge and Aristidou, Petros and Valverde, Gustavo},
  year = {2022},
  month = jul,
  journal = {IEEE Trans. Sustain. Energy},
  volume = {13},
  number = {3},
  pages = {1344--1355},
  publisher = {IEEE},
  issn = {19493037},
  doi = {10.1109/TSTE.2022.3154716},
}

@article{morin2018,
  title = {Coordinated Control of Active Distribution Networks to Help a Transmission System in Emergency Situation},
  author = {Morin, J. and Colas, F. and Dieulot, J. Y. and Grenard, S. and Guillaud, X.},
  year = {2018},
  month = apr,
  journal = {Energy Systems},
  volume = {9},
  pages = {831--851},
  publisher = {Springer},
  issn = {1868-3967},
  doi = {10.1007/s12667-018-0287-7},
}

@article{robitzky2018a,
  title = {Agent-based Identification and Control of Voltage Emergency Situations},
  author = {Robitzky, Lena and Weckesser, Tilman and H{\"a}ger, Ulf and Rehtanz, Christian and Van Cutsem, Thierry},
  year = {2018},
  month = mar,
  journal = {IET Generation, Transmission \& Distribution},
  volume = {12},
  number = {6},
  pages = {1446--1454},
  publisher = {IET},
  issn = {1751-8695},
  doi = {10.1049/iet-gtd.2017.1167},
}

@article{lambrou2021,
  title = {Validation of Voltage Instability Detection and Control Using a Real Power System Incident},
  author = {Lambrou, Charalambos and Mandoulidis, Panagiotis and Vournas, Costas},
  year = {2021},
  month = nov,
  journal = {Energies},
  volume = {14},
  number = {21},
  pages = {7165},
  publisher = {MDPI},
  issn = {19961073},
  doi = {10.3390/en14217165},
}

@article{tran2022a,
  title = {Sparse Identification for Model Predictive Control to Support Long-term Voltage Stability},
  author = {Tran, Minh-Quan and Tran, Trung Thai and Nguyen, Phuong H. and Pemen, Guus},
  year = {2022},
  month = nov,
  journal = {IET Generation Trans \& Dist},
  pages = {1--13},
  issn = {1751-8687, 1751-8695},
  doi = {10.1049/gtd2.12662},
  urldate = {2022-12-30},
  langid = {english},
}

@article{pabonospina2020,
  title = {Power Factor Improvement by Active Distribution Networks during Voltage Emergency Situations},
  author = {Pab{\'o}n Ospina, Luis David and Van Cutsem, Thierry},
  year = {2020},
  month = dec,
  journal = {Electric Power Systems Research},
  volume = {189},
  pages = {106771},
  publisher = {Elsevier},
  issn = {03787796},
  doi = {10.1016/j.epsr.2020.106771},
}

@article{pilatte2019,
  title = {{{TDNetgen}}: {{An}} Open-Source, Parametrizable, Large-Scale, Transmission, and Distribution Test System},
  author = {Pilatte, Nicolas and Aristidou, Petros and Hug, Gabriela},
  year = {2019},
  month = mar,
  journal = {IEEE Systems Journal},
  volume = {13},
  number = {1},
  eprint = {1706.01656},
  pages = {729--737},
  publisher = {IEEE},
  issn = {19379234},
  doi = {10.1109/JSYST.2017.2772914},
  archiveprefix = {arXiv},
}

@article{kroposki2020,
  title = {Autonomous Energy Grids: {{Controlling}} the Future Grid with Large Amounts of Distributed Energy Resources},
  author = {Kroposki, Benjamin and Bernstein, Andrey and King, Jennifer and Vaidhynathan, Deepthi and Zhou, Xinyang and Chang, Chin-Yao and Dall'Anese, Emiliano},
  year = {2020},
  month = nov,
  journal = {IEEE Power and Energy Magazine},
  volume = {18},
  number = {6},
  pages = {37--46},
  publisher = {IEEE},
  issn = {1540-7977},
  doi = {10.1109/MPE.2020.3014540},
}

@article{aristidou2017,
  title = {Contribution of Distribution Network Control to Voltage Stability: {{A}} Case Study},
  author = {Aristidou, Petros and Valverde, Gustavo and Van Cutsem, Thierry},
  year = {2017},
  month = jan,
  journal = {IEEE Trans. Smart Grid},
  volume = {8},
  number = {1},
  pages = {106--116},
  publisher = {IEEE},
  issn = {1949-3053},
  doi = {10.1109/TSG.2015.2474815},
}

\end{document}